\documentclass[superscriptaddress,pra,nofootinbib]{revtex4}

\usepackage{graphicx}
\usepackage{amssymb,amsmath}
\usepackage[utf8]{inputenc}
\usepackage{multirow}
\usepackage{xcolor}
\usepackage{hyperref}
\hypersetup{colorlinks=true,linkcolor=blue,anchorcolor=blue,citecolor=magenta,filecolor=blue,urlcolor=blue,bookmarksnumbered=true}
\usepackage[flushleft]{threeparttable}
\usepackage{mathtools}

\usepackage{amsmath}

\usepackage{dsfont}
\usepackage{amsthm}

\usepackage{enumitem}
\setlist[itemize]{leftmargin=10mm}

\newtheorem{innercustomgeneric}{\customgenericname}
\providecommand{\customgenericname}{}
\newcommand{\newcustomtheorem}[2]{%
  \newenvironment{#1}[1]
  {%
   \renewcommand\customgenericname{#2}%
   \renewcommand\theinnercustomgeneric{##1}%
   \innercustomgeneric
  }
  {\endinnercustomgeneric}
}

\newcustomtheorem{customthm}{Theorem}
\newcustomtheorem{customlemma}{Lemma}

\begin{document}
\title{Multi-nucleon structure and dynamics via quantum computing}
\author{Weijie Du}
\email[Email:]{{\ }duweigy@gmail.com}
\affiliation{Department of Physics and Astronomy, Iowa State University, Ames, Iowa 50010, USA}
\author{James P. Vary}
\affiliation{Department of Physics and Astronomy, Iowa State University, Ames, Iowa 50010, USA}
\date{\today}

\begin{abstract}
We propose a framework for computing the structure and dynamics for second-quantized many-nucleon Hamiltonians on quantum computers. 
We develop an oracle-based Hamiltonian input model that computes the many-nucleon states and nonzero Hamiltonian matrix elements of the many-nucleon system. 
With our Fock-state based input model, we show how to implement the sparse matrix simulation algorithms to calculate the dynamics of the second-quantized many-nucleon Hamiltonian. 
Based on the dynamics simulation methods, we also present the methodology for structure calculations of the many-nucleon system. 
In this work, we provide an explicit circuit design of our input model of the second-quantized Hamiltonian within a direct encoding scheme that maps the occupation of each available single-particle state in the many-nucleon state to the state of specific qubit in a quantum register. 
We analyze our method and provide the asymptotic cost in computing resources for structure and dynamics calculations of many-nucleon systems. 
For pedagogical purposes, we demonstrate our input model with two model problems in restricted model spaces.

\end{abstract}

\maketitle

\vspace{.25in}

\section{Introduction}

{\it Ab initio} calculations are powerful tools for investigating the structure (such as the spectrum and electromagnetic moments) and dynamics (such as the transition probabilities and scattering cross sections) of many-nucleon systems \cite{Carlson:1997qn,Carlson:2014vla,Carlson:2017ebk}. In contrast with many-electron systems that reside in an external potential and that are governed by the long-range Coulomb force, many-nucleon systems are self-bound and governed by multiscale interactions: the strong nuclear interaction at short range, and the electromagnetic interaction between the protons at long range. Constituent nucleons can also interact with external probes via various types of interactions, e.g., the electroweak interaction. Hence, the properties and dynamics of many-nucleon systems are complex and challenging.

Precision {\it ab initio} investigations of many-nucleon systems provide predictive power that complements experiments. However, {\it ab initio} calculations exploit the quantum many-body framework and are numerically demanding as large Hilbert space dimensions are required to accurately account for the multiple scales involved. Indeed, the required Hilbert space dimension scales exponentially with the system size, which makes the {\it ab initio} calculations intractable on world-leadership supercomputers even for simple systems with more than a few nucleons.

Quantum computers exploit the principles of quantum mechanics to avoid the exponential scaling in computing resources \cite{Feynman:1982fey,NielsenANDChuang:2001}. Hence, quantum computing techniques appear to offer a promising cure for the dimensionality curse in {\it ab initio} investigations of many-nucleon systems. 
To date, prototype nuclear many-body calculations have been performed on real-world quantum hardwares \cite{Dumitrescu:2018njn,Roggero:2020sgd,Kiss:2022kkz}.
Various quantum algorithms \cite{Klco:2021lap,Du:2020glq,Baroni:2021xtl,Stetcu:2021cbj,Romero:2022blx} have been proposed for the structure and dymanics investigations for many-nucleon systems on near-term noisy intermediate scale quantum (NISQ) devices \cite{Preskill:2018preskill}.
Future fault-tolerant quantum hardwares promise to open up a fruitful path to revolutionize the investigations in nuclear many-body theory.

In this work, we propose a framework for studying the dynamics and structure of many-nucleon systems described by second-quantized many-nucleon Hamiltonians \cite{Barrett:2013nh,Navratil:2000ww,Navratil:2000gs} on a quantum computer. We focus on the low-energy regime, where a non-relativistic particle number conserving Hamiltonian describes the fermion system. 

The second-quantized many-nucleon Hamiltonian can be expressed as a linear combination of monomials of the ladder operators. For realistic calculations, one performs truncations on the monomials and retains only up to few-body terms in the Hamiltonian. Meanwhile, the retained monomials are further restricted according to the symmetries of the many-nucleon system. Due to such truncations and symmetry restrictions, the second-quantized many-nucleon Hamiltonian is sparse, which makes it appropriate for the applications of the sparse matrix simulation methods \cite{Aharonov:2003aha,Childs:2003am,Berry:2007dwb,AMChilds:2009,DWBerry:2012,AMChilds:2013,Berry:2015prlDWB,Berry:2015IEEE,Low:2017,Low:2019,Low:2018IntPic,Berry:2020}. 

One key question in applying such sparse matrix simulation methods \cite{Aharonov:2003aha,Childs:2003am,Berry:2007dwb,AMChilds:2009,DWBerry:2012,AMChilds:2013,Berry:2015prlDWB,Berry:2015IEEE,Low:2017,Low:2019,Low:2018IntPic,Berry:2020} to the second-quantized Hamiltonian is the appropriate Hamiltonian input model for the second-quantized Hamiltonian \cite{Kirby:2021ajp}. In contrast to the input models proposed for the sparse Hamiltonian matrices, which access the nonvanishing matrix elements via their row and column indices, one deals with the Fock states within the formalism of second quantization. Indeed, the adoption of the sparse matrix simulation methods to the second-quantization formalism necessitates development of new Hamiltonian input models that operate on Fock states directly in order to perform corresponding structure and dynamics calculations.

\begin{figure*}[ht] 
  \centering
  \includegraphics[width=0.30\textwidth]{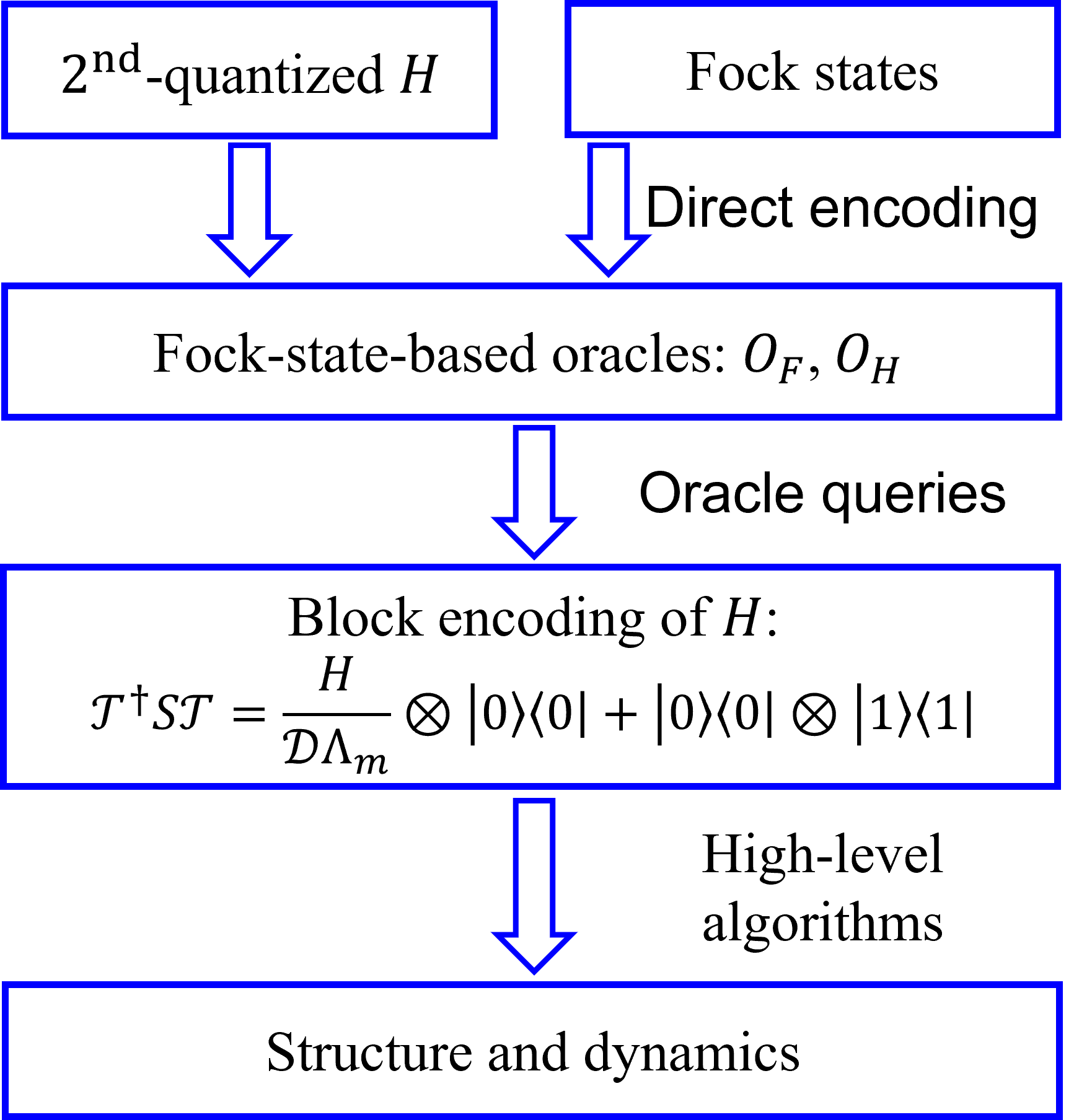}
  \caption{(color online) Sketch of the algorithmic framework. The second-quantized Hamiltonian $H$ of the many-nucleon system is formulated in Eq. \eqref{eq:H_total}. The identity of the block encoding is discussed in Eq. \eqref{eq:TST_relation}. The oracles $O_{\rm F}$ and $O_{\rm H}$ are defined in Eqs. \eqref{eq:O_F_prime} and \eqref{eq:O_H_prime}, respectively.
  } 
  \label{fig:workflow}
\end{figure*}

We develop an oracle-based Hamiltonian input model that is suitable for simulating second-quantized Hamiltonians (see the algorithmic framework in Fig. \ref{fig:workflow}). We adopt the direct encoding (DE) scheme (see details in Sec. \ref{sec:DirectEncoding}) which maps the many-nucleon (Fock) states to binary strings in the quantum registers. Our input model operates directly on the Fock states, or the corresponding binary strings. The oracles of our input model compute 1) the output Fock state based on the corresponding input Fock state; 2) the active single-nucleon bases that are first annhilated and then created in the input Fock state to form the output state; and 3) the few-nucleon kernel based on the active single-nucleon bases, where the kernel eventually contributes to the Hamiltonian matrix element. We achieve this oracle design by incorporating the monomials of the ladder operators of the Hamiltonian, whereby the symmetries of the Hamiltonian are encoded in the oracles. Using oracle queries, we can construct a block encoding of the Hamiltonian \cite{Low:2019,Chakraborty:2018,Lin:2022lectureNote}.

Based on our input model, high-level sparse-matrix simulation algorithms with optimal query complexities can be implemented to simulate the second-quantized many-nucleon Hamiltonians. The structure calculations share the circuit construction of the same time-evolution unitary as that in the simulation algorithms. Therefore, such Hamiltonian simulation algorithms can also be adapted to the structure calculations of many-nucleon system. We analyze the cost of our method for the dynamics and structure calculations for many-nucleon systems, based on the query complexities of the simulation algorithms and the cost of the oracles.

For pedagogical purposes, we illustrate our method with two model problems in restricted model spaces. Whereas the sparse-matrix algorithms are presented extensively in the literature \cite{Aharonov:2003aha,Childs:2003am,Berry:2007dwb,AMChilds:2009,DWBerry:2012,AMChilds:2013,Berry:2015prlDWB,Berry:2015IEEE,Low:2017,Low:2019,Low:2018IntPic,Berry:2020}, we focus on demonstrating the design of our input model. As a future research effort, we will perform dynamics and structure calculations via our method for simple many-nucleon systems with realistic inter-nucleon interactions within restricted model spaces. 

Our work complements the work by Kirby {\it et al.} \cite{Kirby:2021ajp}. We share the theme of developing efficient and precise quantum algorithms for second-quantized Hamiltonians. The following features distinguish our work:  1) different target Hamiltonians and single-particle basis representations; 2) different encoding schemes; and 3) different designs of the input model. While Ref. \cite{Kirby:2021ajp} employs the compact encoding (CE) scheme (see details in Sec. \ref{sec:DirectEncoding}) and log-local operations for controlled arithmetic calculations to improve the gate complexity and qubit cost, which is preferable for the simulations of quantum field theory on long-term quantum computers, we adopt the DE scheme for the problems with fixed particle number and species and design our input model based on primitive gate operations, aiming for straightforward prototype  nuclear structure and dynamics calculations on NISQ devices. With further development, our method can be applied to second-quantized Hamiltonians for systems with bosons and fermions, and with particle creations and annihilations.

This work is organized as follows. 
In Sec. \ref{sec:section_II}, we introduce the elements of many-nucleon calculations, which include the many-nucleon Hamiltonian and our choice of basis. 
In Sec. \ref{sec:DirectEncoding}, we discuss the encoding scheme for many-nucleon calculations on quantum computers. 
In Sec. \ref{sec:section_IV}, we present algorithms for solving the structure and dynamics problems of second-quantized many-nucleon Hamiltonians based on our oracle-based input model that treats the Fock states directly. We also discuss the query complexity of our algorithms. 
In Sec. \ref{sec:oracle_designs}, we show the design of our oracles within the DE scheme, where we also analyze their gate and qubit costs. 
Combined with the query complexity, we present the asymptotic qubit and gate cost of our algorithms in Sec. \ref{sec:resource_section}, where we compare our algorithm with those for simulating a molecular Hamiltonian.
We illustrate our method with two model problems in Sec. \ref{sec:applications}. 
We conclude in Sec. \ref{sec:summary_and_outlook}, where we also provide an outlook.

\section{Many-nucleon system}
\label{sec:section_II}
In this section, we discuss the Hamiltonian of the many-nucleon system. To facilitate the design of our input model, we reformulate the second-quantized many-nucleon Hamiltonian. Finally, we discuss the constructions of the single-particle (SP) basis and the many-nucleon bases.

\subsection{Many-nucleon Hamiltonian}
\subsubsection{First quantization}
The Hamiltonian of the $A$-nucleon system $(A\geq 2)$, which acts only on the internal degrees of freedom, reads \cite{Barrett:2013nh,Navratil:2000ww,Navratil:2000gs}:
\begin{align}
H_A = T_{\rm rel } + V = \sum _{i<j}^A \frac{( \vec{p}_i -\vec{p}_j )^2}{2m_NA} + V_{\rm NN} + V_{\rm NNN} + \cdots , \label{eq:total_case_H_A}
\end{align}
where $m_N$ denotes the nucleon mass, $T_{\rm rel}$ denotes the kinetic energy, and $\vec{p}_i$ denotes the momentum of the $i^{\rm th}$ nucleon. The inter-nucleon interaction $V$ consists of the two-nucleon interaction $ V_{\rm NN} $, the three-nucleon interaction $V_{\rm NNN}$, etc. For the purpose of demonstration, we restrict our discussion to the two-nucleon interaction $V_{NN} $ and ignore the many-nucleon interactions in this work. Generalization of the formalism in this work to the applications with many-nucleon interactions is straightforward.

A popular choice of the SP basis employed in nuclear physics is the three-dimensional harmonic oscillator basis (3DHO) (see, e.g., Ref. \cite{Barrett:2013nh} and references therein). This basis, with appropriate many-body truncation, enables an exact factorization of an eigenfunction of the nuclear system into the “intrinsic” and “center of mass” (CM) components in order to preserve Galilean invariance. With this advantage, the spurious CM excitation due to the adoption of an SP basis can be pushed higher than the physically interesting spectrum by employing a Lipkin-Lawson Lagrange multiplier term \cite{Lipkin:1958zza,Gloeckner:1974sst}. Overall, we have the total Hamiltonian as\footnote{We adopt natural units in this work and take $\hbar = c =1$.}
\begin{align}
H = H_A + \Lambda _{\rm CM} (H_{\rm CM} - 3 \Omega /2) \label{eq:totalHamiltonian},
\end{align}
where the second term is the Lipkin-Lawson term that penalizes the spurious CM excitation with the coefficient $\Lambda _{\rm CM} >0$. $H_{\rm CM}$ is the Hamiltonian of the CM harmonic oscillator, which is defined as
\begin{align}
H_{\rm CM} = T_{\rm CM} + U_{\rm CM} = \frac{\vec{P}^2}{2M_N} + \frac{1}{2} M_N \Omega ^2 \vec{R}^2 .
\end{align}
with $M_N = A m_N $. 
The total momentum of the $A$-nucleon system is $\vec{P} = \sum _{i=1} ^A \vec{p}_i$. $\Omega $ denotes the oscillator energy of the CM harmonic oscillator, of which the zero-point energy is $3\Omega /2$. The position vector of the mass center of the $A$-nucleon system is $\vec{R}= \frac{1}{A} \sum _{i=1}^A \vec{r}_i$. One can readily show that $[H_A, H_{\rm CM}]=0$. 
The operator $ \Lambda _{\rm CM} (H_{\rm CM} - \frac{3}{2} \Omega ) $ in Eq. \eqref{eq:totalHamiltonian} is semi-positive definite.
It penalizes the solutions with CM excitations; the spectrum of $H$ with all the states in the lowest oscillator eigenmode in the CM degree of freedom (with energy $3 \Omega /2$) corresponds to the available set of solutions for the intrinsic motion of the $A$-nucleon system.

\subsubsection{Second-quantized Hamiltonian}
The Hamiltonian $H $ [Eq. \eqref{eq:totalHamiltonian}] consists of three two-body terms [recall that we retain only the two-nucleon interaction $V_{\rm NN}$ in $H_A$ [Eq. \eqref{eq:total_case_H_A}]]. In the formalism of second quantization, we can formulate $H$ as a linear combination of the monomials of ladder operators
\begin{align}
H=  \sum _{p<q,r<s} \langle pq | H | rs \rangle a_p^{\dag} a_q^{\dag} a_s a_r \label{eq:H_total} ,
\end{align}
where $a_p^{\dag}$ and $a_q$ are the fermionic creation and annihilation operators, which obey the anti-commutation relations:
\begin{align}
\{ a_p ^{\dag} ,a_q \} = \delta _{pq}, \ \{ a_p ^{\dag} , a_q ^{\dag} \} = \{a_p , a_q \} = 0 \label{eq:anti_commutation} ,
\end{align}
with the subscript specifying the SP basis states that each specific ladder operator acts on. For example, we have
\begin{align}
a_p ^{\dag} |0 \rangle = |1 \rangle _p ,\ a_p |1 \rangle _p = |0 \rangle ,\ a_p ^{\dag} |1 \rangle _p = 0,\  a_p  |0 \rangle = 0,
\end{align}
where $|0\rangle $ denotes the vacuum for SP state (no occupation), and $|1\rangle _p$ denotes that the $p^{\rm th}$ SP state is occupied. The Hamiltonian matrix element $H_{pqrs} \coloneqq \langle pq | H | rs \rangle $ is 
\begin{align}
H_{pqrs} = T^{\rm rel}_{pqrs} + V^{\rm NN}_{pqrs} + H^{\rm CM}_{pqrs} , \label{eq:second_quantized_H_matrix_element}
\end{align}
where we have
\begin{align}
O_{pqrs} \equiv \langle pq | O  | rs \rangle = \frac{1}{2} \int  \Big[
 \phi _p^{\dag}(1) \phi _q^{\dag}(2) -(p \leftrightarrow q)
\Big] O  \Big[
 \phi _r(1) \phi _s(2) -(r \leftrightarrow s)
\Big] d\vec{r}_1 d\vec{r}_2 ,
\end{align}
with the operator $O$ representing $T_{\rm rel}$, $V_{\rm NN}$, or $(H_{\rm CM} - 3\Omega /2)$.
$\phi _{v}$ ($ v =p,q,r,s$) denotes the wave function of the ${v}^{\rm th}$ SP basis state that specified by the spatial, spin, and isospin variables. The labels ``$1$" and ``$2$" in the parentheses are the nucleon indices. Note that the integrals are over the spatial variables $\vec{r}_i$ ($i=1,\ 2$), while the matrix elements can be spin- and isospin-dependent in general.

For the input Fock state $ | \mathcal{F} \rangle $ and the output Fock state $| \mathcal{G} \rangle $, we can calculate the Hamiltonian matrix element with Eq. \eqref{eq:H_total} as
\begin{align}
\langle \mathcal{G} | H | \mathcal{F} \rangle = \sum _{p<q,r<s} \langle pq | H | rs \rangle  \langle \mathcal{G} | a_p^{\dag} a_q^{\dag} a_s a_r | \mathcal{F} \rangle . \label{eq:concreteME}
\end{align}

\subsubsection{Modified second-quantized Hamiltonian for quantum computing}

We now introduce our modification to the second-quantized Hamiltonian in order to facilitate the design of our Hamiltonian input model in quantum computing. The basic idea is to introduce tags to specify the monomials and their respective coefficients in Eq. \eqref{eq:H_total}. To this end, we first rewrite Eq. \eqref{eq:H_total} as
\begin{align}
{H} = \sum _{P} \sum _{Q} H(Q,P) b_Q^{\dag} b_P , \label{eq:modification_step1}
\end{align} 
where the tag $P \mapsto \{r,s\}$ (with $r<s$) is an integer that indexes the pair of SP bases (order sensitive) to be annihilated, while the tag $Q \mapsto \{p,q\}$ (with $p<q$) is an integer that indexes the pair of SP bases (order sensitive) to be created. 
{The tags $P$ and $Q$ can be taken as positive integers, and we require that $ P = Q $ if and only if $r=p$ and $s=q$.}
We use the tags to specify the pairwise creation operators and annihilation operators as $b_Q^{\dag} = a_p^{\dag} a_q^{\dag} $, $ b_P = a_s a_r $ such that $ b_Q^{\dag} b_P = a_p^{\dag} a_q^{\dag}  a_s a_r $. Correspondingly, $P$ and $Q$ are also employed to index the matrix element (two-nucleon kernel in this work)
\begin{align}
H(Q,P)=  \langle pq | H | rs \rangle .
\end{align}
We note that Eq. \eqref{eq:modification_step1} is equivalent to Eq. \eqref{eq:H_total}. 

Next, we attach (decorate) each term in the right-hand-side of Eq. \eqref{eq:modification_step1} with an additional selection operator ($ |Q \rangle \langle P | $) constructed based on the tags $P$ and $Q$, and obtain
\begin{align}
\mathcal{H} = \sum _{P} \sum _{Q} H(Q,P) b_Q^{\dag} b_P \otimes |Q\rangle \langle P | . \label{eq:H_tot_modified}
\end{align} 
Note that the tags $P$ and $Q$ are encoded in the ancilla registers as the tag states $|P\rangle $ and $|Q\rangle$ in our design of input model, respectively. 
Analogous to the term-selection scheme in the algorithm of linear combination of unitary \cite{Childs:LCU2012}, we can operate on the tag states $|P\rangle $ and $|Q\rangle$ in the ancilla registers to select the corresponding monomial $ b_{Q}^{\dag} b_{P} $ (acting on the Fock states encoded in separate registers) together with the coefficient $ H(Q,P) $.
\footnote{In other words, we introduce the mapping between the tag states $ \{ |Q\rangle , \ |P \rangle \} $ and $ \{ H(Q,P) \ b_Q^{\dag} b_P \} $ in our algorithmic design. This corresponds to the decoration with the selection operator $|Q \rangle \langle P |$ in Eq. \eqref{eq:H_tot_modified}.}

We now illustrate how to compute the Hamiltonian matrix element $ \langle \mathcal{G} | H | \mathcal{F} \rangle $ [Eq. \eqref{eq:concreteME}] with $\mathcal{H}$ via a concrete example. These Fock states are constructed as the tensor products of the elements in the SP basis set, where we take the total number of the SP bases in the set to be $N_{\rm sp}$. The conclusion of this example is shown as Eq. \eqref{eq:equivalence_restore} below.

We note that only a subset of monomials 
\begin{align}
\{ b_Q^{\dag} b_P \} = \{ a_p^{\dag} a_q^{\dag} a_s a_r \ |\ p<q,\ r<s ,\ \text{and} \  0 \leq p,q,s,r \leq N_{\rm sp} -1 \} 
\end{align}
exist in the Hamiltonian according to the symmetries of the Hamiltonian (e.g., the conservation of the baryon number, parity, total angular momentum, etc.). For a pair of SP bases to be annihilated in the input Fock state, only those pairs that satisfy certain criteria can be created, such that the corresponding $ \langle \mathcal{G} | H | \mathcal{F} \rangle $ is nonzero due to the restriction of the symmetries of $H$ (see Sec. \ref{sec:applications} for examples).

We assume the total number of the distinct symmetry-preserving monomials $ b_Q^{\dag} b_P $ to be $\mathcal{D} $, and employ the index $i$ to label each distinct pair of tags as $ (Q_i, P_i ) $ with $ i \in [0,\mathcal{D}-1] $, $ P_i \mapsto \{r_i ,s_i \} $, and $ Q_i \mapsto \{p_i,q_i\} $. According to Eq. \eqref{eq:H_tot_modified}, we can write 
\begin{align}
\mathcal{H} = \sum _{i=0}^{\mathcal{D}-1} H(Q_i,P_i) b_{Q_i}^{\dag} b_{P_i} \otimes |Q_i\rangle \langle P_i | . \label{eq:H_modified_H_prime}
\end{align} 

Then, for the input Fock state $| \mathcal{F} \rangle $ with the tag state $|P_i\rangle $ and the output Fock state $| \mathcal{G} \rangle $ with the tag state $|Q_i \rangle $, the contribution of the two-nucleon matrix element with the active SP bases labeled by $P_i$ and $Q_i$ to the many-nucleon matrix element is computed as 
\begin{align}
\langle \mathcal{G}, Q_i | \mathcal{H} | \mathcal{F}, P_i \rangle = \Big( \langle \mathcal{G} | \langle Q_i | \Big) \Bigg[ \sum _{j=0}^{\mathcal{D}-1}  H(Q_j,P_j) b_{Q_j}^{\dag} b_{P_j} \otimes |Q_j\rangle \langle P_j | \Bigg] \Big( | \mathcal{F} \rangle | P_i \rangle \Big) =  H(Q_i,P_i)  \langle  \mathcal{G} | b_{Q_i}^{\dag} b_{P_i} | \mathcal{F} \rangle , \label{eq:H_modified_XXXXL}
\end{align}
where $ H(Q_i,P_i) = \langle p_i q_i | H | r_i s_i \rangle $ is the specific two-nucleon matrix element. The factor $ \langle \mathcal{G} | b_{Q_i}^{\dag} b_{P_i} | \mathcal{F} \rangle $ accounts for the weight that results from the anti-commutation relations [Eq. \eqref{eq:anti_commutation}]; it can be $\pm 1$ or 0 (e.g., when $| \mathcal{F} \rangle $ and $| \mathcal{G} \rangle $ differ for more than two SP bases in their respective compositions). 

Finally, one sums over all the $\mathcal{D}$ pairs of $ ( Q_i,P_i ) $ to enumerate the possible two-nucleon kernels that could contribute to the Hamiltonian matrix element 
\begin{align}
\sum _{i=0}^{\mathcal{D}-1} \langle \mathcal{G}, Q_i | \mathcal{H} | \mathcal{F}, P_i \rangle = \langle \mathcal{G} | H | \mathcal{F} \rangle . \label{eq:equivalence_restore}
\end{align}
This reproduces the result of Eq. \eqref{eq:concreteME} in the case with $N_{\rm sp}$ SP bases.

\subsection{SP basis and many-nucleon basis}
\label{sec:basisSpace}

``{\it Ab initio}" nuclear theory addresses the nuclear structure and the dynamics problems based on the best available inter-nucleon interactions using a quantum many-body framework that respects all the known symmetries of nuclear systems \cite{Maris:2012du}. In {\it ab initio} nuclear structure and dynamics calculations, one specifies the SP basis by a set of quantum numbers. With a set of SP bases, one constructs the many-nucleon bases (or Fock states), which enable the construction of the matrix representation of the many-nucleon Hamiltonian. With the Hamiltonian matrix constructed, the resulting eigenvalue problem is solved numerically to obtain structure observables (e.g., eigenenergies) and dynamical quantities (e.g., cross sections). Due to the nature of the quantum many-body theory, {\it ab initio} calculations are recognized to be computationally hard: the dimension of the Hilbert space increases exponentially with the number of SP bases. Even with world leadship supercomputers, current ``{\it ab initio}" nuclear structure and dynamics calculations are limited to light nuclear systems with restricted number of constituent nucleons and SP bases \cite{Abe:2021sky}. One then seeks for the quantum advantage to facilitate the {\it ab initio} investigations of increasingly complex nuclear systems.

In this work, we construct the SP basis $|\beta \rangle $ as follows. We adopt the 3DHO basis for the spatial degree of freedom. The quantum number for the 3DHO basis is the radial quantum number $n$ and the orbital angular momentum $l$. The projection of $l$ is $m_l$. The corresponding excitation quanta of 3DHO basis state is $2n + l$. Meanwhile, we have the spin part of the wave function $\chi _{S m_s}$, where the spin of the nucleon $S$ is taken to be $\frac{1}{2}$ and the spin projection $m_s$ is taken to be  $\pm \frac{1}{2}$. The total angular momentum $j$ is coupled from the orbital angular momentum $l$ and the spin $S$, with $m_j = m_l + m_s $ being the projection of $j $. Finally, we include the isospin part of the wave function $\chi _{T \tau}$, where $T =\frac{1}{2} $ for the isospin doublet (proton and neutron) and the isospin projection is $ {\tau} $ ($+1/2$ for protons and $-1/2$ for neutrons). Overall, we have the SP basis $|\beta \rangle$ to be labeled by the set of quantum numbers $\{n, l, S, j, m_{j}, T, {\tau}\}$.\footnote{Alternative choices of basis can be adopted. For example, one can adopt the momentum basis for the spatial degree of freedom, which is frequently used in the research of quantum field theories.}  In the following, we will omit the labels of the spin $S$ and isospin $T$ (which take the constant values) for brevity.

We can construct the many-nucleon basis based on the SP basis set $\mathbb{S}$. In particular, we index the $N_{\rm sp} $ SP bases in the basis set as $ \mathbb{S} = \{ |\beta _0 \rangle ,\ |\beta _1 \rangle ,\ \cdots ,\ |\beta _{N_{\rm sp}-1} \rangle \} $. A many-nucleon basis can be written as
\begin{align}
| \mathcal{F} \rangle = | \alpha _0, \alpha _1, \cdots , \alpha _{A -1} \rangle  = a^{\dag}_{\alpha _0} a^{\dag}_{\alpha _1}  \cdots a^{\dag}_{\alpha _{A -1}} | 0 \rangle ,
\end{align}
where $ | 0  \rangle $ denotes the vacuum. Each $| \alpha _i \rangle $ (with $i=0, 1, \cdots , A-1$) represents a unique and exclusive element in the set $ \mathbb{S} $ according to the Pauli principle. Meanwhile, we also require that the ordering of $ |\alpha _0 \rangle , \ | \alpha _1 \rangle ,\ \cdots ,\ | \alpha _{A -1} \rangle $ preserves the ordering of $\mathbb{S}$. This regulation of the ordering results from the anticommutation relations [Eq. \eqref{eq:anti_commutation}] and is important to retain the relative phases upon the action of the fermionic ladder operators. 

We can construct the many-nucleon bases based on the set $ \mathbb{S} $. In general, the number of the many-nucleon bases that can be constructed for an $A$-nucleon system with $ N_{\rm sp} $ available SP bases is $N_{\rm mp} = \tbinom{N_{\rm sp}}{A}$. One can also index these many-nucleon bases and obtain the many-nucleon basis set as
\begin{align}
\mathbb{M} = \{ | \mathcal{F}_0 \rangle, \ | \mathcal{F}_1 \rangle , \ \cdots , \ | \mathcal{F}_{N_{\rm mp} -1 } \rangle    \} .
\end{align}

\section{Encoding scheme}

\label{sec:DirectEncoding}

Various encoding schemes can be employed to map a many-nucleon state to a state of qubits in a quantum register. In this work, we choose to employ the DE scheme for this mapping.

In the DE scheme, we implement a quantum register containing $N_{\rm sp}$ qubits, where each qubit corresponds to a particular SP basis state in the set $ \mathbb{S} = \{ |\beta _0 \rangle ,\ |\beta _1 \rangle ,\ \cdots ,\ |\beta _{N_{\rm sp}-1} \rangle \} $ with elements fixed in order. While the locations of the qubits are fixed (this preserves the order of SP bases in the set $ \mathbb{S} $), the state of each qubit represents the occupation of the corresponding SP state in the many-nucleon state: if the SP state is occupied (vacant), the corresponding qubit is in the state $|1 \rangle $ ($ |0 \rangle $). With this scheme, a particular many-nucleon state $| \mathcal{F} \rangle $ in the set $\mathbb{M} $ is represented by a unique and order-sensitive binary string. 

The number of qubits required by the DE scheme equals the number of the SP bases, and is independent of the number of nucleons in the system. As an example, provided the SP basis set $\mathbb{S}= \{| \beta _0 \rangle, | \beta _1 \rangle,| \beta _2 \rangle,| \beta _3 \rangle,| \beta _4 \rangle,| \beta _5 \rangle,| \beta _6 \rangle,| \beta _7 \rangle  \}$ ($N_{\rm sp}=8$), it takes $N_{\rm sp} =8$ qubits for the DE scheme to encode the particular five-nucleon ($A=5$) state
\begin{align}
|\beta _1, \beta _3, \beta _4, \beta _6, \beta _7  \rangle = a^{\dag}_{\beta _1} a^{\dag}_{\beta _3} a^{\dag}_{\beta _4} a^{\dag}_{\beta _6} a^{\dag}_{\beta _7} | 0\rangle , \label{eq:example_many_body_state}
\end{align}
as the binary string $| 01011011 \rangle $ on the qubit register, where we obtain the binary string by 1) arranging these 8 SP bases in the set $\mathbb{S}$ from left to right; and 2) recording the occupation of each state.
The total number of occupations corresponds to the nucleon number $A$. Two more detailed examples can be found in Sec. \ref{sec:applications}. 

We comment that the DE scheme presented here follows directly the Jordan-Wigner encoding scheme \cite{JW:1928,DAbrams:1997,Somma:2002}, which maps the occupations of fermionic SP states into a string of binaries. The application of the DE scheme enables us to develop the many-nucleon theories on quantum computers following the well-established routine adopted in the many-body theories via classical computing, such as the full-configuration interaction approach (see, e.g., Ref. \cite{Jensen:2017} and references therein).

We note that the many-nucleon state can also be encoded in other schemes as well. One alternative is the CE scheme \cite{Kreshchuk:2020dla,Kirby:2021ajp}. Compared to the DE scheme where the occupations of all SP states are encoded as a binary string (whether they are occupied or not), we can record only the indices (and/or the corresponding quantum numbers) of the occupied SP states in qubits. Respecting the anticommutation rule, one also requires that the order of the indices (of the occupied states) recorded in respective quantum registers preserve that of the SP bases in the set $\mathbb{S}$: this can be achieved by the reordering procedure shown in Ref. \cite{Kirby:2021ajp}, which would necessitate controlled arithmetic operations on quantum computers.

In general, the cost of the qubit resources via the CE scheme scales better than that of the DE scheme. In particular, the CE scheme would take $A$ quantum registers, each containing $ \lceil \log _2 N_{\rm sp} \rceil $ qubits, to encode an $A$-nucleon basis state that is constructed based on $N_{\rm sp}$ SP bases, where each register records one index of the corresponding occupied SP state. In this case, the total number of qubits required is $ A \lceil \log _2 N_{\rm sp} \rceil $, which scales better than $N_{\rm sp}$ via the DE scheme.

The price for achieving a better qubit cost via the CE scheme is the circuit complexity, which requires log-local operations for the controlled arithmetic operations on quantum computers \cite{Kirby:2021ajp}. In order to achieve the straightforward prototype structure/dynamics calculations on the NISQ devices, we proceed with the DE scheme in this work: while the qubit cost is less favorable than the CE scheme, the circuit design of the Hamiltonian input model is more straightforward within the DE scheme.

\section{Sparse Hamiltonian problems on quantum computer}
\label{sec:section_IV}

In this section, we first review some of the sparse matrix simulation algorithms for simulating the time-dependent and time-independent Hamiltonians. 
Then, we discuss how these algorithms can be implemented to simulate the second-quantized Hamiltonian, where we devise the necessary input model to access the Hamiltonian matrix in the Fock-state representation in terms of the block-encoding scheme. 
Finally, we propose a framework for solving structure problems based on the second-quantized Hamiltonian.

\subsection{Notation of matrix norms}
For clarification, we first summarize the notation of different norms of the matrix $B$, where $B$ is Hermitian and $B \in \mathbb{C}^{N_{\rm dim} \times N_{\rm dim}}$ with $N_{\rm dim} $ being the matrix dimension. We denote the spectral norm of $B $ as $|| B ||$. The induced 1-norm of $B$ is defined as
\begin{align}
||B||_1 \coloneqq \max _j \sum _{k=0}^{N_{\rm dim}-1} | B_{jk} |,
\end{align}
where $j \in [0, N_{\rm dim}-1]$ and $k \in [0, N_{\rm dim}-1]$ are the row and column indices, respectively.

The max norm of $B$ is defined as the largest matrix element of $B$ in absolute value, i.e.,
\begin{align}
|| B ||_{\rm max} \coloneqq \max _{j,k} |B_{jk}| .
\end{align}
The above three norms satisfy the inequality [$Lemma$ 1 in Ref. \cite{Childs:2010Lim}]
\begin{align}
|| B ||_{\rm max} \leq || B|| \leq || B ||_1 \leq N_{\rm dim} || B ||_{\rm max} .
\end{align}
Especially, when $B$ is $d$ sparse, i.e., there are at most $d$ nonzero entries in any row and column of $B$, the above inequality can be written as \cite{Childs:2010Lim}
\begin{align}
|| B ||_{\rm max} \leq || B|| \leq || B ||_1 \leq d || B ||_{\rm max} . \label{eq:norm_convention}
\end{align}

\subsection{Input model and simulation algorithms for sparse Hamiltonian matrices}
\label{sec:normal_oracle_model}

Efficient sparse Hamiltonian simulation algorithms depend on efficient input models of the Hamiltonian matrix. One of the most widely used input models specifies the $d$-sparse Hamiltonian matrix $H$ via two types of oracles that can be implemented in terms of elementary unitary operations. Originally defined in the quantum-walk-based algorithms \cite{AMChilds:2009,DWBerry:2012}, the first type of the oracle computes the locations (in terms of indices) of the nonzero matrix elements. This oracle is termed as the ``enumerator oracle". It is typically defined as 
\begin{align}
O'_{\rm F}|j, i \rangle = | j , f(j,i) \rangle , \label{eq:O_F_primeXX}
\end{align}
with $j \in \{0,1,2, \cdots , N_{\rm dim} -1 \} $ and $i \in \{0,1, \cdots , d-1 \} $. The function $f(j,i)$ gives the column index of the $i^{\rm th}$ nonzero Hamiltonian matrix element in the $j^{\rm th}$ row. 

The second oracle, referred to as the ``matrix-element oracle", calculates the matrix element when provided the indices computed by the enumerator oracle. The matrix-element oracle can be defined as
\begin{align}
O'_{\rm H} | j, k , 0 \rangle = | j, k , H_{jk}\rangle , \label{eq:O_H_primeXX}
\end{align}
After $ O'_{\rm H} $ functions, the matrix element $ H_{jk} $ is stored in the quantum register (initialized in the $|0 \rangle $ state as denoted by $0$ in the left-hand side of the above equation) in the binary form with some desired precision. 
It is worth noting that $O'_{\rm H}$ operates only when the $O'_{\rm F}$ computes the pair of indices that correspond to the nonzero matrix element.

With $O(1)$ queries to the enumerator and matrix-element oracles, one can construct the isometry for the discretized quantum walk ($Lemma$ 4 in Ref. \cite{DWBerry:2012})
\begin{align}
\mathcal{T}_0 = \sum _{j=0}^{ N_{\rm dim}-1 } |j \rangle | \phi _j \rangle \langle j | , \label{eq:T_0_standard_def}
\end{align}
where
\begin{align}
| \phi _j \rangle = \sqrt{ \frac{\rho}{||H||_1 } } \sum _{k=0}^{ N_{\rm dim} -1} \sqrt{ H_{jk}^{\ast} } |k \rangle |0 \rangle + \sqrt{1- \frac{\rho \sigma _j}{||H||_1} } | \zeta _j \rangle |1 \rangle , 
\end{align}
with $|\zeta _j \rangle $ being some superposition of the $| k \rangle $ and $\sigma _j = \sum _{k=0}^{N_{\rm dim}-1 } |H_{jk} | $. $\rho \in (0,1]$ is a parameter that can be tuned to obtain a {\it lazy} quantum walk \cite{DWBerry:2012}. 

Based on the isometry $\mathcal{T}_0 $, the quantum walk operator is defined as \cite{DWBerry:2012}
\begin{align}
W_0 \coloneqq i S' (2 \mathcal{T}_0 \mathcal{ T}_0^{\dag} - \mathds{I}) , \label{eq:walk_Operator}
\end{align}
where the swap operator $S' $ is defined such that 
\begin{align}
\langle j | \langle \phi _j | S' | k \rangle | \phi _k \rangle = \frac{\rho }{||H ||_1} H_{jk} .
\end{align} 
It can also be proved that \cite{DWBerry:2012}
\begin{align}
\mathcal{T}_0^{\dag} S' \mathcal{T}_0 = \frac{\rho }{||H ||_1} H \otimes |0 \rangle \langle 0 | + | \cdot \rangle \langle \cdot | \otimes |1 \rangle \langle 1 | , \label{eq:block_encoding_1}
\end{align}
where the first term is related to the Hamiltonian; it can be projected onto by taking the ancilla to be in the state $|0\rangle $. The second term is orthogonal to the first term, where $| \cdot \rangle \langle \cdot | $ denotes some operator irrelevant to the Hamiltonian simulation. Equation \eqref{eq:block_encoding_1} is a block-encoding of the Hamiltonian $H$ \cite{Low:2019,Lin:2022lectureNote}.

The oracle-based sparse matrix input model is also adopted in other well-known quantum simulation algorithms. For example, qubitization employs the two queries to the $ O'_{\rm F} $ oracle, and one query to the $O'_{\rm H} $ oracle to construct the isometries ($Lemma$ 6 in Ref. \cite{Low:2019})
\begin{align}
\mathcal{T}_1 =& \sum _j | \psi _j \rangle \langle 0 |_a \langle j |_s , \label{eq:T_1_standard_def} \\
\mathcal{T}_2 =& \sum _k | \chi _k \rangle \langle 0 |_a \langle k |_s , \label{eq:T_2_standard_def}
\end{align}
where 
\begin{align}
| \psi _ j \rangle =& \sum _{p \in F_j} \frac{ | p \rangle _{a_3}}{\sqrt{d}} \Bigg( \sqrt{ \frac{H_{pj}}{||H||_{\rm max}} }|0 \rangle _{a_1} + \sqrt{ 1-  \frac{ | H_{pj} | }{||H||_{\rm max}} } |1 \rangle _{a_1}
\Bigg) | 0 \rangle _{a_2} |j \rangle _s , \label{eq:xxx_isometry} \\ 
\langle \chi _ k | =& \sum _{k\in F_k} \frac{ \langle p | _{s}}{\sqrt{d}}
 \Bigg( \sqrt{ \frac{H_{kp}}{||H||_{\rm max}} } \langle 0 | _{a_2} + \sqrt{ 1-  \frac{ | H_{kp} | }{||H||_{\rm max}} } \langle 1 | _{a_2}
\Bigg) | 0 \rangle _{a_1} |j \rangle _{a_3} , \label{eq:yyy_isometry}
\end{align}
with $\langle \chi _ k | \psi _j \rangle = \frac{H_{kj}}{d || H || _{\rm max}} $. $F_j = \{ f(j,k) \}_{k \in [d]}$ denotes the set of the column indices of all the nonzero entries in the $j^{\rm th}$ row. 

The construction of $\mathcal{T}_1$ and $\mathcal{T}_2$ follows that of $\mathcal{T}_0$ (see $Lemma$ 4 in Ref. \cite{DWBerry:2012}). Here one chooses the parameter $\rho $ such that 
\begin{align}
\frac{\rho}{||H||_1} = \frac{1}{d ||H||_{\rm max} } .
\end{align}
In doing so, one replaces $ ||H||_1 $ by the quantity $||H ||_{\rm max} $. This is useful in the cases where $||H||_1$ is not known exactly, replacing of $||H||_1$ by some better known quantities (e.g., $||H||_{\rm max}$ in this case) is helpful. We see from Eqs. \eqref{eq:xxx_isometry} and \eqref{eq:yyy_isometry} that: 1) the amplitudes for the $|0 \rangle _{a_1}$ state and the $|0\rangle _{a_2}$ state are at most 1; and 2) the single-qubit states in the parentheses correspond to the simple rotations of the qubits $a_1$ and $a_2$ from the $|0 \rangle$ states, respectively.

The $d$-sparse Hamiltonian $H $ can be accessed via the block encoding of Hamiltonian as
\begin{align}
(\langle G|_a \otimes \mathds{I}_s) \mathcal{T}_2^{\dag} \mathcal{T}_1 ( |G \rangle _a \otimes \mathds{I}_s ) = \frac{H}{d||H||_{\rm max}} , \label{eq:qubitization_SparseMat}
\end{align}
with $|G\rangle =|0 \rangle _{a_1} |0 \rangle _{a_2}|0 \rangle _{a_3} $. For the dynamics simulation with time-independent $d$-sparse Hamiltonian $H$, the quantum signal processing takes\footnote{We adopt the typical convention in computer science in this work. For any functions $w$ and $v$, $w\in \Theta (v)$ denotes that $w$ is asymptotically upper and lower bounded by multiples of $v$, while $w\in \mathcal{O} (v)$ indicates that $w$ is asymptotically upper bounded by $v$, and $w\in o (v)$ indicates that $w/v \rightarrow 0$ in the asymptotic limit.} 
\begin{align}
\mathcal{O}\Bigg( d || H ||_{\rm max} t + \frac{\log (\frac{1}{\epsilon })}{\log \log (\frac{1}{\epsilon })} \Bigg) \label{eq:calls_qubitization}
\end{align}
queries to the oracles [Corollary 15 in Ref. \cite{Low:2019}]. It is noteworthy that this scaling is optimal in the simulation error $\epsilon $ [Theorem 1.2 in Ref. \cite{AMChilds:2013}], and the simulation time $t$ [according to the no-fast-forwarding theorem \cite{DWBerry:2012}].

Moreover, the oracle-based sparse matrix input model is adopted in the rescaled Dyson-series (RDS) algorithm \cite{Berry:2020} to simulate the time-dependent Hamiltonians $H(t)$ with the $L^1$-norm scaling. Besides $\mathcal{O}'_{\rm F}$ and $\mathcal{O}'_{\rm H}$, the RDS algorithm employs two additional oracles to rescale the Hamiltonian depending on its instantaneous max-norm during the evolution. These additional oracles are \cite{Berry:2020}
\begin{align}
\mathcal{O}_{\rm var} | \varsigma , z \rangle =& | \varsigma , z \oplus f^{-1}(\varsigma ) \rangle , \label{eq:O_var} \\
\mathcal{O}_{\rm norm} | t, z \rangle =& | t, z \oplus ||H (t)||_{\rm max} \rangle \label{eq:O_norm_query} , 
\end{align}
where $ \mathcal{O}_{\rm var} $ implements the inverse of changing variable and $ \mathcal{O}_{\rm norm} $ computes the max-norm. $f(t)$ is defined as
\begin{align}
f(t) \coloneqq \int _0^t ||H(t')||_{\rm max} dt' , \label{eq:L1_max_norm}
\end{align} 
where the evolution is taken to be from the initial time $t_i=0$ to the final time $t_f=t$, without loss of generality. As $ f(t) $ increases monotonically, one can implement binary search to compute $f^{-1}(\varsigma) $ up to precision $ \bar{\delta} $ using $\mathcal{O} (\log ({t}/{\bar{\delta}}) )$ queries to $f $. As long as $ ||H (t)||_{\rm max} $ can be efficiently computed for any time during the evolution, one can efficiently implement the $ \mathcal{O}_{\rm var} $ and $ \mathcal{O}_{\rm norm}  $ oracles.

Overall, with the rescaled Hamiltonian ${H}( f^{-1}(\varsigma))/|| H(f^{-1}(\varsigma) )  ||_{\rm max}$ for a rescaled total evolution time of $||H|| _{\rm max,1} \coloneqq \int _0^t  || H(t') ||_{\rm max} dt' $, the RDS algorithm takes 
\begin{align}
O\Bigg( \widetilde{\tau } \frac{\log (\frac{  \widetilde{\tau }  }{\epsilon})}{\log \log (\frac{ \widetilde{\tau } }{\epsilon })} \Bigg) \label{eq:L1_scaling_original}
\end{align}
oracle queries of $\mathcal{O}'_{\rm F}$, $\mathcal{O}'_{\rm H}$, $ \mathcal{O}_{\rm var}  $, and $\mathcal{O}_{\rm norm} $ to simulate $H(t)$ for time $t$ within error $\epsilon$ \cite{Berry:2020}. Here $\widetilde{\tau } $ depends on the $L^1$-norm of the $|| H(t') ||_{\rm max} $ during the evolution, i.e.,
\begin{align}
\widetilde{\tau } \coloneqq d \int _{0}^{t} || H(t') ||_{\rm max} dt' .
\end{align}
The RDS algorithm achieves a near optimal scaling of the oracle queries with respect to the evolution time $t$, and is optimal in the scaling of the oracle queries with respect to the simulation error $\epsilon $.

\subsection{Fock-state-based input model and algorithms for structure and dynamics}
\label{sec:input_model_Fock_state_based}

In Sec. \ref{sec:normal_oracle_model}, we review the oracle-based input model that accesses the sparse Hamiltonian matrix elements via their indices. We also review the efficient simulation algorithms that were developed based on this input model, and perform efficient simulations for both time-dependent and time-independent Hamiltonians. However, the above input model can become inefficient when one treats second-quantized Hamiltonians for many-nucleon systems, where one deals with the Fock states instead of the row and column indices and it would be complicated to switch between the indices and the Fock states in many-body calculations \cite{Kirby:2021ajp}. 

In this section, we develop an input model for the second-quantized Hamiltonian. This input model is constructed based on the Fock states employing the idea of Ref. \cite{Kirby:2021ajp}. Compared to the input model reviewed in Sec. \ref{sec:normal_oracle_model} that accesses the nonzero Hamiltonian matrix elements via their row and column indices, we seek to restore the natural connection between the Fock states and the matrix elements of the second-quantized Hamiltonian in our input model. Here, we first introduce our definitions of the oracles for the Fock states, and then the construction of the isometry by $\mathcal{O}(1)$ queries to these oracles. The isometry is implemented to block-encode the many-nucleon Hamiltonian. Our Fock-state-based input model for the second-quantized Hamiltonian can be implemented with those efficient simulation algorithms discussed in Sec. \ref{sec:normal_oracle_model}. 

\subsubsection{Oracle definitions}

We define our enumerator oracle as 
\begin{align}
O_{\rm F} | \mathcal{F} \rangle | i\rangle  | 0 \rangle |0 \rangle | 0 \rangle | 0 \rangle = | \mathcal{F} \rangle | {0}\rangle | P_i \rangle | \mathcal{F}'_i \rangle | Q_i \rangle | y_i (\mathcal{F}, P_i, Q_i ) \rangle , \label{eq:O_F_prime} 
\end{align}
with $ y_i(\mathcal{F}, P_i, Q_i )  = 0  $ or $ 1 $. That is, provided the input $A$-nucleon (Fock) state $ | \mathcal{F} \rangle $ and the index $i\in [0, \mathcal{D}-1] $, the enumerator oracle computes: 1) the pair of SP states $\{r_i,s_i\}  $ tagged by $ P_i $ to be annihilated in $ | \mathcal{F} \rangle $ (to form $ | \mathcal{F} _{P_i} \rangle $); 2) the pair of SP states $\{p_i,q_i\} $ tagged by $Q_i $ to be created in $ | \mathcal{F}_{P_i} \rangle $ (to form $ | \mathcal{F} _i' \rangle $); 3) the output Fock state $ | \mathcal{F}_i' \rangle $; and 4) the function $ y_i (\mathcal{F}, P_i, Q_i ) $. Here $ \mathcal{D} $ denotes the total number of the different combinations of $P_i$ and $Q_i$, where each pair is indexed by $i$. The $O_{\rm F}$ oracle also uncompute the ancilla register that encode the index $i$, which is initialized as $|0\rangle $.

We remark that: 1) if $ |y_i(\mathcal{F}, P_i, Q_i ) \rangle  = | 0 \rangle $, then $ | \mathcal{F} _i' \rangle | Q_i \rangle $ connects to $ | \mathcal{F} \rangle |P_i \rangle $ via the term $ H(Q_i,P_i) b_{Q_i}^{\dag} b_{P_i} |Q_i\rangle \langle P_i | $ in Eq. \eqref{eq:H_modified_H_prime}, which, in principle,\footnote{This is determined by the value of the corresponding kernel $ H(Q_i,P_i) $ up to some precision $\bar{\delta }$.} results in a nonvanishing two-body kernel $\langle \mathcal{F}'_i , Q_i | \mathcal{H} | \mathcal{F} , P_i \rangle$; 2) otherwise, if $ |y_i(\mathcal{F}, P_i, Q_i ) \rangle  = | 1 \rangle $, then $ | \mathcal{F} _i' \rangle  | Q_i \rangle $ does not connect to $ | \mathcal{F} \rangle |P_i \rangle $ via $  H(Q_i,P_i) b_{Q_i}^{\dag} b_{P_i} |Q_i\rangle \langle P_i | $ and the two-body kernel $\langle \mathcal{F}'_i , Q_i | \mathcal{H} | \mathcal{F} , P_i \rangle =0 $. It is also noted that the index $i$ here does not number the location of the sparse matrix element; it labels different terms in Eq. \eqref{eq:H_modified_H_prime}, each of which corresponds to a distinct monomial $b_{Q_i}^{\dag} b_{P_i} = a^{\dag}_{p_i} a^{\dag}_{q_i} a_{s_i} a_{r_i}$.

We define the matrix-element oracle $O_{\rm H}$ as
\begin{align}
O_{\rm H} | \mathcal{F} \rangle | P_i \rangle  | \mathcal{F}'_i \rangle | Q_i \rangle | 0 \rangle =  | \mathcal{F} \rangle | P_i \rangle  | \mathcal{F}'_i \rangle | Q_i \rangle |  \mathcal{H}(\mathcal{F}'_i,Q_i; \mathcal{F}, P_i)   \rangle  , \label{eq:O_H_prime} 
\end{align}
where $ \mathcal{H}(\mathcal{F}'_i,Q_i; \mathcal{F}, P_i) = \langle \mathcal{F}'_i , Q_i | \mathcal{H} | \mathcal{F} , P_i \rangle $ is defined in Eq. \eqref{eq:H_modified_XXXXL} and its relation to the $A$-nucleon Hamiltonian matrix element is shown in Eq. \eqref{eq:equivalence_restore}. Indeed, the $O_{\rm H}$ oracle takes the input from $O_{\rm F}$ and it operates only when $ | y_i (\mathcal{F}, P_i, Q_i ) \rangle = |0 \rangle $ (i.e., the corresponding $\mathcal{H}(\mathcal{F}'_i,Q_i; \mathcal{F}, P_i)$ is, in principle, nonvanishing). This will always be the case for us in this work.

\subsubsection{Isometry construction}
\label{sec:isometry_construction_fock_state_based}

In comparison to the standard definitions of the isometry defined in Eqs. \eqref{eq:T_0_standard_def}, \eqref{eq:T_1_standard_def}, and \eqref{eq:T_2_standard_def}, we define the isometry $ \mathcal{T} $ for our matrix input model that is based on the Fock states as
\begin{align}
\mathcal{T} =& \sum _b \sum _{| \mathcal{F} \rangle } | \mathcal{F} \rangle \langle \mathcal{F} | \otimes |b \rangle \langle b| \otimes | \phi _{\mathcal{F}, b} \rangle , \label{eq:isometry_T_modified}
\end{align}
where we also introduce a single-qubit ancilla $|b \rangle $ with $b=0,1$. We note that $|b \rangle $ plays the role of the ancilla state that flags the signal operator (Hamiltonian) in the formalism of the qubitization \cite{Low:2017,Low:2019}. Based on the values of $b$, $ | \phi _{\mathcal{F}, b} \rangle  $ is defined as
\begin{align}
| \phi _{\mathcal{F}, 0} \rangle =&  \sqrt{ \frac{1}{\mathcal{D} \Lambda _m} }  \sum _{i \in \mathcal{I}(\mathcal{F})}  \Bigg[\sqrt{ \langle \mathcal{F}_i', Q_i | \mathcal{H} | \mathcal{F} , P_i \rangle }  | {0} \rangle | P_i \rangle |\mathcal{F}'_i \rangle | Q_i \rangle | y_i \rangle | 0 \rangle |0 \rangle \Bigg] 
				    + \sqrt{ 1- \frac{ \sigma _{\mathcal{F}}}{ \mathcal{D} \Lambda _m }  } | \zeta _{\mathcal{F}} \rangle |1 \rangle , \label{eq:equivalent_to_A7} \\
| \phi _{\mathcal{F}, 1} \rangle =&  |0 \rangle |0 \rangle |0 \rangle | 0 \rangle | 0 \rangle | 0 \rangle |1 \rangle , \label{eq:equivalent_to_A2}
\end{align}
where we define
\begin{align}
| \zeta _{\mathcal{F}} \rangle =  \sqrt{\frac{1}{ 1- \frac{ \sigma _{\mathcal{F}}}{\mathcal{D} \Lambda _m  } }}  \sum _{i=0}^{\mathcal{D} -1} \Bigg[ \sqrt{ 1 -  \frac{ | \langle \mathcal{F}_i', Q_i | \mathcal{H} | \mathcal{F} , P_i \rangle | }{\Lambda _m } }  | { 0} \rangle | P_i \rangle |\mathcal{F}'_i \rangle | Q_i \rangle | y_i \rangle | 0 \rangle  \Bigg]
\end{align}
with $ \sigma _{\mathcal{F}} = \sum _{i=0}^{ \mathcal{D} -1} |\langle \mathcal{F}_i', Q_i | \mathcal{H} | \mathcal{F} , P_i \rangle |$. The parameter $\Lambda _m$ is defined as
\begin{equation}
    \Lambda _m \geq \max _{i} | \langle \mathcal{F}_i', Q_i | \mathcal{H} | \mathcal{F} , P_i \rangle  |  = \max _{i} |\langle p_iq_i |H| r_i s_i \rangle | . 
\end{equation}
$\mathcal{I}(\mathcal{F})$ denotes the set of indices $i$ for which $\langle \mathcal{F}_i', Q_i | \mathcal{H} | \mathcal{F} , P_i \rangle \neq 0$ and $y_i = y_i(\mathcal{F},P_i,Q_i) = 0$.

The isometry $\mathcal{T}$ defines the mapping
\begin{align}
\mathcal{T} | \mathcal{F} \rangle | b \rangle = | \mathcal{F} \rangle | b \rangle | \phi _{\mathcal{F}, b} \rangle . \label{eq:isometry_T_prime}
\end{align}
It can be proved that isometry $\mathcal{T} $ [Eq. \eqref{eq:isometry_T_modified}] can be implemented with $\mathcal{O}(1)$ queries to $O_{\rm F}$ and $O_{\rm H}$ defined in Eqs. \eqref{eq:O_F_prime} and \eqref{eq:O_H_prime} (see {\it Lemma} \ref{construction_of_T} in Appendix \ref{sec:isometry_construction} for the proof).

Analogous to Ref. \cite{DWBerry:2012}, we find that the isometry $\mathcal{T} $, together with the corresponding swap operator $S$, defines the block encoding of the second-quantized Hamiltonian as (see the proof in Appendix \ref{sec:relation_to_H_matrix_element}) 
\begin{align}
\mathcal{T}^{\dag} S \mathcal{T} = \frac{1 }{\mathcal{D} \Lambda _m } H \otimes |0 \rangle \langle 0 | + |0 \rangle \langle 0|  \otimes |1 \rangle \langle 1 | . \label{eq:TST_relation}
\end{align}
It follows that, with $\widetilde{ \lambda }_j$ and $|\lambda _j \rangle $ being the eigenvalue and the corresponding eigenvector of the scaled Hamiltonian $\widetilde{H}=\frac{ 1 }{\mathcal{D} \Lambda _m } H$, we have
\begin{align}
\mathcal{T}^{\dag} S \mathcal{T} |\lambda _j \rangle |0 \rangle = \Big[ \frac{1}{\mathcal{D} \Lambda _m } H \otimes |0 \rangle \langle 0 | + |0 \rangle \langle 0|  \otimes |1 \rangle \langle 1 | \Big] |\lambda _j \rangle |0 \rangle =  \widetilde{\lambda } _j  |\lambda _j \rangle |0 \rangle . \label{eq:walk_Hamiltonian_eigenequation}
\end{align}

We can also rewrite Eq. \eqref{eq:TST_relation} as
\begin{align}
(\mathds{1} \otimes  \langle 0 | ) (\mathcal{T}^{\dag} S \mathcal{T} ) (\mathds{1}  \otimes  | 0 \rangle  ) = \frac{1 }{\mathcal{D} \Lambda _m } H , \label{eq:qubitization_formula}
\end{align}
where $\mathds{1} $ denotes the unit operator acting on the signal register that encodes the Fock states, while $|0 \rangle $ and $\langle 0|$ act on the ancilla register that encodes $| b \rangle $. We note that Eq. \eqref{eq:qubitization_formula} is related to Eq. \eqref{eq:qubitization_SparseMat}. This can be seen by noticing that: 1) the isometries  $S\mathcal{T}$ and $\mathcal{T}$ take the roles of $\mathcal{T}_1$ [Eq. \eqref{eq:T_1_standard_def}] and $\mathcal{T}_2$ [Eq. \eqref{eq:T_2_standard_def}], respectively; 
and 2) the sparsity $d$ of the Hamiltonian matrix is replaced by the number of the monomials (of the ladder operators), $\mathcal{D}$, in the second-quantized Hamiltonian, where we have $\mathcal{D} \in \mathcal{O}(N_{\rm sp}^4)$ for the many-nucleon Hamiltonian that contains at most two-body terms \cite{Babbush:2018bubbush}.

In particular, we can compute the Hamiltonian matrix element for the second-quantized Hamiltonian as [Eq. \eqref{eq:walk_identity}]
\begin{equation}
\langle \mathcal{F} | \langle 0| ( \mathcal{T}^{\dag}S \mathcal{T} ) | \mathcal{G} \rangle | 0 \rangle =  \frac{ 1 }{\mathcal{D} \Lambda _m } \langle   \mathcal{F} | H | \mathcal{G} \rangle , \label{eq:qubitization_formula_1}
\end{equation}
with the action of the Fock states $|\mathcal{F} \rangle $ and $|\mathcal{G} \rangle $.

\subsection{Simulating the second-quantized many-nucleon Hamiltonian}
\label{sec:dynamics_calculation}

Our input model can be directly implemented to high-level algorithms to simulate the second-quantized Hamiltonians with optimal and near-optimal oracle complexities with respect to the simulation error and simulation time.

As for simulating the time-independent second-quantized Hamiltonian, we can implement our input model with the quantum signal processing \cite{Low:2017,Low:2019}. Following Corollary 15 in Ref. \cite{Low:2019}, it takes
\begin{align}
\mathcal{O}\Bigg( \mathcal{D} \Lambda _m  t + \frac{\log (\frac{1}{\epsilon })}{\log \log (\frac{1}{\epsilon })} \Bigg)
\label{eq:qubitization_oracle_scaling}
\end{align}
queries to the $O_{\rm F}$ and $O_{\rm H}$ oracles to simulate the time-independent Hamiltonian $H$ for time $t$ within error $\epsilon $.
Compared to Eq. \eqref{eq:calls_qubitization}, we note that the scaling of the simulation time in the oracle complexity is replaced by $\mathcal{D} \Lambda _m $. This can be understood by comparing Eq. \eqref{eq:qubitization_SparseMat} and Eq. \eqref{eq:qubitization_formula}, where the scalings of the block-encoded Hamiltonians are $\frac{1}{d||H||_{\rm max}}$ and $\frac{1}{\mathcal{D}\Lambda _m}$ within the frameworks of the first- and second-quantization, respectively.

Our input model can also be directly implemented into the RDS algorithm \cite{Berry:2020}. It follows from Theorem 10 in Ref. \cite{Berry:2020} that one can simulate the evolution of the time-dependent second-quantized Hamiltonian $H=H(t)$ for time $t$ using 
\begin{align}
O\Bigg( \widetilde{\tau }' \frac{\log (\frac{  \widetilde{\tau }'  }{\epsilon})}{\log \log (\frac{ \widetilde{\tau }' }{\epsilon })} \Bigg) 
\label{eq:RDS_algorithm_query_complexity}
\end{align}
oracle queries of $\mathcal{O}_{\rm F}$ and $\mathcal{O}_{\rm H}$, as well as the compatible $ \mathcal{O}_{\rm var}  $ and $\mathcal{O}_{\rm norm} $,\footnote{Here we define $ \mathcal{O}_{\rm var}  $ and $\mathcal{O}_{\rm norm} $ according to $\Lambda _{\rm m}=\Lambda _m (t)$ instead of the max-norm of the instantaneous Hamiltonian $||H(t)||_{\rm max}$ in Eqs. \eqref{eq:O_var}, \eqref{eq:O_norm_query} and \eqref{eq:L1_max_norm}. Following the analysis in Ref. \cite{Berry:2020} which deals with $||H(t)||_{\rm max}$, we assume $\Lambda _{\rm m}(t)$ can then be efficiently computed so it to be straightforward to implement $ \mathcal{O}_{\rm var}  $ and $\mathcal{O}_{\rm norm} $ during the simulation.} within error $\epsilon $. $\widetilde{\tau }' $ depends on the $L^1$-norm of $ \Lambda _m (t) $ during the evolution, i.e.,
\begin{align}
\widetilde{\tau }' \coloneqq \mathcal{D} \int _{0}^{t} \Lambda_m(t') dt' . \label{eq:L1_scaling_changed}
\end{align}
$\Lambda _m (t') $ is defined based on the instantaneous Hamiltonian as  $\Lambda _m (t') \geq \max _i | \langle p_iq_i | H(t') | r_i s_i \rangle  | $ for $i \in [0,\mathcal{D}-1]$ and $t'\in [0,t]$, where the time-dependent kernel is defined as Eq. \eqref{eq:H_modified_XXXXL} with the time dependence explicitly shown. We also note that the scaling in $\widetilde{\tau } = d \int _0^t ||H(t')||_{\rm max} dt' $ in Eq. \eqref{eq:L1_scaling_original} is substituted by Eq. \eqref{eq:L1_scaling_changed}, as a consequence of the change in the scaling of the block-encoded Hamiltonian, from $\frac{1}{d||H(t')||_{\rm max}}$ to $\frac{1}{\mathcal{D} \Lambda _m(t')}$, in our input model [Eq. \eqref{eq:qubitization_formula}].

\subsection{Structure problems of the many-nucleon Hamiltonian}
\label{sec:structure_calculation}

Besides simulating the dynamics of the many-nucleon systems, one is also interested in solving structure problems which involve spectra and other observables. 
In such structure problems, one confronts constructing functions of the Hamiltonian $U(H)$ (e.g., $\exp [-iH t ]$) and the controlled version of $U(H)$, as those in dynamics simulations.
The Hamiltonian input model is a key ingredient shared between the structure and dynamics problems.

A state-of-the-art approach to solve the spectra is the adiabatic state preparation (ASP) \cite{Farhi:2001Science} together with the quantum phase estimation (QPE) [see Ref. \cite{Albash:2018RMP} and references therein]. As the major idea, this approach prepares the ground state of the Hamiltonian $H$ by evolving the ground state of a simple reference Hamiltonian $H_{\rm ref}$ via a parameterized adiabatic path defined by the time-dependent Hamiltonian 
\begin{align}
\bar{H} (t') = g(t') H_{\rm ref} + [1 - g(t')] H, 
\end{align}
where the $ g(t') $ is a smoothly behaved scalar function defined in the domain of $t' \in [0,t]$, with $ g(0) =1 $ and $ g(t) = 0 $. In this way, one is guaranteed to obtain the ground state of $H$ according to the well-known adiabatic theorem \cite{Messiah:1962QM,Farhi:2001Science}. The consequent QPE algorithm \cite{Kitaev:1995,DAbrams:1997,Abrams:1998pd,NielsenANDChuang:2001} acts on the ground state of $H$ to obtain the corresponding ground state energy of $H$. In principle, this approach of eigensolver can be generalized to the excited states in cases where there is no degeneracy or near-degeneracy for $\bar{H} (t')$. To improve efficiency, one could modify the adiabatic path by introducing appropriate reference Hamiltonian and/or perturbations to avoid level crossing \cite{Farhi:2011QiC,Wecker:2015pra,Du:2021ctr}.

In this adiabatic approach, one can employ the RDS algorithm to perform the ASP based on the time-dependent Hamiltonian $\bar{H}(t') $ and implement the quantum signal processing to build the controlled version of the function $ e^{i2\pi \gamma (H) }$ for the QPE \cite{Guzik:2005Sci}, where one can define, e.g., $ \gamma (H) \coloneqq x H + y$ with $x$ and $y $ being scalars such that $ || \gamma (H) || <1 $.

The ASP-QPE, in general, necessitates a deep circuit. Alternatively, one can adopt the non-adiabatic approaches, such as the quantum cooling algorithms \cite{GGC:2014,Choi:2020pdg,Qian:2021wya,Bee-Lindgren:2022nqb}, where it is shown that the non-adiabatic approaches can be exponentially faster than the ASP-QPE. 

One example of the non-adiabatic approaches is the Rodeo algorithm \cite{Choi:2020pdg,Qian:2021wya,Bee-Lindgren:2022nqb}. It is shown in the Rodeo algorithm can be utilized to solve the spectrum of $H$. One can also prepare the eigenstates of $H$ by proper parameter settings of the Rodeo algorithm,\footnote{There are subtleties regarding the choice of the input, e.g., the preparation of proper initial state \cite{DLeePC}, which is beyond the scope of this work.} which enables solving other observables and transition kernels, via e.g., the Hadamard test, the Swap test, or other methods \cite{Mitarai:2019,Siwach:2022ugy}, which will be the foci of further work. 

The elementary building block of the Rodeo algorithm consists of a few copies of the controlled evolution unitary $e^{-iH \Delta }$ (with $ \Delta$ being the standard deviation of a set of Gaussian random variables), which can be constructed utilizing the quantum signal processing based on the Hamiltonian input model.
As expected from evolving a time-independent second-quantized Hamiltonian [Eq. \eqref{eq:qubitization_oracle_scaling}], the asymptotic query complexity to the $O_{\rm F}$ and $O_{\rm H}$ in applying the Rodeo algorithm for the structure calculation based on the second quantized Hamiltonian is 
\begin{align}
\mathcal{O}\Bigg( \mathcal{D} \Lambda _m  \Delta + \frac{\log (\frac{1}{\epsilon })}{\log \log (\frac{1}{\epsilon })} \Bigg), \label{eq:rodeo_with_qubitization}
\end{align}
with $\epsilon $ being the error.

Our input model can also be implemented with other promising algorithms for structure calculations. We do not aim for a thorough review of these algorithms in this work and interested readers are referred to Refs. \cite{Lin:2022lectureNote,Lin:2020Linlin,Chuang:2021chuangIssac,Dong:2022Linlin} and references therein.

\section{Oracle designs}
\label{sec:oracle_designs}

In this work, we restrict our discussion on the second-quantized many-nucleon Hamiltonian [Eq. \eqref{eq:H_total} or, equivalently, Eq. \eqref{eq:H_tot_modified}] that includes only the two-body terms. Based on this Hamiltonian, we introduce our designs for the enumerator oracle [Eq. \eqref{eq:O_F_prime}] and the matrix-element oracle [Eq. \eqref{eq:O_H_prime}] based on the DE scheme [Sec. \ref{sec:DirectEncoding}]. We also analyze the complexity of implementing these oracles. It is worth noting that our prototype oracle design can also be generalized to many-nucleon Hamiltonians that include more than two-body terms in a straightforward manner.

\subsection{Enumerator oracle}
\label{sec:O_F_oracle_sec}

\begin{figure*}[ht] 
  \centering
  \includegraphics[width=0.90\textwidth]{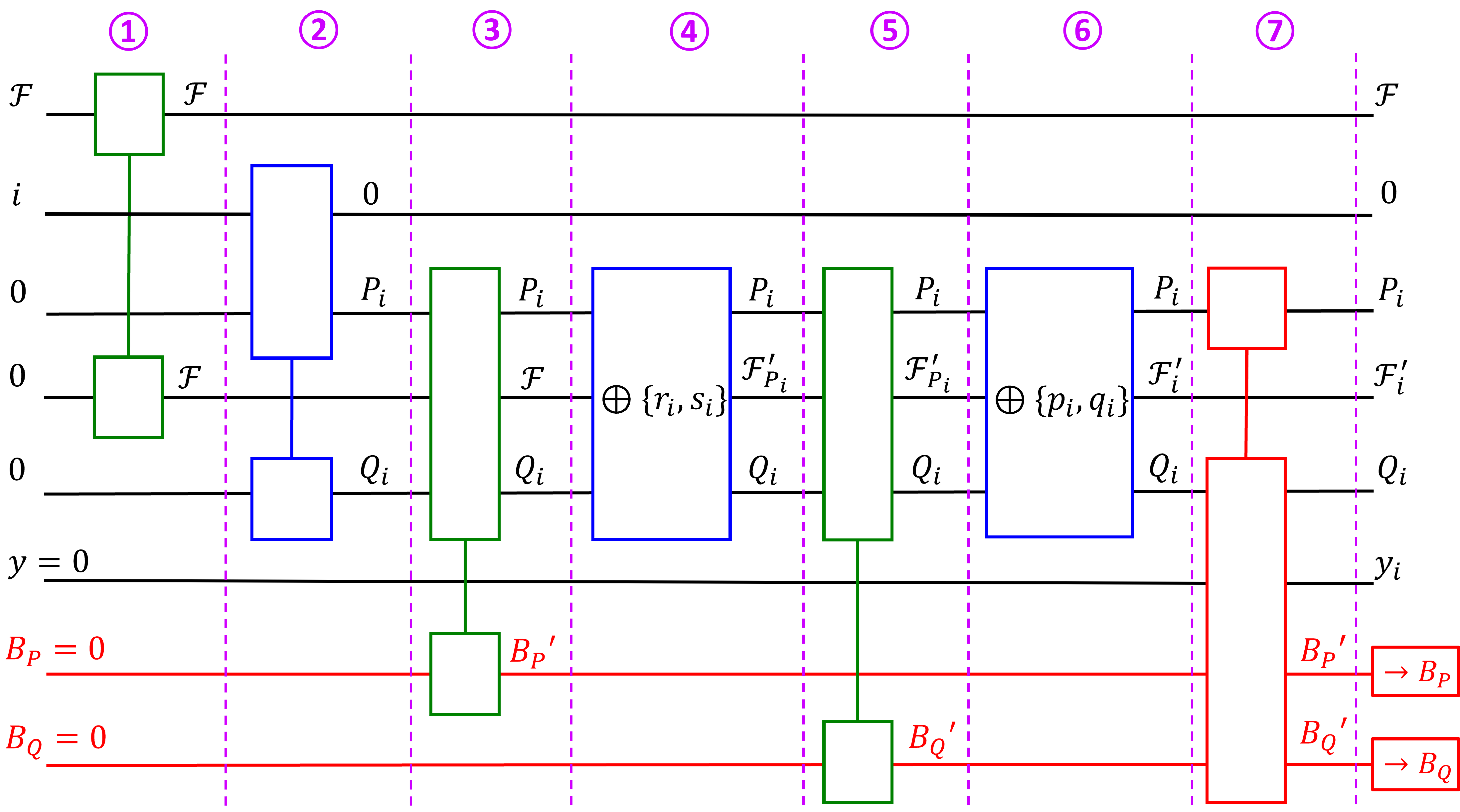}
  \caption{(color online) The construction of the $O_{\rm F}$ oracle. See the text for details.} 
  \label{fig:OF}
\end{figure*}

We define the enumerate oracle $O_{\rm F}$ [Eq. \eqref{eq:O_F_prime}] that takes the input many-nucleon state $| \mathcal{F} \rangle $ and the index $i$ { (encoded in an ancilla register)}, and compute the quantities: 1) the output state $| \mathcal{F}_i' \rangle $; 2) the tags $P_i$ and $Q_i$ that marks the active SP states; 3) the function $y_i(\mathcal{F}, P_i, Q_i)$ that marks the possible error message in the construction.
We construct the $O_{\rm F}$ as follows, where the procedures of the construction is shown in Fig. \ref{fig:OF}. 

\paragraph*{Step 1: Duplicate the input Fock state.} 
Provided the input many-nucleon state, we duplicate the Fock state $| \mathcal{F} \rangle $ to a second register. After a sequence of operations, the copy of $| \mathcal{F} \rangle $ will eventually produce the output state $ | \mathcal{F}_i' \rangle $ at the end of the implementation. Note that duplicating a general quantum state in quantum computing is not possible. However, for the specific case of duplicating the Fock state $| \mathcal{F} \rangle $ (represented by a sequence of 0's and 1's), this operation can be performed. In our case [also in the case treated in Ref. \cite{Kirby:2021ajp}], this duplication can be implemented, for example, using a sequence of qubit-wise CNOT gates duplicating the input $| \mathcal{F} \rangle $ from the first register to the second register, which is initialized in the all-zero state. 

\paragraph*{Step 2: Identify the active SP states.} Based on the index $i$, we compute the tags of the active pairs of SP states, $P_i$ and $Q_i$, in the input and output many-nucleon state, respectively. This can be achieved by the iterations with the classically precomputed data table following the ideas in Refs. \cite{Kirby:2021ajp,RBabbush:2016}. After the iteration, we access the tags $P_i$ and $Q_i$ and store them in the corresponding qubit registers, which are initialized in the $|0\rangle $ states, as tag states $|P_i\rangle $ and $|Q_i \rangle $, respectively. { After we compute the tags $P_i$ and $Q_i$ via $i$, we also uncompute the ancilla register that encodes the index $i$ with the pair of tags $(P_i, Q_i)$. In the following steps (as well as the steps in the $O_{\rm H}$ oracles), it is sufficient to proceed with $(P_i, Q_i)$, which is uniquely determined by the index $i$.}

\paragraph*{Step 3: Identify the active SP states to be removed in $ | \mathcal{F} \rangle $ and check their occupancy.}
Given $P_i  $ and controlled by $Q_i$, we determine the active SP states $\{r_i,s_i\}$ (with $r_i<s_i$) of which the occupancies should be removed in the copy of $| \mathcal{F} \rangle $ (recall that each $ P_i  $ tags a pair of SP states $ \{r_i,s_i\}$). However, before the removal of occupancies, we need to check if the pair of states $ \{r_i,s_i \} $ are indeed occupied in $| \mathcal{F} \rangle $. To achieve this, we employ an ancilla qubit $B_P$, which is initialized in the $|0\rangle $ state, to flag the error cases; we require that $| B_P \rangle $ remains in $|0 \rangle $ only when the SP bases $\{r_i ,s_i \} $ are both occupied in the $| \mathcal{F} \rangle $; and we flip $| B_P \rangle $ to be $|1\rangle $ if else. After this step, the state of the ancilla qubit $B_P$ is denoted as $|B_P' \rangle $.

\paragraph*{Step 4: Flip the active SP states tagged by $P_i $ in the copy of $| \mathcal{F} \rangle $.}
Controlled on the tags $P_i $ and $Q_i$, we flip the qubits which correspond to the $r_i^{\rm th}$ and $s_i^{\rm th}$ SP states in the copy of $| \mathcal{F} \rangle $. This could contain undesired operations where we could add on the occupation(s) on the target SP state(s), instead of removing the occupation(s), which results in undesired many-nucleon states. However, the error messages stored in the ancilla state $| B_P' \rangle $ shall help to distinguish such undesired operations. As we shall discuss below (see Step 7), such undesired states will not contribute to the Hamiltonian matrix element eventually. After this step, the copy of $| \mathcal{F} \rangle $ becomes $| \mathcal{F}_{P_i}' \rangle $ as shown in Fig. \ref{fig:OF}.

\paragraph*{Step 5: Identify the active SP states to be added in $| \mathcal{F}_{P_i}' \rangle $ and check their occupancy.} 
Similar to Step 3, we determine the active SP states $\{p_i,q_i\}$ (with $p_i<q_i$) with the tag $Q_i $ (recall that we use $Q_i  $ to tag such a pair SP states, i.e., $ \{p_i,q_i\} $). The occupations of this pair of SP states should be created in the intermediate many-nucleon state $| \mathcal{F}_{P_i}' \rangle $. However, we need to first check that the $p_i^{\rm th}$ and $q_i^{\rm th}$ SP states are both vacant in $| \mathcal{F}_{P_i}' \rangle $ controlled by the tags $P_i$ and $Q_i$. If either or neither of them are vacant, we shall record this error message in the ancilla qubit $ B_Q $. In particular, we initialize $| B_Q \rangle $ in the $|0\rangle $: 1) it remains in the $|0 \rangle $ state only if both $p_i^{\rm th}$ and $q_i^{\rm th}$ SP states are vacant; and 2) $| B_Q \rangle $ is flipped to $|1 \rangle $ otherwise. After this step, the state of the ancilla qubit $B_Q $ is denoted as $| B_Q' \rangle $.

\paragraph*{Step 6: Flip the active SP states tagged by $Q_i  $.}
Similar to Step 4, we flip the $p_i^{\rm th} $ and $q_i^{\rm th} $ SP states (tagged by $Q_i $) in the intermediate many-nucleon state $| \mathcal{F}_{P_i}' \rangle $ controlled by the tags $P_i$ and $Q_i$. Again, this could include undesired operations where either or both the pair of SP states are occupied, which results in undesired output many-nucleon states. However, the error message stored in $| B_Q' \rangle $ shall help to distinguish such contributions (see Step 7 below). After this step, we obtain the output many-nucleon state $| \mathcal{F}'_i \rangle $.

\paragraph*{Step 7: Compile the error messages.} 
Controlled by the tags $P_i$ and $Q_i$, we generate the desired many-nucleon state $| \mathcal{F}'_i \rangle $ only when both $| B_P' \rangle $ and $| B_Q' \rangle $ remain in the $|0 \rangle $ states (same as they are initialized). If we have mistakenly removed the occupation(s) in the vacant SP state(s), or created the occupation(s) in the occupied state(s), we would generate undesired output many-nucleon state, which produces vanishing matrix element $ \langle \mathcal{F}'_i, Q_i | \mathcal{H} | \mathcal{F}, P_i \rangle  $ [Eq. \eqref{eq:H_modified_XXXXL}]. In this step, we would then flip the qubit $ y  $ that is initialized in $|0 \rangle $ to $|1\rangle $ if either or both of $|B _P' \rangle $ and $|B _P' \rangle $ are in the $|1 \rangle $ states. After this operation, the qubit state $|y \rangle $ is denoted by $| y_i \rangle  =|y_i(\mathcal{F},P_i,Q_i) \rangle $. We remark that the output many-nucleon state $| \mathcal{F}'_i \rangle $ is valid only when $|y_i \rangle =|0\rangle $. The $| \mathcal{F}'_i \rangle $ would, in principle, produce nonvanishing two-body kernel $ \langle \mathcal{F}'_i, Q_i | \mathcal{H} | \mathcal{F}, P_i \rangle  $ with the input state $| \mathcal{F} \rangle $, and the tags $(P_i,Q_i)$.

With the above steps, we construct the output many-nucleon state $| \mathcal{F}'_i \rangle $ based on the input state $| \mathcal{F} \rangle $ and the index $i $. We also identify the active SP bases (tagged by $P_i$ and $Q_i$) that contribute to the integral to evaluate the Hamiltonian matrix element (see in the follow-up matrix-element oracle). In addition, we monitor the errors in the $O_{\rm F}$ oracle via $| y_i \rangle  $. After the construction, we uncompute the ancilla states $ | B'_P \rangle $ and $ | B'_Q \rangle $ by reversing the construction. 

This completes the construction of $O_{\rm F}$ oracle.

\subsection{Matrix-element oracle}
\label{sec:OH_oracle_sec}

\begin{figure*}[!ht] 
  \centering
  \includegraphics[width=0.90\textwidth]{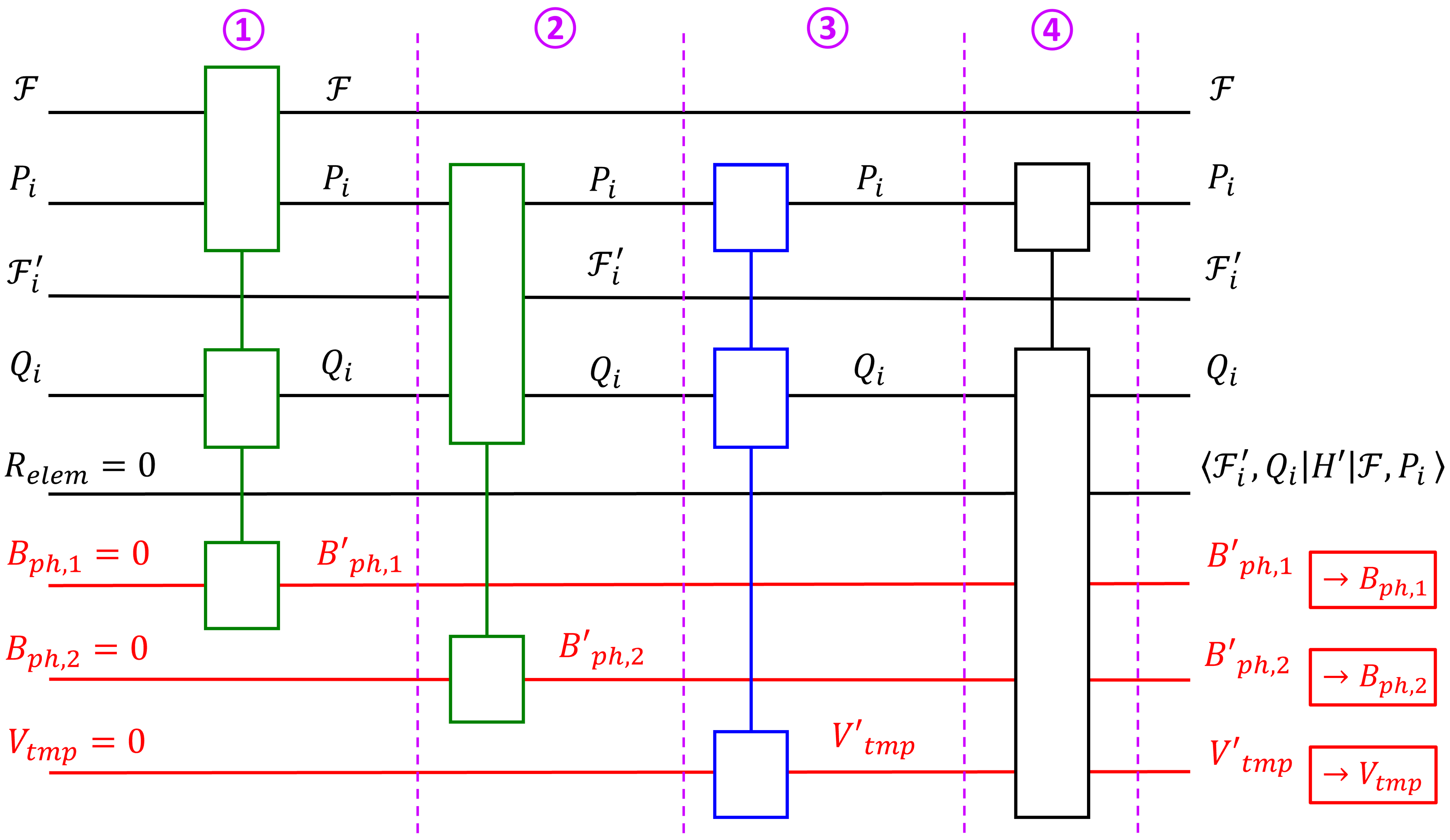}
  \caption{(color online) The sketch of the $O_{\rm H}$ oracle. See the text for details.} 
  \label{fig:OH}
\end{figure*}

We now construct the $O_{\rm H}$ oracle according to the definition in Eq. \eqref{eq:O_H_prime}. Taking the input from the $O_{\rm F}$ oracle [Eq. \eqref{eq:O_F_prime}], which are $|\mathcal{F} \rangle $, $| P_i \rangle $, $|\mathcal{F}_i' \rangle$, $| Q_i \rangle $, and  $|y_i \rangle $, the $O_{\rm H}$ operates controlled on $|y_i \rangle = |0 \rangle $, and it computes the kernel $\langle \mathcal{F}'_i , Q_i | \mathcal{H} | \mathcal{F} ,P_i \rangle $. 

In practice, we divide the task of the $O_{\rm H}$ oracle into two parts according to Eq. \eqref{eq:H_modified_XXXXL}: 1) computing the factor that determines the phase, namely, $\langle \mathcal{F}'_i | b_{Q_i}^{\dag} b_{P_i} | \mathcal{F} \rangle  $; and 2) obtaining the value of the entry $\langle Q_i |H |P_i \rangle $. We elect to compute the overall phase $c_{i,1} \cdot c_{i,2} $ by separately computing $ b_{P_i} | \mathcal{F} \rangle = c_{i,1} | \bar{\mathcal{F} }_i \rangle $ and $ b_{Q_i} | \mathcal{F}'_i \rangle = {c_{i,2}} | \bar{\mathcal{F} }_i \rangle  $, with the reference state $ | \bar{\mathcal{F} }_i \rangle $ satisfying $ \langle \bar{\mathcal{F} }_i  | \bar{\mathcal{F} }_i \rangle =1  $. That is, we design the $O_{\rm H}$ oracle to evaluate the two-body kernel as
\begin{align}
\langle \mathcal{F}'_i , Q_i | \mathcal{H} | \mathcal{F} ,P_i \rangle = c_{i,1} c_{i,2} H(Q_i,P_i) = c_{i,1} c_{i,2} \langle p_i q_i | H | r_i s_i \rangle . \label{eq:computation_formula_OH} 
\end{align}
As described in Fig. \ref{fig:OH}, the $O_{\rm H}$ is constructed as follows.

\paragraph*{Step 1: Compute $ c_{i,1} $ based on $| \mathcal{F} \rangle $ and $P_i$.}
Controlled by the tag states $|P_i\rangle $ and $|Q_i \rangle$, we compute $ c_{i,1}$ provided the input many-nucleon state $| \mathcal{F} \rangle $ and the tag $P_i$. We first determine the active SP bases $\{r_i, s_i\}$ (with $r_i < s_i$) based on the tag $P_i$. Then, we compute the total number of occupations, $N^{P_i}_{\rm ocp}$, for the the $(r_i +1)^{\rm th} , \ (r_i + 2)^{\rm th} , \cdots , (s_i -2)^{\rm th}, (s_i -1)^{\rm th} $ SP bases of $| \mathcal{F} \rangle $. We encode $ {\rm mod} (N^{P_i}_{\rm ocp}, 2)$ in the ancilla qubit $ B_{{\rm ph},1}  $, which is initialized as $ |0 \rangle $: 1) if $ {\rm mod} (N^{P_i}_{\rm ocp}, 2) =0 $, $ |B_{{\rm ph},1} \rangle $ remains in the $|0 \rangle $ state; and 2) if $ {\rm mod} (N^{P_i}_{\rm ocp}, 2) = 1 $, we flip the $| B_{{\rm ph},1} \rangle $ to be in the $|1 \rangle $ state. This can be achieved, for example, by a sequence of CNOT gates, where each of the qubits that encode the $(r_i +1)^{\rm th} , \ (r_i + 2)^{\rm th} , \cdots , (s_i -2)^{\rm th}, (s_i -1)^{\rm th} $ SP bases in $| \mathcal{F} \rangle $ serves as the control bit and the target bit is the ancilla $| B_{{\rm ph},1} \rangle $. After this step, the state of the ancilla qubit $ B_{{\rm ph},1} $ is denoted as $| B'_{{\rm ph},1} \rangle $. We note that the desired quantity $ c_{i,1}  $ is encoded in $ |B'_{{\rm ph},1} \rangle  $. Indeed, we can access $ c_{i,1}  $ with the identity $ Z |B'_{{\rm ph},1} \rangle = c_{i,1} |B'_{{\rm ph},1} \rangle = (-1)^{{\rm mod} (N^{P_i}_{\rm ocp}, 2)} |B'_{{\rm ph},1} \rangle $, where $Z$ denotes the Pauli-Z gate. 

\paragraph*{Step 2: Compute $ c_{i,2} $ based on $| \mathcal{F}_i' \rangle $ and $Q_i$.} This step is similar to {\it Step 1}. We determine the active SP bases $\{p_i, q_i\} $ ($p_i < q_i$) with the tag $Q_i$. Then, we compute the number of occupations $ N^{Q_i}_{\rm ocp} $ and encode $ {\rm mod} (N^{Q_i}_{\rm ocp}, 2)$ in the ancilla qubit $ B_{{\rm ph},2} $, which is initialized as $|0 \rangle $: 1) the ancilla state $| B_{{\rm ph},2} \rangle $ remains in the $|0 \rangle $ state if $ {\rm mod} (N^{Q_i}_{\rm ocp}, 2) =0 $; and 2) we flip the $| B_{{\rm ph},2} \rangle $ to $|1 \rangle $ if $ {\rm mod} (N^{Q_i}_{\rm ocp}, 2) = 1 $. After this step, the state of the ancilla qubit $ B_{{\rm ph},2}  $ is denoted as $| B'_{{\rm ph},2} \rangle $. The desired quantity $ c_{i,2} $ can be computed as $ Z |B'_{{\rm ph},2} \rangle = c_{i,2} |B'_{{\rm ph},2} \rangle =  (-1)^{{\rm mod} (N^{Q_i}_{\rm ocp}, 2)} |B'_{{\rm ph},2} \rangle $.

\paragraph*{Step 3: Compute $ H(Q_i,P_i)  $ based on $P_i$ and $Q_i$.} Provided the tags $P_i$ and $Q_i$, we compute the entry $ H(Q_i,P_i) $. We approach this calculation via the database approach. In particular, we apply the value of $P_i$ and $Q_i$ to iterate with a classically precomputed lookup table, in which process we pick the entry $ H(Q_i,P_i) $ with the tags $P_i$ and $Q_i$. This value of the entry is then recorded in the ancilla register $ V_{\rm tmp} $ that is initialized in the state $| V_{\rm tmp} \rangle = |0 \rangle $ with the proper representation and scaling [see Eqs. \eqref{eq:single_qubit_rot} and  \eqref{eq:def_sqrt}]. After this step, we denote the state of $ V_{\rm tmp}  $ as $| V'_{\rm tmp} \rangle $, which encodes the matrix element.

\paragraph*{Step 4: Compute $ \langle \mathcal{F}'_i , Q_i | \mathcal{H} | \mathcal{F} ,P_i \rangle $.} Controlled on the tags $P_i$ and $Q_i$, we compute $ \langle \mathcal{F}'_i , Q_i | \mathcal{H} | \mathcal{F} ,P_i \rangle $ with the phase $c_{i,1}$ encoded in $| B'_{{\rm ph},1} \rangle $,  $c_{i,2}$ encoded in $| B'_{{\rm ph},2} \rangle $, and the entry $\langle Q_i |H |P_i \rangle $ encoded in $| V'_{\rm tmp}\rangle $. Combining the results encoded in these three ancilla registers, we can compute $ \langle \mathcal{F}'_i , Q_i | \mathcal{H} | \mathcal{F} ,P_i \rangle $ according to Eq. \eqref{eq:computation_formula_OH}. The result is encoded in the multi-qubit register $| R_{\rm elem} \rangle $ that is initialized in the $|0 \rangle $ state. This can be achieved, for example, by a sequence of CNOT gates (which copy the bitwise data from $| V'_{\rm tmp} \rangle $ to $| R_{\rm elem} \rangle $, and extract the phases from $| B'_{{\rm ph},1} \rangle $ and $ | B'_{{\rm ph},2} \rangle $ to the most significant qubit of $| R_{\rm elem} \rangle $ to be the overall sign).

With the above steps, we generate the two-body kernel $ \langle \mathcal{F}'_i , Q_i | \mathcal{H} | \mathcal{F} ,P_i \rangle $ with the input $| \mathcal{F} \rangle $, $|P_i \rangle $, $| \mathcal{F}'_i \rangle $, and $|Q_i \rangle $ controlled on $|y_i(\mathcal{F},P_i,Q_i) \rangle =0$. After these operations, we need to uncompute the ancillas by reversing the above operations. 

This completes the construction of the $O_{\rm H}$ oracle.

\subsection{Complexity analysis of oracle implementation}
\label{sec:complex_analysis}

We analyze the asymptotic gate cost of the oracles $O_{\rm F}$ and $O_{\rm H}$ according to the steps outlined in Sec. \ref{sec:O_F_oracle_sec} and Sec. \ref{sec:OH_oracle_sec}. Our analysis of the gate count for each step of the $O_{\rm F}$ and $O_{\rm H}$ is given in terms of the elementary one-qubit and two-qubit gates \cite{NielsenANDChuang:2001} and the controlled version of them as well.\footnote{Alternatively, one can also analyze the gate complexity in terms of log-local operations as in Ref. \cite{Kirby:2021ajp}, where such log-local operations can be compiled into primitive gates \cite{Vedral:1996,Abhari:2014} that are hardware specific.} The asymptotic qubit cost is also provided for each oracle.

\subsubsection{Analysis of the $O_{\rm F}$ oracle}
The gate analysis of each step in the construction of the $O_{\rm F}$ oracle is as follows. 
\paragraph*{Step 1.} It takes $ N_{\rm sp} $ gates to achieve the set of pairwise operations that are controlled by one set of $N_{\rm sp} $ qubits and act on the other set of $N_{\rm sp}$ qubits. 

\paragraph*{Step 2.} In this step, the upper bound of the gate count is $ \widetilde{\mathcal{O}}(N^4_{\rm sp}) $.\footnote{In the following context, we employ the notation of $\widetilde{\mathcal{O}}$ to denote the suppression of the logarithmic components in the corresponding upper bound. For example, such logarithmic components can result from compiling multi-controlled gates into corresponding local gates in this work.} This scaling can be estimated by the possible number of the pairs ($P_i, Q_i$), i.e., $\mathcal{D} \in \mathcal{O}(N^4_{\rm sp})$. 
Indeed, there are at most $N_{\rm sp}^2$ choices for the active SP states (each tagged by $P_i$) in the input many-nucleon (Fock) state and, likewise, $N_{\rm sp}^2$ different combinations of the active SP states (each tagged by $Q_i$) in the output Fock state. { The uncomputation of the ancilla register that encodes the index $i$ via ($P_i, Q_i$) is expected to take the same asymptotic gate cost, i.e., $ \widetilde{\mathcal{O}}(N^4_{\rm sp})$.} Note, however, that this gate scaling can be much less than $ \widetilde {\mathcal{O} }(N_{\rm sp}^4) $ for realistic calculations due to the properties of the Hamiltonian such as conservation of parity and total angular momentum projection. 

\paragraph*{Step 3.} It takes $ \widetilde{\mathcal{O}}(1)$ gates to check the occupations and then record the possible error message in the ancilla for the pair of SP bases $ \{r_i,s_i \}$ specified by $P_i$. We have $\mathcal{O}(N_{\rm sp}^2)$ such $P_i$'s. Controlled by $Q_i$, which can take $\mathcal{O}(N_{\rm sp}^2)$ different values, it takes $\widetilde{ \mathcal{O} } (N_{\rm sp}^4)$ gates to accomplish this step for all the possible cases with specific $| \mathcal{F} \rangle $ and $P_i$. 

\paragraph*{Step 4.} We need $\widetilde{ \mathcal{O} }(N_{\rm sp}^4)$ gates for this step. In particular, there are $\mathcal{O}(N_{\rm sp}^2)$ combinations of the active pair of the SP bases. For each combination $\{r_i,s_i \}$ specified by $ P_i $, we need $\widetilde{ \mathcal{O} }(1)$ gates to flip the qubits that correspond to $r_i ^{\rm th}$ and $s_i^{\rm th}$ SP states. 
While this operation is controlled by the tag $Q_i$ (which can take $\mathcal{O}(N_{\rm sp}^2)$ different values), the total gate count for this step are upper bounded by $\widetilde { \mathcal{O} } (N_{\rm sp}^4)$.

\paragraph*{Step 5 and Step 6.} The operations these two steps are, in essence, identical to those in {\it Step 3} and {\it Step 4}, respectively. By analogous analysis, we obtain the upper bound of the total gate count for {\it Step 5} and {\it Step 6}, which is $\widetilde{ \mathcal{O} } (N_{\rm sp}^4)$.

\paragraph*{Step 7.} In this step, we compile the error message. This takes $\mathcal{O}(1)$ gates for a specific choice of $(P_i,Q_i)$. As there are $\mathcal{O}(N_{\rm sp}^4)$ different pairs of tags $(P_i,Q_i)$, the overall gate count for this step scales as $\mathcal{O}(N_{\rm sp}^4)$.

In sum, it takes $\widetilde{ \mathcal{O} }(N_{\rm sp}^4) $ gates to complete the $O_{\rm F}$ oracle, where we have included the gate count to uncompute the ancillas. It is worth noting that, this gate cost of $O_{\rm F}$ directly scales with the number of monomials of the ladder operators in the Hamiltonian, as shown in {\it Step 2} above. We also note that, with the $O_{\rm F}$ oracle, the quantum computer operates each input specified by the many-nucleon state $| \mathcal{F} \rangle $ and the index $i$ in parallel, which is referred to as the ``quantum parallelism".

We analyze the qubit cost for the $O_{\rm F}$ oracle. In particular, we need $2N_{\rm sp}$ qubits to encode the input and output Fock states via the direct encoding scheme. On the other hand, we need to store the following identity in terms of binary strings: 1) the index $i$ of $N^4_{\rm sp} $ possible values (which takes $4\log N_{\rm sp}$ qubits); 2) the tag $P_i $ of $N^2_{\rm sp} $ possible values (which takes $2\log N_{\rm sp}$ qubits); and 3) the tag $Q_i $ of $N^2_{\rm sp} $ possible values (which takes $2\log N_{\rm sp}$ qubits). In addition, we also need three qubits to process and store the error messages. Overall, the upper bound of the qubit count for the $O_{\rm F}$ oracle are $ \mathcal{O}( 2N_{\rm sp} + 8\log N_{\rm sp} +3 )= \widetilde{\mathcal{O}}(N_{\rm sp} ) $. 

\subsubsection{Analysis of the $O_{\rm H}$ oracle}
The asymptotic gate scaling of each step in the $O_{\rm H}$ oracle is as follows. 

\paragraph*{Step 1.} For one single choice of $ P_i $ that tags the pair of SP bases $\{r_i,s_i\}$, we need $\widetilde{ \mathcal{O} } (N_{\rm sp})$ gates to count the occupation for the relevant qubits that encode the $r_i +1 , \ r_i + 2 , \cdots , s_i -2, s_i -1 $ SP bases. As there are $\mathcal{O} (N^4_{\rm sp}) $ different $(P_i, Q_i)$'s, we need at most $\widetilde{ \mathcal{O} } (N^5_{\rm sp})$ gates for this step. 

\paragraph*{Step 2.} This step precedes with the same procedures as in {\it Step 1}. By analogous analysis, we find that this step also takes $\widetilde{\mathcal{O}}(N^5_{\rm sp})$ gates. 

\paragraph*{Step 3.} We use the pairwise tags $P_i $ and $Q_i$ to obtain the entry $\langle Q_i |H |P_i \rangle $ from the precomputed database.
In practice, the $O_{\rm H}$ oracle provides this entry in some finite precision $\xi$, using $ \mathcal{O}(\log \frac{1}{\xi}) $ qubits for the output (we assume $\xi \ll \epsilon $ so the imprecision of this entry does not affect the analysis as in Ref. \cite{DWBerry:2012}). With the tags $P_i $ and $Q_i$, it takes $\widetilde{ \mathcal{O} } \big( \log \frac{1}{\xi} \big) $ gates to copy $\langle Q_i |H |P_i \rangle $ from the database to the ancilla register $ V_{\rm tmp} $ in terms of binaries. Since there are $ \mathcal{D} \in \mathcal{O}(N^4_{\rm sp})$ distinct pairs of $(P_i,Q_i)$, it takes $\widetilde{\mathcal{O}} ( N^4_{\rm sp} ) $ gates to complete this step. Note that here we have further suppressed the factor $ \log \frac{1}{\xi} $ that is a constant determined by the precision of the input database.

\paragraph*{Step 4.} To get the output two-body kernel $ \langle \mathcal{F}'_i , Q_i | \mathcal{H} | \mathcal{F} ,P_i \rangle $, it takes $\mathcal{O}(\log \frac{1}{\xi} ) $ gates to copy the bit strings from the ancilla $| V'_{\rm tmp} \rangle $ to the $| R_{\rm elem} \rangle $, and $\mathcal{O}(1)$ gates to extract the corresponding phases from $| B'_{{\rm ph},1} \rangle $ and $| B'_{{\rm ph},2} \rangle $ to $| R_{\rm elem} \rangle $. It is noted that this part of the circuit is shared by all the other calculations with different input to the $O_{\rm H}$ oracle.

Including the asymptotic gate count to uncompute the ancillas (which doubles the asymptotic gate scaling for the above steps), we obtain the asymptotic scaling of gate count to be $\widetilde{ \mathcal{O} }( N^5_{\rm sp} )$ for the complete design of the $O_{\rm H}$ oracle. 

As for the asymptotic qubit cost, we notice that the Fock states $|\mathcal{F} \rangle $ and $|\mathcal{F}_i'\rangle $, the tag states $| P_i \rangle $ and $| Q_i \rangle $ are inherited from the $O_{\rm F}$ oracle, of which the qubit cost has been already counted. While one can store the matrix elements up to the precision $\xi$ in the precomputed database with $ \mathcal{O}(\log \frac{1}{\xi}) $ qubits, the additional qubit cost arises from the registers $ R_{\rm elem}$, $V_{\rm tmp}$, $B_{{\rm ph},1}$ and $B_{{\rm ph},2}$. 
We note that it takes $\mathcal{O}(\log \frac{1}{\xi})$ qubits to store the matrix element obtained from the database with precision up to $\xi$, while $B_{{\rm ph},1}$ and $B_{{\rm ph},2}$ are both single-qubit registers.
Therefore, the number of ancilla qubits for the $\mathcal{O}_{\rm H}$ oracle scales as $\mathcal{O}(\log \frac{1}{\xi})$, which is of the order $\mathcal{O}(1)$ for the input matrix elements of finite precision.

\section{Summary of the asymptotic qubit and gate cost}
\label{sec:resource_section}

In Sec. \ref{sec:oracle_designs}, we present the enumerator oracle $O_{\rm F}$ and the matrix-element oracle $O_{\rm H}$ that operate with the Fock states within the DE scheme. 
We present in Sec. \ref{sec:input_model_Fock_state_based} the construction of the isometry $\mathcal{T}$ via $\mathcal{O}(1)$ oracle queries to $O_{\rm F}$ and $O_{\rm H}$, and the input model to access the many-nucleon Hamiltonian via the block encoding scheme. High-level Hamiltonian simulation algorithms, such as the quantum signal processing \cite{Low:2017,Low:2019} and the RDS \cite{Berry:2020} (reviewed in Sec. \ref{sec:normal_oracle_model}), can function with the Fock-state-based input model. In this section, we analyze the asymptotic gate and qubit cost for the simulation algorithms with our Fock-state-based input model. We also show the comparison between our framework (algorithm) for simulating the time-independent Hamiltonian with previous works that simulate the molecular Hamiltonian.

We start with the asymptotic scaling of the qubit cost for the input model. Based on the discussions in Sec. \ref{sec:complex_analysis}, the $O_{\rm F}$ oracle dominates the qubit cost, of which the asymptotic scaling is $\widetilde{ \mathcal{O} } (N_{\rm sp})$. 
We note that this qubit scaling is dominated by the qubit resources necessary to encode the input and output Fock states within the DE scheme. 
This scaling is independent of the number of nucleons in the system; it only depends on the size of the SP basis set $ \mathbb{S} = \{ |\beta _0 \rangle ,\ |\beta _1 \rangle ,\ \cdots ,\ |\beta _{N_{\rm sp}-1} \rangle \} $ as discussed in Sec. \ref{sec:DirectEncoding}.

As for the asymptotic scaling of the gate cost for the input model, we find that the implementation of the $O_{\rm H}$ oracle dominates the gate cost, which scales as $ \widetilde{ \mathcal{O} } (N_{\rm sp}^5)$.
Therefore, the asymptotic gate cost for executing $\mathcal{O}(1)$ oracles scales as $\widetilde{ \mathcal{O} } (N_{\rm sp}^5)$ in constructing the isometry $\mathcal{T}$ [Eq. \eqref{eq:isometry_T_modified}].\footnote{We remark that, in constructing $\mathcal{T}$, the oracles $O_{\rm F}$ and $O_{\rm H}$ operate controlled on the ancilla qubit $b$ being in the state $|b\rangle = |0 \rangle $ according to the discussion in Sec. \ref{sec:isometry_construction_fock_state_based} (or Sec. \ref{sec:isometry_construction}). However, this does not affect the asymptotic gate cost $\widetilde{ \mathcal{O} } (N_{\rm sp}^5)$ for the input model.}

High-level simulation algorithms can be implemented with our Fock-state-based input model to solve both the dynamics and structure problems of many-nucleon systems.
Based on the discussion in Sec. \ref{sec:input_model_Fock_state_based}, if our input Hamiltonian is time-independent, then the algorithm of quantum signal processing \cite{Low:2017,Low:2019} can perform the Hamiltonian simulation with the scaling of the oracle queries being Eq. \eqref{eq:qubitization_oracle_scaling}. 
Multiplied by the gate cost of the oracles, the overall asymptotic gate cost for simulating the time-independent Hamiltonian is
\begin{align}
\widetilde{ \mathcal{O} } \Bigg( N_{\rm sp}^5 \Bigg( \mathcal{D} \Lambda _m  t + \frac{\log (\frac{1}{\epsilon })}{\log \log (\frac{1}{\epsilon })} \Bigg) \Bigg) . \label{eq:gate_cost_QSP}
\end{align}
We further suppress the logarithmic component and rewrite the above equation as
\begin{align}
\widetilde{\mathcal{O}} \big( N_{\rm sp}^5 \mathcal{D} \Lambda _m t  \big) .
\end{align}

For simulating the time-dependent Hamiltonians, the RDS algorithm \cite{Berry:2020} provides the query complexity [Eq. \eqref{eq:RDS_algorithm_query_complexity}] that is optimal with respect to the simulation error and near optimal with respect to the simulation time.
We expect that the dominant gate cost is from the $O_{\rm H}$ oracle,\footnote{We assume the other two oracles $O_{\rm var}$ and $O_{\rm norm}$ can be implemented efficiently as assumed in Ref. \cite{Berry:2020}.} and the total asymptotic gate cost for evolving the time-dependent Hamiltonian from $0$ to $t$ via the RDS algorithm based on our input model is
\begin{align}
\widetilde{ \mathcal{O} } \Bigg( N_{\rm sp}^5 \Bigg( \widetilde{\tau }' \frac{\log (\frac{  \widetilde{\tau }'  }{\epsilon})}{\log \log (\frac{ \widetilde{\tau }' }{\epsilon })} \Bigg) \Bigg) ,
\end{align}
with $ \widetilde{\tau }' = \mathcal{D} \int _{0}^{t} \Lambda_m(t') dt' $ defined  in Eq. \eqref{eq:L1_scaling_changed}. 
We further suppress the logarithmic component and the overall asymptotic gate cost for the simulation can be rewritten as 
\begin{align}
\widetilde{\mathcal{O}} \Big( N_{\rm sp}^5 \mathcal{D} \int _{0}^{t} \Lambda_m(t') dt' \Big) .
\end{align}

Similar to the analysis of simulating the time-independent Hamiltonian with our input model [Eq. \eqref{eq:gate_cost_QSP}], we can compute the asymptotic gate cost for the structure calculation [Sec. \ref{sec:structure_calculation}] via the Rodeo algorithm to be 
\begin{align}
\widetilde{ \mathcal{O} } \Bigg( N_{\rm sp}^5 \Bigg( \mathcal{D} \Lambda _m  \Delta + \frac{\log (\frac{1}{\epsilon })}{\log \log (\frac{1}{\epsilon })} \Bigg) \Bigg) ,
\end{align}
where we have applied Eq. \eqref{eq:rodeo_with_qubitization} and taken into account that the dominant gate cost results from the $O_{\rm H}$ oracle, which scales as $\widetilde{ \mathcal{O}} (N^5_{\rm sp})$.  Recall also that $\Delta $ is some constant from the discussion in Sec. \ref{sec:structure_calculation}. With further suppression of the logarithmic factors, this asymptotic gate cost can be rewritten as
\begin{align}
\widetilde{\mathcal{O}} \big( N_{\rm sp}^5 \mathcal{D} \Lambda _m \Delta  \big)
\end{align}

To sum, the asymptotic gate cost for the dynamics and structure calculations based on our Fock-state-based input model is 
\begin{align}
\widetilde{ \mathcal{O}} (N_{\rm sp}^9 \Upsilon ), \label{eq:general_scaling}
\end{align}
where we take $\Upsilon =  \Lambda _m t$ for simulating time-independent Hamiltonian, $\Upsilon = \int _{0}^{t} \Lambda_m(t') dt' $ for simulating time-dependent Hamiltonian, and $\Upsilon = \Lambda _m \Delta $ for structure calculations. In Eq. \eqref{eq:general_scaling}, we have also taken into account that the total number of the monomials of the ladder operators $\mathcal{D}$ scales as $ \mathcal{O}(N_{\rm sp}^4)$ for the second-quantized Hamiltonian that includes only two-body terms [Eqs. \eqref{eq:H_total} and \eqref{eq:H_modified_H_prime}].

We expect that this scaling is loose. In particular, $\mathcal{D}$ is restricted by the properties of the Hamiltonians in practical applications. This can be understood by the limited model problems in Sec. \ref{sec:applications} below. 
One would expect $\mathcal{D} = \tbinom{6}{2} \tbinom{6}{2} = 900$ for the simplified pairing Hamiltonian with 6 SP bases, but there are indeed $\mathcal{D}=9$ monomials of the ladder operators [Eq. \eqref{eq:H_pair_temp}] that satisfy the properties of the Hamiltonian. As for the other example of the four-neutron problem with 12 SP bases, one also see the reduction in $\mathcal{D} $ due to the symmetries of the Hamiltonian. On the other hand, one also expects that the gate cost of the $O_{\rm F}$ and $O_{\rm H}$ reduces with $\mathcal{D} $ as they are designed according to the momomials of the ladder operators in the second-quantized Hamiltonian.

\subsection{Comparison to previous works}
We compare our algorithm with previous works of Hamiltonian simulations in quantum chemistry, which deals with the second-quantized molecular Hamiltonian that includes at most two-body terms, or the corresponding first-quantized Hamiltonian that involves at most two-body operators. Note that the two-body terms (operators) dominates the gate cost over what results from the one-body terms in the Hamiltonian. 

For this comparison, we first evaluate our asymptotic gate cost for simulating the time-independent Hamiltonian based on Eq. \eqref{eq:general_scaling}.  
We note that the absolute value of the Hamiltonian matrix element scales as $\Theta (N_{\rm sp}) $.\footnote{With the 3DHO basis, the absolute value of the Hamiltonian matrix element is dominated by the matrix element of the $T^{\rm rel}_{pqrs}$ and $H_{pqrs}^{\rm CM}$ in Eq. \eqref{eq:second_quantized_H_matrix_element}, while $V_{pqrs}^{\rm NN}$ diminishes with increasing radial quantum number $n$ of the 3DHO basis. Both $T^{\rm rel}_{pqrs}$ and $H_{pqrs}^{\rm CM}$ scales as $\Theta (N_{sp}) $.} Therefore, we take $\Lambda _m \in \Theta (N_{\rm sp})$ and estimate the gate cost of our framework to be $ \widetilde{\mathcal{O}} \big( N_{\rm sp}^{10} t  \big)  $. We note that this gate cost is loose as we have not included the restrictions of the properties of the target Hamiltonians; a tighter bound for the calculations of complex nuclei is suspected to be $ \widetilde{\mathcal{O}} \big( N_{\rm sp}^8 t  \big)  $.

As in quantum chemistry, the tightest known bound of the gate cost for simulating the second-quantized molecular Hamiltonian via arbitrarily high-order Trotter formula is $\widetilde{ \mathcal{O}} (N_{\rm sp}^8 t/\epsilon ^{o(1) } ) $ \cite{Berry:2007dwb,Wiebe:2011wbe,RBabbush:2016}. With significantly more practical Trotter decomposition, the best known gate cost scales as $\widetilde{ \mathcal{O}} (N_{\rm sp}^9 \sqrt{t^3/\epsilon}) $ \cite{Hastings:2015hast,RBabbush:2016}. The scaling of our gate complexity with $N_{\rm sp}$ is close to that of the Trotter-based methods. 
While the algorithm based on the arbitrarily high-order Trotter formula also presents an optimal scaling with the simulation time according to the no fast-forwarding theorem [Theorem 3 in Ref. \cite{Berry:2007dwb}], our algorithm exhibits an exponential improvement in precision over these Trotter-based algorithms.

Meanwhile, Ref. \cite{RBabbush:2016} introduces a so-called ``database" algorithm. This algorithm represents the molecular Hamiltonian as a weighted sum of $\mathcal{O}(N^4_{\rm sp})$ local unitaries, and approximates the time-evolution by the truncated Taylor series approach \cite{Berry:2015prlDWB}. It accesses the Hamiltonian matrix elements with a classically precomputed database of the molecular integrals. The asymptotic gate count of the database algorithm [Eq. (46) in Ref. \cite{RBabbush:2016}] scales as 
\begin{align}
{\mathcal{O}} \Bigg( N_{\rm sp}^4 \widetilde{\Lambda} t \frac{\log (\frac{ N_{\rm sp} t }{\epsilon })}{\log \log (\frac{N_{\rm sp} t }{\epsilon })} \Bigg) = \widetilde{\mathcal{O}} \big(  N^8_{\rm sp} t  \big),
\end{align}
with the normalization factor $ \widetilde{\Lambda} \in \mathcal{O}(N_{\rm sp}^4 )$. Our asymptotic gate cost is close to that of the database algorithm, with an extra factor of $N^2_{\rm sp}$. 

Ref. \cite{RBabbush:2016} also introduces the ``on-the-fly" algorithm. 
This algorithm computes the two-electron integrals via the ``so-called" integrand oracle which is designed based on the discretization of space in Riemann integration. Within the framework of the truncated Taylor series \cite{Berry:2015prlDWB}, the on-the-fly algorithm presents the gate cost to be $ \widetilde{\mathcal{O}} \big(  N^5_{\rm sp} t  \big) $. 
Meanwhile, Ref. \cite{Kirby:2021ajp} introduces the algorithm for simulating second-quantized Hamiltonians via the Fock-state-based input model employing the CE scheme. 
The authors design their input model based on the controlled arithmetic operations that can be realized via log-local operations on quantum computers. With the algorithms of Ref. \cite{Kirby:2021ajp}, the complexity for simulating $A$-electron quantum chemistry Hamiltonian (including two-body terms at most) is reported to be $ \widetilde{\mathcal{O}} \big( A^2 N^4_{\rm sp} t  \big) $ in terms of log-local operations.\footnote{Such log-local operations can be compiled into hardware-specific primitive gates \cite{Vedral:1996,Abhari:2014}.} Moreover, Ref. \cite{Babbush:2018bubbush} reports an algorithm that adopts a compressed configuration-interaction matrix representation for simulating the first-quantized molecular Hamiltonian based on the truncated Taylor series approach. 
With the application of the configuration-interaction matrix representation, the Slater-Condon rules \cite{Slater:1929,Condon:1930} are explicitly enforced for the Hamiltonian matrix elements that are computed on the fly. The gate cost of this algorithm is $ \widetilde{\mathcal{O}} \big( A^2 N^3_{\rm sp} t  \big) $. Compared to these algorithms, the gate cost of our algorithm has a worse scaling in the factor of $N_{\rm sp}$. We remark that our gate cost is loose due to the fact that the properties of the target Hamiltonians are enforced implicitly to the input model (which reduces $\mathcal{D}$). We also comment that our algorithm is designed on the basis of the elementary gate operations aiming for straightforward prototype applications on quantum computer. The gate cost of our algorithm can be further improved by incorporating the (controlled-) arithmetic operations and the on-site evaluations of the two-body kernels that contribute to the Hamiltonian matrix elements.

\section{Model problems}
\label{sec:applications}

For pedagogical purposes, we present the applications of our method to solve the dynamics and structure problems in nuclear physics with two model problems.
In both problems, we retain only the two-body terms in the corresponding second-quantized Hamiltonians and restrict our discussions to truncated model spaces.
Whereas the execution of the high-level sparse-matrix simulations algorithms are extensively discussed in Refs. \cite{Low:2019,Berry:2020,Choi:2020pdg,Qian:2021wya,Bee-Lindgren:2022nqb}, we will focus on explaining the design of our input model. 
We remark that the discussions of the simple model problems in this section can be generalized to more complex cases, e.g., those containing many-body terms and larger model-space size, whereas the corresponding analyses for such general calculations are summarized in Sec. \ref{sec:resource_section}.

\subsection{Pairing Hamiltonian}
\label{sec:pairing_Hamiltonian}
We start with a simple Hamiltonian for a many-nucleon system, where we neglect the kinetic energy of the nucleons and consider only the pairing interaction $V_{\rm pairing}$ between the nucleons. The Hamiltonian of the many-nucleon system is defined as 
\begin{align}
H_{\rm pair} = V_{\rm pairing} = g \sum  _p \sum _r a_{p,-}^{\dag} a_{p,+}^{\dag} a_{r,+} a_{r,-} , \label{eq:H_pair_Hamilotonian}
\end{align}
where the operator $a_{p,\pm}^{\dag} $ creates a SP state labeled by the quantum numbers $(n_p, l_p, j_p, \pm m_{j,p}, \tau _p )$, while the operator $ a_{r,\pm} $ annihilates a SP state labeled by the quantum numbers $(n_r, l_r, j_r, \pm m_{j,r}, \tau _r)$. 
$g$ is the coupling constant of the interaction; we assume $g $ to be real without loss of generality.
We assume that $H_{\rm pair} $ is independent of the isospin (i.e., the species of the nucleons).
Each single term $ a_{p,-}^{\dag} a_{p,+}^{\dag} a_{r,+} a_{r,-} $ in $H_{\rm pair} $ operates on two pairs of SP bases, where each SP basis within a pair differs from its partner only in the total angular momentum  projection.

\begin{table*}[!ht] 
\caption{The restricted SP basis set for a single-species three-nucleon system. See text for details.}
\begin{tabular}{c c c c c }
\hline \hline
SP basis (qubit)  & $\ n\ $ & $\ l\ $ & $\ 2j\ $ & $\ 2m_j\ $  \\ 
\hline
${\bf 0} $ &  $0 $ & $0 $ & $1$ & $-1$  \\ 
${\bf 1} $ &  $0 $ & $0 $ & $1$ & $+1$  \\ 
${\bf 2} $ &  $1 $ & $0 $ & $1$ & $-1$  \\ 
${\bf 3} $ &  $1 $ & $0 $ & $1$ & $+1$  \\ 
${\bf 4} $ &  $2 $ & $0 $ & $1$ & $-1$  \\ 
${\bf 5} $ &  $2 $ & $0 $ & $1$ & $+1$  \\ 
\hline \hline
\end{tabular} 
\label{tab:basisPairingHamiltonian}
\end{table*}
For the purpose of demonstration, we consider a three-nucleon system  ($A=3$) with a single species (either neutrons only or protons only) within a restricted set of SP bases shown in Table \ref{tab:basisPairingHamiltonian}. In particular, we retain the six SP bases with 1) the principle quantum number $n\leq 2$; 2) the orbital angular momentum $l=0$; 3) the total angular momentum $j= \frac{1}{2}$; and 4) the projection of the total angular momentum $m_j = \pm \frac{1}{2}$. We omit the quantum number of the spin for each SP basis in Table \ref{tab:basisPairingHamiltonian}, which is understood to be $\frac{1}{2}$. We also omit the quantum numbers for the isospin as we have assumed that $H_{\rm pair} $ is isospin independent. Each SP basis is labeled and mapped to a distinct qubit in a quantum register (index shown in the first column in Table \ref{tab:basisPairingHamiltonian}), whereas the state of each qubit denotes the occupancy of the corresponding SP basis (recall that the state $|1 \rangle $ denotes occupied and $|0 \rangle $ denotes vacant). To get some more intuition, we can rewrite the pairing Hamiltonian $ H_{\rm pair} $ [Eq. \eqref{eq:H_pair_Hamilotonian}] within this restricted basis space as
\begin{align}
H_{\rm pair} = g \Big[ & \underbrace{ a_0^{\dag} a_1^{\dag} a_1 a_0 }_{i=0} + \underbrace{ a_2^{\dag} a_3^{\dag} a_1 a_0 }_{ i=1 } + \underbrace{ a_4^{\dag} a_5^{\dag} a_1 a_0 }_{i=2} + \underbrace{ a_0^{\dag} a_1^{\dag} a_3 a_2 }_{i=3} + \underbrace{ a_2^{\dag} a_3^{\dag} a_3 a_2}_{i=4} + \underbrace{ a_4^{\dag} a_5^{\dag} a_3 a_2 }_{i=5} + \underbrace{ a_0^{\dag} a_1^{\dag} a_5 a_4 }_{i=6} + \underbrace{ a_2^{\dag} a_3^{\dag} a_5 a_4 }_{i=7} + \underbrace{ a_4^{\dag} a_5^{\dag} a_5 a_4 }_{i=8} \Big]
, \label{eq:H_pair_temp}
\end{align}
where we employ instead a subscript to denote the index of the SP basis [Table \ref{tab:basisPairingHamiltonian}] that each ladder operator applies on. 

For this three-nucleon system in the restricted SP basis set (Table \ref{tab:basisPairingHamiltonian}), we have $\tbinom{6}{3}=20$ three-nucleon states in total. In particular, there are 1) 9 states with $M_J = -\frac{1}{2}$; 2) 9 states with $M_J = +\frac{1}{2}$; 3) 1 state with $M_J = -\frac{3}{2}$; and 4) 1 state with $M_J = +\frac{3}{2}$. We sort these three-nucleon states in Table \ref{tab:mJ_scheme_pairing_model}. We can also solve for the matrix elements of the pairing Hamiltonian within the set many-nucleon bases shown in Table \ref{tab:mJ_scheme_pairing_model}. For example, in terms of the three-nucleon states $\{ | 0,1,3 \rangle , | 0,1,5 \rangle , | 0,3,5 \rangle , | 1,2,3 \rangle , | 1,2,5 \rangle , | 1,3,4 \rangle , | 1,4,5 \rangle , | 2,3,5 \rangle , | 3,4,5 \rangle \} $, we can write the matrix of the pairing Hamiltonian in the $M_J=+\frac{1}{2} $ subspace as
\begin{align}
H_{\rm pair}\big(M_J= + \frac{1}{2} \big) = g
\begin{pmatrix}
1 & 0 & 0 & 0 & 0 & 0 & 0 & 0 & 1 \\ 
0 & 1 & 0 & 0 & 0 & 0 & 0 & 1 & 0 \\ 
0 & 0 & 0 & 0 & 0 & 0 & 0 & 0 & 0 \\ 
0 & 0 & 0 & 1 & 0 & 0 & 1 & 0 & 0 \\ 
0 & 0 & 0 & 0 & 0 & 0 & 0 & 0 & 0 \\ 
0 & 0 & 0 & 0 & 0 & 0 & 0 & 0 & 0 \\ 
0 & 0 & 0 & 1 & 0 & 0 & 1 & 0 & 0 \\ 
0 & 1 & 0 & 0 & 0 & 0 & 0 & 1 & 0 \\ 
1 & 0 & 0 & 0 & 0 & 0 & 0 & 0 & 1
\end{pmatrix} , \label{eq:matrix_rep_pairing_Hamiltonian}
\end{align}
whereas the three-nucleon states with different $M_J$'s do not connect to each other via the action of the pairing Hamiltonian (i.e., vanishing Hamiltonian matrix element), as the pairing Hamiltonian preserves the quantum number $ M_J $.

\begin{table*}[!ht] 
\caption{The three-nucleon states sorted according to $M_J$. Only the indices of the occupied SP states are recorded, where the quantum numbers of each SP state are in Table \ref{tab:basisPairingHamiltonian}. For example, the three-nucleon state $|0,1,3 \rangle $ is equivalent to $|110100 \rangle $ in notations.}
\begin{tabular}{cccccccccc}
\hline \hline
$M_J$  & \multicolumn{9}{c}{three-nucleon state}               \\
\hline
$+3/2$ & $\ |1,3,5\rangle \ $ &  --  &  --  &  --  &  --   &   -- &   -- &  --   &  --   \\
$+1/2$ & $\ |0,1,3\rangle \ $ & $\ |0,1,5\rangle \ $ & $\ |0,3,5\rangle \ $ & $\ |1,2,3\rangle \ $ & $\ |1,2,5\rangle \ $ & $\ |1,3,4\rangle \ $ & $\ |1,4,5\rangle \ $ & $\ |2,3,5\rangle \ $ & $\ |3,4,5\rangle \ $ \\
$-1/2$ & $\ |0,1,2\rangle \ $ & $\ |0,1,4\rangle \ $ & $\ |0,2,3\rangle \ $ & $\ |0,2,5\rangle \ $ & $\ |0,3,4\rangle \ $ & $\ |0,4,5\rangle \ $ & $\ |1,2,4\rangle \ $ & $\ |2,3,4\rangle \ $ & $\ |2,4,5\rangle \ $ \\
$-3/2$ & $\ |0,2,4\rangle \ $ &  --  &  --  &  --  &  --   &   --  &  --   & --   & --  \\ \hline  \hline
\end{tabular}
\label{tab:mJ_scheme_pairing_model}
\end{table*}

To facilitate the oracle design, we rewrite Eq. \eqref{eq:H_pair_temp} according to Eq. \eqref{eq:H_tot_modified} as 
\begin{align}
\mathcal{H}_{\rm pair} = \sum _{i=0}^{8} H_{\rm pair}(Q_i,P_i) b_{Q_i}^{\dag} b_{P_i} \otimes |Q_i\rangle \langle P_i |, \label{eq:H_pair_temp_primed}
\end{align} 
with $ H_{\rm pair}(Q_i,P_i) = \langle p_i q_i | H_{\rm pair} | r_i s_i \rangle $. The index $i$ ($i=0,1,\cdots , 8$) labels different combinations of the ladder operators. The tag $P_i \mapsto \{r_i, s_i \}$ with $r_i <s_i$ ($Q_i \mapsto \{p_i,q_i \}$ with $p_i<q_i$) denotes the pair of SP states labeled by $r_i$ and $s_i$ ($p_i$ and $q_i$) on which the annihilation (creation) operators act. We recall that $b^{\dag}_{Q_i}b_{P_i} = a_{p_i}^{\dag} a_{q_i}^{\dag} a_{s_i} a_{r_i} $. With $i=2$, for example, we have $b^{\dag}_{Q_2}b_{P_2} = a_4^{\dag} a_5^{\dag} a_1 a_0$. In Table \ref{tab:P_Q_mapping}, we enumerate all the combinations of the tags $P_i $ and $Q_i$. In addition, it is noted that $ H_{\rm pair}(Q_i,P_i) = g $ for all the cases in Eq. \eqref{eq:H_pair_temp_primed}.

\begin{table*}[!ht] 
\caption{Distinct combinations of the tags $P_i \mapsto \{r_i,s_i\}$ ($r_i <s_i$) and $Q_i \mapsto \{p_i,q_i \}$ ($p_i<q_i$) for the three-nucleon problem described by the pairing Hamiltonian  $\mathcal{H} _{\rm pair}$ [Eq. \eqref{eq:H_pair_temp_primed}] in the restricted SP bases set [Table \ref{tab:basisPairingHamiltonian}]. $r_i$ and $s_i$ denote the indices of the SP bases operated by the corresponding annihilation operators, while $p_i$ and $q_i$ denote the indices of SP bases operated by the corresponding creation operators.
}
\begin{tabular}{c c c}
\hline \hline
$i$ & $P_i \mapsto \{r_i, s_i \}$ & $Q_i \mapsto \{p_i, q_i \}$ \\ 
\hline
0 & $P_0 \mapsto \{ 0,1 \}$ & $Q_0 \mapsto \{ 0,1 \} $ \\ 
1 & $P_1 \mapsto \{ 0,1 \}$ & $Q_1 \mapsto \{ 2,3 \} $ \\ 
2 & $P_2 \mapsto \{ 0,1 \}$ & $Q_2 \mapsto \{ 4,5 \} $ \\ 
3 & $P_3 \mapsto \{ 2,3 \}$ & $Q_3 \mapsto \{ 0,1 \} $ \\ 
4 & $P_4 \mapsto \{ 2,3 \}$ & $Q_4 \mapsto \{ 2,3 \} $ \\ 
5 & $P_5 \mapsto \{ 2,3 \}$ & $Q_5 \mapsto \{ 4,5 \} $ \\ 
6 & $P_6 \mapsto \{ 4,5 \}$ & $Q_6 \mapsto \{ 0,1 \} $ \\ 
7 & $P_7 \mapsto \{ 4,5 \}$ & $Q_7 \mapsto \{ 2,3 \} $ \\ 
8 & $P_8 \mapsto \{ 4,5 \}$ & $Q_8 \mapsto \{ 4,5 \} $ \\ 
\hline \hline
\end{tabular} 
\label{tab:P_Q_mapping}
\end{table*}

We show the work flow of the $O_{\rm F}$ oracle [Eq. \eqref{eq:O_F_prime}] with an example. Without loss of generality, we start with the input state $| \mathcal{F} \rangle = |0,1,3 \rangle = | 110100 \rangle $ and the index $i=2$. It is understood that the $O_{\rm F}$ operates on all the choices of the $ | \mathcal{F} \rangle $ and $i$ simultaneously. According to the discussion in Sec. \ref{sec:O_F_oracle_sec}, the $O_{\rm F}$ oracle executes as
\begin{enumerate}
\item copy the input state $ | \mathcal{F} \rangle = |0,1,3 \rangle  $. In the following, we operate on the copy of $ | \mathcal{F} \rangle$;

\item compute the tags $P_2 \mapsto \{0,1\}$ and $Q_2 \mapsto \{4,5\}$ based on $i=2$ by the iterations with the classically precomputed look-up table [Table \ref{tab:P_Q_mapping}];
\item check the occupations of the $0^{\rm th}$ and $1^{\rm st}$ SP bases in $ | \mathcal{F} \rangle $ by checking the states of the corresponding ($0^{\rm th}$ and $1^{\rm st}$) qubits (recall that $|0\rangle $ denotes that the SP state is ``vacant", while $|1 \rangle $ denotes that the SP state being ``occupied"). In this case, we find that both the $0^{\rm th}$ and $1^{\rm st}$ are occupied in $ | \mathcal{F} \rangle $. Therefore, we do not flip the corresponding ancilla qubit $|B_P\rangle= |0 \rangle$, and we end up having $|B_P' \rangle= |0 \rangle$;

\item annihilate the occupations on the $0^{\rm th}$ and $1^{\rm st}$ SP states by flipping the corresponding qubits from $|1\rangle $ (occupied) to $|0 \rangle $ (vacant) controlled by the tags $P_2$ and $Q_2$. The copy of $ | \mathcal{F} \rangle = |0,1,3 \rangle  $ becomes $|\mathcal{F}_{P_2} \rangle = |3\rangle = | 000100 \rangle$;

\item check the occupancies of the $4^{\rm th}$ and $5^{\rm th}$ SP states in $|\mathcal{F}_{P_2} \rangle $. This is done by checking the states of the corresponding qubits. In this case, both of the SP states are not occupied (the corresponding qubits are in the $|0\rangle $ states) and we do not flip the ancilla qubit $| B_Q \rangle =|0\rangle $. After this procedure, the ancilla is denoted as $| B'_Q \rangle =|0\rangle $;

\item create the occupations on the $4^{\rm th}$ and $5^{\rm th}$ SP states in $|\mathcal{F}_{P_2} \rangle $ by flipping the corresponding qubits from $|0 \rangle $ to $|1 \rangle $ controlled by the tags $P_2$ and $Q_2$. This produces the output state $|\mathcal{F}_2'\rangle =|3,4,5 \rangle = |000111\rangle$;

\item compile the error message in the ancillas controlled by $(P_i, Q_i)$. As both $|B_P' \rangle $ and $|B_Q' \rangle$ are in the state $|0\rangle $, we have $|a_2\rangle =|0\rangle $. This means that $\mathcal{F}_2'=|3,4,5 \rangle = |000111\rangle$ is a desired three-nucleon state that shall, in principle, produce non-vanishing contribution to the matrix element $  \langle \mathcal{F}'_2 , Q_2 | \mathcal{H} _{\rm pair} | \mathcal{F} ,P_2 \rangle  $. 
\end{enumerate}
Overall, we have the input to the $O_{\rm F}$ oracle to be $|\mathcal{F} \rangle =|0,1,3\rangle $ and $i =2 $, and we get the output to be $|\mathcal{F} \rangle =|0,1,3\rangle $, $i=2  $, $| P_2 \rangle $, $| \mathcal{F}'_2 \rangle = |3,4,5 \rangle $, $| Q_2 \rangle $, and $|a_2 \rangle = |0\rangle $.

Next, we discuss the implementation of the $O_{\rm H}$ oracle for this case. Controlled on $| a_2 \rangle =|0 \rangle $, the $O_{\rm H}$ functions with the input $ |\mathcal{F} \rangle =|0,1,3\rangle  $, $ | P_2 \rangle $, $| \mathcal{F}'_2 \rangle = |3,4,5 \rangle $, $|Q_2 \rangle $; it produces the matrix element $  \langle \mathcal{F}'_2 , Q_2 | \mathcal{H}_{\rm pair} | \mathcal{F} ,P_2 \rangle  $. In this simple model, we have 
\begin{align}
\langle \mathcal{F}'_2 , Q_2 | \mathcal{H}_{\rm pair} | \mathcal{F} ,P_2 \rangle = g,
\end{align}
without involving the calculations of phases. The reason for the ``phase-independent" matrix element is that each pair of the fermionic creation (annihilation) operators acts on adjacent SP states; in essence, this makes the pair of operators behave as if they are the ladder operators for bosons. That is, the product of the phase factors in Eq. \eqref{eq:computation_formula_OH} $c_{i,1} c_{i,2} = +1 $ for this case (and also for the other cases where the $O_{\rm H}$ operates). 

Having illustrated the work flow of the $O_{\rm F}$ and $O_{\rm H} $, we present concrete examples to show 1) how the isometry $\mathcal{T}$ is constructed [Sec. \ref{sec:isometry_construction}], and how $\mathcal{T}^{\dag}S\mathcal{T}$ is related to the Hamiltonian [Eqs. \eqref{eq:identity_Quantum_walk} and \eqref{eq:walk_identity}]. 

According to the formalism in Appendix \ref{sec:quantumWalk_Appd_A}, we can set $\Lambda _m \geq | g |$ and take $\mathcal{D}=9$. We perform more calculations with the input $ |\mathcal{F} \rangle =|0,1,3\rangle $ for the rest index $i$'s following the same approach as presented above. In particular, with the input $ |\mathcal{F} \rangle =|0,1,3\rangle $ and $i = 0$, the output of the $O_{\rm F}$ oracle are: 1) $ |\mathcal{F} \rangle =|0,1,3\rangle $; 2) $ i=0 $; 3) $P_0 \mapsto \{0,1\}$; 4) $| \mathcal{F}'_0 \rangle = |0,1,3 \rangle $; 5) $Q_0 \mapsto \{0,1\}$; and 6) $|a_0 \rangle = |0\rangle $. Controlled on $|a_0 \rangle = |0\rangle $, the $O_{\rm H}$ oracle functions, which results the matrix element $  \langle \mathcal{F}'_0 , Q_0 | \mathcal{H}_{\rm pair} | \mathcal{F} ,P_0 \rangle = g $. On the other hand, for the rest cases with $i=\ 1,\ 3,\ 4,\ 5,\ 6,\ 7,\ 8$, we end up with $|y_i \rangle = |1\rangle $, which flags the error message. In these cases, the $O_{\rm H}$ oracle is not activated. 

Having enumerated all the possible cases with $| \mathcal{F} \rangle =|0,1,3\rangle $ and $i=0,1,\cdots , 8$, we have for Eq. \eqref{eq:equivalent_to_A7} that 
\begin{align}
\mathcal{T}| 0,1,3 \rangle | 0 \rangle =& \sqrt{ \frac{1 }{ \mathcal{D} \Lambda _m } } \Big[ \sqrt{g} | 0,1,3 \rangle | 0 \rangle | 0 \rangle |P_0 \rangle | 0,1,3 \rangle |Q_0 \rangle | 0 \rangle |0\rangle |0\rangle \nonumber \\
& \ \ \ \ \ \ \ \ \ \  +  \sqrt{g} | 0,1,3 \rangle | 0 \rangle | { 0} \rangle |P_2 \rangle | 3,4,5 \rangle |Q_2 \rangle | 0 \rangle |0\rangle |0\rangle \Big] + | \perp _{ |0,1,3 \rangle } \rangle , \label{eq:T_1_relation}
\end{align}
where $ | \perp _{\{0,1,3\} } \rangle  $ denotes the terms that are orthogonal to the first two terms in the above equation. We note that $ | \perp _{\{0,1,3\} } \rangle  $ includes, in part, those terms that are from the cases with $|y_i \rangle = |1\rangle $ ($i=\ 1,\ 3,\ 4,\ 5,\ 6,\ 7,\ 8$). Meanwhile, we also have the complementary equation Eq. \eqref{eq:equivalent_to_A2} 
\begin{align}
\mathcal{T}| 0,1,3 \rangle | 1 \rangle =& | 0,1,3 \rangle | 1 \rangle |0 \rangle |0 \rangle |0 \rangle | 0 \rangle | 0 \rangle | 0 \rangle |1 \rangle . 
\end{align}

We repeat the calculations with the cases with the input being $ |\mathcal{G} \rangle =|2,3,5\rangle $ and $i=0,1,\cdots , 8$ and obtain
\begin{align}
\mathcal{T}|2,3,5 \rangle | 0 \rangle =& \sqrt{ \frac{1 }{\mathcal{D} \Lambda _m } } \Big[ \sqrt{g} | 2,3,5 \rangle | 0 \rangle | { 0} \rangle |P_3 \rangle | 0,1,5 \rangle |Q_3 \rangle | 0 \rangle |0\rangle |0\rangle \nonumber \\
& \ \ \ \ \ \ \ \ \ \  +  \sqrt{g} | 2,3,5  \rangle | 0 \rangle | { 0} \rangle |P_4 \rangle | 2,3,5 \rangle |Q_4 \rangle | 0 \rangle |0\rangle |0\rangle \Big] + | \perp _{ |2,3,5 \rangle } \rangle ,
\end{align}
where only the cases with $i=3$ and $i=4$ (see in Table \ref{tab:P_Q_mapping}) yield $|a_3\rangle =|0\rangle $ and $|a_4 \rangle =|0\rangle $, respectively. Again, $ | \perp _{ |2,3,5 \rangle } \rangle  $ denotes the terms that are orthogonal to the first two terms in the above equation. The complementary equation holds according to Eq. \eqref{eq:equivalent_to_A2} as
\begin{align}
\mathcal{T}| 2,3,5 \rangle | 1 \rangle =& | 2,3,5 \rangle | 1 \rangle |0 \rangle |0 \rangle |0 \rangle | 0 \rangle | 0 \rangle | 0 \rangle |1 \rangle .
\end{align}

For the case with the input being $|\mathcal{G}\rangle = |3,4,5\rangle $ and $i=0,1,\cdots , 8$, we have
\begin{align}
\mathcal{T}|3,4,5 \rangle | 0 \rangle =& \sqrt{ \frac{1 }{\mathcal{D} \Lambda _m } } \Big[ \sqrt{g} | 3,4,5 \rangle | 0 \rangle | { 0} \rangle |P_6 \rangle | 0,1,3 \rangle |Q_6 \rangle | 0 \rangle |0\rangle |0\rangle \nonumber \\
& \ \ \ \ \ \ \ \ \ \  +  \sqrt{g} |  3,4,5  \rangle | 0 \rangle | { 0} \rangle |P_8 \rangle |  3,4,5 \rangle |Q_8 \rangle | 0 \rangle |0\rangle |0\rangle \Big] + | \perp _{ |3,4,5 \rangle } \rangle ,
\end{align}
where only the cases with $i=6$ and $i=8$ yield $|a_6\rangle =|0\rangle $ and $|a_8 \rangle = |0\rangle $. The complementary equation holds 
\begin{align}
\mathcal{T}| 3,4,5 \rangle | 1 \rangle =& | 3,4,5 \rangle | 1 \rangle |0 \rangle |0 \rangle |0 \rangle | 0 \rangle | 0 \rangle | 0 \rangle |1 \rangle .
\end{align}

As for the case with the input being $|\mathcal{G}\rangle = |0,1,5\rangle $ and $i=0,1,\cdots , 8$, we have
\begin{align}
\mathcal{T}|0,1,5 \rangle | 0 \rangle =& \sqrt{ \frac{1 }{\mathcal{D} \Lambda _m } } \Big[ \sqrt{g} | 0,1,5 \rangle | 0 \rangle | { 0} \rangle |P_0 \rangle | 0,1,5 \rangle |Q_0 \rangle | 0 \rangle |0\rangle |0\rangle \nonumber \\
& \ \ \ \ \ \ \ \ \ \  +  \sqrt{g} |  0,1,5 \rangle | 0 \rangle | { 0} \rangle |P_1 \rangle |  2,3,5 \rangle |Q_1 \rangle | 0 \rangle |0\rangle |0\rangle \Big] + | \perp _{ |0,1,5 \rangle } \rangle ,
\end{align}
where only the cases with $i=0$ and $i=1$ yield $|a_0\rangle =|0\rangle $ and $|a_1 \rangle = |0\rangle $. According to Eq. \eqref{eq:equivalent_to_A2}, we also have the complementary equation 
\begin{align}
\mathcal{T}| 0,1,5 \rangle | 1 \rangle =& | 0,1,5 \rangle | 1 \rangle |0 \rangle |0 \rangle |0 \rangle | 0 \rangle | 0 \rangle | 0 \rangle |1 \rangle .
\end{align}

We can readily check [see detailed calculations in Eq. \eqref{eq:matrix_element_QuaWalk}], for example,
\begin{align}
\langle 0,1,3 | \langle 0| \mathcal{T} ^{\dag} S \mathcal{T} | 0,1,3  \rangle |0 \rangle = \langle 0,1,3 | H_{\rm pair} | 0,1,3 \rangle = g', \ & 
\langle 3,4,5 | \langle 0| \mathcal{T} ^{\dag} S \mathcal{T} | 0,1,3  \rangle |0 \rangle  = \langle 3,4,5 | H_{\rm pair} | 0,1,3 \rangle = g', \\ 
\langle 2,3,5 | \langle 0| \mathcal{T} ^{\dag} S \mathcal{T} | 0,1,5  \rangle |0 \rangle = \langle 2,3,5 | H_{\rm pair} | 0,1,5 \rangle = g', \ &
\langle 2,3,5 | \langle 0| \mathcal{T} ^{\dag} S \mathcal{T} | 2,3,5  \rangle |0 \rangle  = \langle 2,3,5 | H_{\rm pair} | 2,3,5 \rangle = g', \\
\langle 0,1,3 | \langle 0| \mathcal{T} ^{\dag} S \mathcal{T} | 2,3,5  \rangle |0 \rangle = \langle 0,1,3 | H_{\rm pair} | 2,3,5 \rangle = 0, \ & 
\langle 2,3,5 | \langle 0| \mathcal{T} ^{\dag} S \mathcal{T} | 0,1,3  \rangle |0 \rangle = \langle 2,3,5 | H_{\rm pair} | 0,1,3 \rangle = 0,
\end{align}
with $g' \coloneqq \frac{ 1 }{\mathcal{D} \Lambda _m} g$. The definition of the swap operator $S$ is shown in Eq. \eqref{eq:swap_operation}. In computing the above quantities, it is understood from the mapping shown in Table \ref{tab:P_Q_mapping} that
\begin{align}
\langle P_{i} | Q_j \rangle =  
\begin{cases}
1 , \ \text{for } \ i=0, \  1, \ 2, \ \text{and} \ j = 0,\ 3,\ 6 ,\\ 
1 , \ \text{for } \ i=3, \  4, \ 5, \ \text{and} \ j = 1,\ 4,\ 7 , \\
1 , \ \text{for } \ i=6, \  7, \ 8, \ \text{and} \ j = 2,\ 5,\ 8 , 
\end{cases}
\end{align}
where $P_i $ and $ Q_j  $ take the same value if and only if they are mapped to the same pairs of SP bases. On the other hand, we have $\langle P_{i} | Q_j \rangle = 0$ for the other combinations of $i$ and $j$, where $ P_{i} $ and $Q_j $ tag different pairs of SP bases.

We can also compute the matrix elements 
\begin{align}
\langle \mathcal{F} | \langle b| \mathcal{T} ^{\dag} S \mathcal{T} | \mathcal{G}  \rangle |b' \rangle = \langle \mathcal{G} | \langle b'| \mathcal{T} ^{\dag} S \mathcal{T} | \mathcal{F}  \rangle |b \rangle = 0,
\end{align}
with $(b,b')=(0,1), \ (1,0)$ and $(1,1)$ for $|\mathcal{F} \rangle ,\ |\mathcal{G} \rangle \neq |0\rangle$. 

The above procedures can also be adopted to computed the other pairing Hamiltonian matrix elements for the many-nucleon basis set $\{ | 0,1,3 \rangle , | 0,1,5 \rangle , | 0,3,5 \rangle , | 1,2,3 \rangle , | 0,1,5 \rangle , | 1,3,4 \rangle , | 1,4,5 \rangle , | 2,3,5 \rangle , | 3,4,5 \rangle \} $. It is straightforward to show that $H_{\rm pair}\big(M_J= + \frac{1}{2} \big) $ [Eq. \eqref{eq:matrix_rep_pairing_Hamiltonian}] can be reproduced based on the isometry $\mathcal{T}$ and the block-encoding scheme [Eqs. \eqref{eq:qubitization_formula} and \eqref{eq:qubitization_formula_1}]. The construction of the full-configuration-interaction Hamiltonian with the complete set of the three-nucleon states [Table \ref{tab:basisPairingHamiltonian}] follows suit.

\subsection{Four-neutron system}

We now switch to consider a more complex problem: the input model to access the Hamiltonian of a four-neutron system ($A=4$). Again, we retain only the two-body terms in the Hamiltonian, and restrict our discussion within a limited SP basis set that consists of the SP bases in the $1s_{1/2}0d_{3/2}0d_{5/2}$ valence space with $ 2n+l \leq 2 $. This restricted basis set is enumerated in Table \ref{tab:SPbasis_4n}. 
While the procedures are analogous to those presented above, we elucidate the work flow of the $O_{\rm H}$ oracle via the four-neutron example, as the $O_{\rm H}$ oracle functions in a less trivial manner than that in the pairing model example. On the other hand, this problem provides a prototype of the input model for realistic many-nucleon Hamiltonians \cite{Barrett:2013nh,Navratil:2000ww,Navratil:2000gs}.

\begin{table*}[!ht] 
\caption{The SP bases in the $1s_{1/2} 0d_{3/2} 0d_{5/2}$ valance space with $ 2n+l \leq 2 $. The quantum numbers of each SP basis are presented, whereas the spin and isospin quantum numbers are understood to be $\frac{1}{2}$. Each SP basis is indexed and mapped to a distinct qubit (second column). The occupation of each SP basis is represented by the state of the qubit: the $|0 \rangle $ ($|1\rangle $) state of the qubit denotes that the corresponding SP bases is vacant (occupied).}
\label{tab:SPbasis_4n}
\begin{tabular}{ccccccc}
\hline \hline
\multicolumn{1}{l}{}        & SP basis (qubit) & $\ n\ $ & $\ l \ $ & $\ 2j \ $ & $\ 2m_j\ $ & $\ 2\tau \ $ \\ \hline 
\multirow{2}{*}{$1s_{1/2}$} & {\bf 0}                & $\ 1 \ $   & $\ 0 \ $   & $\ 1 \ $    & $\  -1\ $     & $\ -1 \ $      \\
                            & {\bf 1}                & $1$   & $0$   & $1$    & $+1 $    & $-1 $     \\ \hline
\multirow{4}{*}{$0d_{3/2}$} & {\bf 2}                & $ 0$   & $2$  & $3 $   & $-3$     & $-1$      \\
                            & {\bf 3}                & $0$   & $2$   & $3$    & $-1$     & $-1 $     \\
                            & {\bf 4}                & $0$   & $2 $  & $3$    & $+1$     & $-1 $     \\
                            & {\bf 5}                & $0$   & $2$   & $3$    & $+3$     & $-1$      \\ \hline
\multirow{6}{*}{$0d_{5/2}$} & {\bf 6}                & $0$   & $2$   & $5 $   & $-5$     & $-1 $     \\
                            & {\bf 7}                & $0 $  & $2$   & $5$    & $-3$     & $-1$      \\
                            & {\bf 8}                & $0$   & $2$   & $5$    & $-1$     & $-1$      \\
                            & {\bf 9}                & $0 $  & $2$   & $5$    & $+1$     & $-1$      \\
                            & {\bf 10}               & $0 $  & $2 $  & $5 $   & $+3$     & $-1$      \\
                            & {\bf 11}               & $0 $  & $2$   & $5 $   & $+5$     & $-1$      \\ \hline \hline
\end{tabular}
\end{table*}

With the SP bases set shown in Table \ref{tab:SPbasis_4n}, there are $\tbinom{12}{4} = 495 $ four-neutron states in total, where $81$ of them are of $M_J=0$, $72$ of them are of $M_J=+2 \ (-2)$,  $60$ of them are of $M_J=+4 \ (-4)$, $39$ of them are of $M_J=+6 \ (-6)$, $24$ of them are of $M_J=+8 \ (-8)$, $9$ of them are of $M_J=+10 \ (-10)$, and $3$ of them are of $M_J=+12 \ (-12)$. As the Hamiltonian preserves the total $M_J$, the Hamiltonian does not connect four-neutron states with different $M_J$'s.

Next, we consider the possible combinations of the ladder operators $a^{\dag}_p a^{\dag}_q a_s a_r $ in Eq. \eqref{eq:H_total}. The total number of such combinations is $ \tbinom{12}{2} \tbinom{12}{2} = 4356 $. However, most of them result in vanishing matrix elements as such combinations violate the symmetries of the Hamiltonian (recall that we require the projection of the total angular momentum $M_J$ to be preserved as the baryon number, parity, and total isospin projection are trivially preserved in this example). Indeed, the total number of the symmetry preserving combination of the ladder operators is $ 640$ in this model problem.
We can then sort the symmetry preserving combinations of $a^{\dag}_p a^{\dag}_q a_s a_r =b^{\dag}_Q b_P $, and generate a lookup table [Table \ref{tab:P_Q_mapping_4n_problem}] that lists these combinations. 
\begin{table*}[!ht] 
\caption{Distinct combinations of the tags $P_i \mapsto \{r_i,s_i\}$ ($r_i <s_i$) and $Q_i \mapsto \{p_i,q_i \}$ ($p_i<q_i$) for the four-neutron problem described by the Hamiltonian  $ \mathcal{H} _{\rm 4n}$ [Eq. \eqref{eq:H_4n_temp_primed}] in the restricted SP bases set shown in Table \ref{tab:SPbasis_4n}. $r_i$ and $s_i$ ($p_i$ and $q_i$) denote the indices of the SP bases operated by the corresponding annihilation (creation) operators.}
\begin{tabular}{c c c}
\hline \hline
$i$ & $P_i \mapsto \{r_i, s_i \}$ & $Q_i \mapsto \{p_i, q_i \}$ \\ 
\hline
0 & $P_0 \mapsto \{ 0,1 \}$ & $Q_0 \mapsto \{ 0,1 \} $ \\ 
1 & $P_1 \mapsto \{ 0,2 \}$ & $Q_1 \mapsto \{ 0,2 \}$ \\ 
2 & $P_2 \mapsto \{ 0,5 \}$ & $Q_2 \mapsto \{ 0,5 \}$ \\ 
3 & $P_3 \mapsto \{ 1,2 \}$ & $Q_3 \mapsto \{ 1,2 \} $ \\ 
4 & $P_4 \mapsto \{ 1,5 \}$ & $Q_4 \mapsto \{ 1,5 \}$ \\ 
5 & $P_5 \mapsto \{ 2,5 \}$ & $Q_5 \mapsto \{ 2,5 \}$ \\ 
6 & $P_6 \mapsto \{ 0,1 \}$ & $Q_6 \mapsto \{ 3,4 \}$ \\ 
$\vdots $ & $\vdots $         & $\vdots $ \\ 
639 &  $ P_{639} \mapsto \{ 10, 11 \} $  &  $ Q_{639} \mapsto \{ 10, 11 \} $  \\ 
\hline \hline
\end{tabular} 
\label{tab:P_Q_mapping_4n_problem}
\end{table*}

According to these symmetry preserving combinations of the ladder operators, we can rewrite the Hamiltonian according to Eq. \eqref{eq:H_modified_H_prime} as 
\begin{align}
\mathcal{H}_{\rm 4n} = \sum _{i=0}^{639} H_{\rm 4n}(Q_i,P_i) b_{Q_i}^{\dag} b_{P_i} \otimes |Q_i\rangle \langle P_i |, \label{eq:H_4n_temp_primed}
\end{align}
where $H_{\rm 4n}$ is the Hamiltonian of the four-neutron system and $H_{\rm 4n}(Q_i,P_i) = \langle p_i q_i | H_{\rm 4n} | r_i s_i \rangle$. For this case, we take $\mathcal{D} = 640$ and $\Lambda _m \geq | H_{\rm 4n}(Q_i,P_i) | $ for $ i=0,1,\cdots , \mathcal{D}-1 $.

Provided the four-neutron state to be $| \mathcal{F} \rangle = | 0,1,2,5 \rangle $,  we construct the isometry $ \mathcal{T}| \mathcal{F} \rangle |0 \rangle $ of the form Eq. \eqref{eq:App_Psi_5} following the similar procedures shown in Sec. \ref{sec:pairing_Hamiltonian}.
For example, we can take $i=2$ with $P_2 \mapsto \{0,5\}$ and $Q_2 \mapsto \{0,5\}$ together $| \mathcal{F} \rangle $ as the input to the $O_{\rm F}$ oracle. The corresponding output of the $O_{\rm F}$ oracle are: $ | 0,1,2,5 \rangle $, $0$, $ | P_2 \rangle $, $| \mathcal{F} ' _{2}  \rangle = | 0,1,2,5 \rangle $, $ | Q_2 \rangle $, $|a_2 \rangle = |0 \rangle $.

The $O_{\rm H}$ oracle [Eq. \eqref{eq:O_H_prime}] functions controlled $|a_2 \rangle = |0 \rangle $ to compute the matrix element
\begin{align}
\langle \mathcal{F}'_i , Q_i | \mathcal{H}_{\rm 4n} | \mathcal{F} ,P_i \rangle = c_{i,1} c_{i,2} H_{\rm 4n}(Q_i,P_i) , \label{eq:computation_formula_OH_1}
\end{align}
with $i=2$. With the procedures shown in Sec. \ref{sec:OH_oracle_sec}, $O_{\rm H}$ oracle executes as
\begin{enumerate}
\item Provided the input $| \mathcal{F} \rangle = | 0,1,2,5 \rangle = | 1110010000000 \rangle $, $P_2 \mapsto \{ 0,5 \}$, and $Q_2 \mapsto \{ 0,5 \}$, create a sequence of CNOT gates, each of which is controlled by the $1^{\rm st}$, $2^{\rm nd}$, $3^{\rm rd}$, and $4^{\rm th}$ qubits (these qubits correspond to the occupations of the $1^{\rm st}$, $2^{\rm nd}$, $3^{\rm rd}$, and $4^{\rm th}$ SP bases, respectively) and acts on the ancilla $B_{\rm ph,1}$ that is initialized as $|0\rangle $. As the $1^{\rm st}$, $2^{\rm nd}$ qubits are in the state $ |1 \rangle $ and the $3^{\rm rd}$, and $4^{\rm th}$ qubits are in the state $ |0 \rangle $, the ancilla remains in the $|0\rangle $ state, i.e., $| B'_{\rm ph,1} \rangle =|0\rangle$. Therefore, the desired phase is $c_{2,1}=+1$ for the action of $b_{P_2}| \mathcal{F} \rangle $;

\item Similarly, provided $| \mathcal{F}'_2 \rangle = | 0,1,2,5 \rangle = | 1110010000000 \rangle $, $P_2 \mapsto \{ 0,5 \}$, and $Q_2 \mapsto \{ 0,5 \}$ and the ancilla $B_{\rm ph,2}$ initialized in the $|0\rangle $ state, we have $| B'_{\rm ph,2} \rangle =|0\rangle$ after the operations of a sequence of CNOT gates. This means that $c_{2,2}=+1$ for the action of $b_{Q_2}| \mathcal{F}_2' \rangle $;

\item Provided the tags $P_2 \mapsto \{ 0,5 \}$ and $Q_2 \mapsto \{ 0,5 \}$, we pick the matrix element $ H_{\rm 4n}(Q_2,P_2) $ via the iteration with a classically precomputed database. This matrix element is recorded in the ancilla register as $| V'_{\rm tmp} \rangle $;

\item The matrix element is computed based on the ancilla states $| B'_{\rm ph,1} \rangle $, $| B'_{\rm ph,2} \rangle $ and $| V'_{\rm tmp} \rangle $, according to Eq. \eqref{eq:computation_formula_OH_1} as $ \langle \mathcal{F}'_2 , Q_2 | \mathcal{H}_{\rm 4n} | \mathcal{F} ,P_2 \rangle = (+1)(+1) H_{\rm 4n}(Q_2,P_2) $. The matrix element is encoded in the $R_{\rm elem}$ register and it is the output of the $ O_{\rm H}$ oracle.

\end{enumerate}
In an analogous manner, we can calculate all the other cases with $| \mathcal{F} \rangle = | 0,1,2,5 \rangle $ and $i=0,1 , 3,\cdots 639$. These results can be used to check the validity of the isometry $\mathcal{T}$ and our input model with a minimal example. In particular, we can write Eq. \eqref{eq:equivalent_to_A7} with these results as
\begin{align}
\mathcal{T}| \mathcal{F} \rangle | 0 \rangle = \sqrt{ \frac{1 }{\mathcal{D} \Lambda _m } } \sum _{k=0}^{5} \Big[  \sqrt{\langle \mathcal{F}'_k , Q_k | \mathcal{H}_{\rm 4n} | \mathcal{F} ,P_k \rangle } | \mathcal{F} \rangle | 0 \rangle | 0 \rangle |P_k \rangle | \mathcal{F}_k' \rangle |Q_k \rangle | a_k \rangle |0\rangle |0\rangle \Big] + | \cdot \rangle \label{eq:diagonal_4n}
\end{align}
where $| \cdot \rangle $ denotes the other terms in Eq. \eqref{eq:equivalent_to_A7}. 
Based on Eq. \eqref{eq:diagonal_4n}, we can compute the diagonal Hamiltonian matrix element
\begin{align}
\langle \mathcal{F} | \langle 0 | \mathcal{T} ^{\dag} S \mathcal{T} | \mathcal{F} \rangle |0 \rangle = \frac{ 1 }{\mathcal{D} \Lambda _m } \langle \mathcal{F}  | H_{\rm 4n} | \mathcal{F} \rangle ,
\end{align}
according to Eq. \eqref{eq:walk_identity}. It is also straightforward to check 
\begin{align}
\langle \mathcal{F} | \langle b | \mathcal{T} ^{\dag} S \mathcal{T} | \mathcal{F} \rangle |b' \rangle = 0,
\end{align}
for $(b,b')=(1,0),\ (0,1)$ and $(1,1)$. The above two equations combine to
\begin{align}
\langle \mathcal{F} | \langle b| ( \mathcal{T}^{\dag}S \mathcal{T} ) | \mathcal{F} \rangle | b' \rangle = \frac{ 1 }{\mathcal{D} \Lambda _m} \langle   \mathcal{F} | H_{\rm 4n} | \mathcal{F} \rangle \delta _{b,0} \delta _{b',0} .
\end{align}
With analogous calculations, we can show that the input model can access the full-configuration-interaction Hamiltonian matrix within the basis representation constructed by the SP basis set shown in Table \ref{tab:SPbasis_4n}. 

\section{Summary and outlook}
\label{sec:summary_and_outlook}

In this work, we propose a framework to solve the many-nucleon structure and dynamics on quantum computers. 
We work with the second-quantized Hamiltonian and develop the oracle-based Hamiltonian input model that treats directly the Fock states. 
We can implement high-level sparse Hamiltonian simulation algorithms, such as the quantum signal processing and the rescaled Dyson series algorithm with our Fock-state-based input model to simulate the dynamics of many-nucleon systems. 
The method for dynamics simulation can also be implemented to solve the many-nucleon structure problems.

We first discuss the elements of the many-nucleon calculations. We start with the second-quantized many-nucleon Hamiltonian that retains only two-nucleon terms and introduce a convenient notation of the second-quantized Hamiltonian for quantum computing. We also discuss the our choice of the single-particle (SP) basis, which consists of the spatial, spin, and isospin degrees of freedom. The SP bases are used to construct the many-nucleon states. 

Next, we discuss the encoding scheme that maps the many-nucleon state to the state of the quantum register. In particular, we employ the direct encoding scheme in this work. The direct encoding scheme records every SP basis employed to construct the many-nucleon states, whether they are occupied or not. In this way, the direct encoding scheme encodes each many-nucleon state in terms of a binary string. Though the qubit count is not optimal in the direct encoding scheme (compared to that of the compact encoding scheme), the corresponding circuit design of the Hamiltonian input model is more straightforward.

Then, we discuss our input model for the second-quantized many-nucleon Hamiltonian. 
While the well-known sparse matrix input models access the nonzero Hamiltonian matrix elements based on their row and column indices, our Hamiltonian input model functions with the many-nucleon (Fock) states. 
In particular, we define the enumerator oracle that computes the output Fock state based on the input Fock state, while we also track all the pairs of the SP bases that contribute to the matrix element in the input and output Fock states, respectively. We define the matrix element $O_{\rm H}$ oracle that computes the nonvanishing elements with the input and output Fock states and the pairs of SP states computed by $O_{\rm F}$. We then show the definition of the isometry $\mathcal{T}$ that can be constructed using $\mathcal{O}(1)$ oracles queries of these two oracles and the connection between the isometry and the Hamiltonian.

Based on our input model, we propose a framework for solving the dynamics and structure problems of the second-quantized many-nucleon Hamiltonian. In particular, our input model serves as the basic input unit for those well-known high-level algorithms, i.e., the quantum signal processing \cite{Low:2019} and the rescaled Dyson series \cite{Berry:2020}, to obtain optimal query complexities (with respect to the simulation time and error) for simulating time-independent and time-dependent Hamiltonians. In both cases, we provide the asymptotic query complexities for the simulations. On the other hand, we can directly implement the framework of dynamics simulation to solve structure problems as they share the same component of the time-evolution unitary. In particular, we propose to implement the framework together with the Rodeo algorithm \cite{Choi:2020pdg,Qian:2021wya,Bee-Lindgren:2022nqb} to solve for spectrum and other observables, where we also provided the corresponding query complexity.

We introduce our oracle constructions for the second-quantized Hamiltonian that are based on the direct encoding scheme. In particular, we discuss the details of our design of the $O_{\rm F}$ and $O_{\rm H}$ oracles. We also analyze the asymptotic qubit cost and gate count for two oracles. Jointly, the asymptotic qubit cost of the $O_{\rm F}$ and $O_{\rm H}$ oracles is $\widetilde{ \mathcal{O} } (N_{\rm sp})$, while their asymptotic gate cost is $\widetilde{ \mathcal{O} }(N^5_{\rm sp})$.

Based on the gate cost of the oracles, we can evaluate the asymptotic gate cost for dynamics simulation and structure calculation within our framework. This gate cost is $\widetilde{ \mathcal{O}} (N_{\rm sp}^5 \mathcal{D} \Upsilon ) $, with $\mathcal{D} \in \mathcal{O} (N_{\rm sp}^4) $ being the number of monomials of the ladder operators in the second-quantized Hamiltonian that includes at most two-body terms. For simulating a time-dependent Hamiltonian, we take $\Upsilon = \int _{0}^{t} \Lambda_m(t') dt' $ with $ \Lambda _m (t) \geq \max _i | \langle p_iq_i | H(t) | r_i s_i \rangle | $ and $i \in [0, \mathcal{D}-1]$. We take $\Upsilon =  \Lambda _m t$ for the simulation of a time-independent Hamiltonian and $\Upsilon = \Lambda _m \Delta $ for structure calculations, with $ \Lambda _m  \geq \max _i | \langle p_iq_i | H | r_i s_i \rangle | $. 

Our asymptotic gate cost for simulating a general time-independent many-nucleon Hamiltonian that retains only the two-nucleon terms is $\widetilde{ \mathcal{O}} (N_{\rm sp}^{10} t)$. This can be compared with previous works of simulating a molecular Hamiltonian in quantum chemistry, where the objective Hamiltonian is time-independent and retains up to two-electron terms, where one notes that the two-electron terms result in the major gate cost. We find that the scaling of our gate cost in $N_{\rm sp}$ is close to that of the Trotter-based methods \cite{Berry:2007dwb,Wiebe:2011wbe,Hastings:2015hast}. However, our algorithm is exponentially more precise than these Trotter-based algorithms.

Our asymptotic gate cost is close to that of the database algorithm introduced in Ref. \cite{RBabbush:2016}, which scales as $\widetilde{\mathcal{O}} \big(  N^8_{\rm sp} t  \big) $. Our gate cost scales worse than the on-the-fly algorithm in Ref. \cite{RBabbush:2016}, the algorithm in Ref. \cite{Kirby:2021ajp} based on compact encoding, and the algorithm in Ref. \cite{Babbush:2018bubbush} that adopts a compressed configuration-interaction matrix representation. These algorithms report better scalings of the asymptotic gate cost in $ N_{\rm sp}$ than ours. 

We comment that our estimation of the asymptotic gate cost is conservative as the restrictions of the Hamiltonian properties are not taken into account in the evaluation. As for the applications of complex nuclei, a tighter upper bound of the gate cost is suspected to be $ \widetilde{\mathcal{O}} \big( N_{\rm sp}^7 t  \big)  $. Our gate cost can be further improved by incorporating the (controlled-) arithmetic operations and the on-the-fly evaluations of the two-body kernels as in Refs. \cite{RBabbush:2016,Kirby:2021ajp,Babbush:2018bubbush}. As for this work, we design our algorithm based on the elementary gate operations aiming for straightforward prototype nuclear structure and dynamics calculations on quantum computers.

For pedagogical purposes, we apply our method to two model problems in nuclear physics within restricted basis spaces, where only the two-body terms are retained in respective Hamiltonians. We illustrate the design of our oracle-based input model that is based on Fock states. Implementing our Fock-state-based input model with the high-level sparse matrix simulation algorithms, we can perform prototype structure and dynamics calculations for these model problems on near-term noisy intermediate scale quantum devices. Generalization to complex and realistic many-nucleon problems is straightforward.

Going forward, we plan to explore the applications of our framework to perform dynamics simulations and structure calculations of simple systems such as the simplified pairing models described in this work. These calculations serve as benchmark tests for more complicated applications, such as the four-nucleon calculations in a restricted model space. We also plan to improve the performance of our input model. While better options of basis set may reduce the counts of the terms in the second-quantized Hamiltonian \cite{Babbush:2018BasisChoice}, the incorporation of simple arithmetic operations \cite{Vedral:1996,Abhari:2014} would also be promising. In addition, we will further develop the current framework to treat more general second-quantized Hamiltonians for systems including both bosons and fermions with particle creations and annihilations.

\section{Acknowledgments}
We acknowledge fruitful discussions with William M. Kirby, Peter J. Love, Michael Kreshchuk, Pieter Maris, and Chao Yang. WD thanks Morten Hjorth-Jensen and Dean Lee for valuable suggestions. This work was supported by the U.S. Department of Energy under Grants No. DE-SC0018223 (SciDAC4/NUCLEI), DE-SC0023495 (SciDAC5/NUCLEI), DE-SC0023707 (NuHaQ), and No. DE-FG02-87ER40371.

\appendix 

\section{Hamiltonian input model}
\label{sec:quantumWalk_Appd_A}

\subsection{Isometry construction}
\label{sec:isometry_construction}
For completeness and self-consistency, we present the construction of the isometry $\mathcal{T} $ that performs the mapping defined in Eq. \eqref{eq:isometry_T_prime}.
\begin{customlemma}{1}\label{construction_of_T}
For an input state $| \mathcal{F} \rangle | b \rangle $, where $| \mathcal{F} \rangle $ denotes the many-nucleon state and $b \in \{0,1 \}$, the isometry $ \mathcal{T} $ defined in Eq. \eqref{eq:isometry_T_modified} can be implemented with $\mathcal{O}(1)$ queries to $O_{\rm F}$ and $O_{\rm H}$ defined in Eqs. \eqref{eq:O_F_prime} and \eqref{eq:O_H_prime}. (This Lemma follows $Lemma$ 1 in Ref. \cite{Kirby:2021ajp} and $Lemma$ 4 in Ref. \cite{DWBerry:2012}.)
\end{customlemma}

\begin{proof}
The isometry $\mathcal{T}$ maps the state $| \mathcal{F} \rangle | b \rangle  $ to $| \mathcal{F} \rangle | b \rangle | \phi _{ \mathcal{F} ,b } \rangle $ as shown in Eq. \eqref{eq:isometry_T_prime}. We give the proof by showing the construction of $\mathcal{T}$ as follows.

Provided the many-nucleon state $|\mathcal{F} \rangle $, we initialize the input state to be 
\begin{align}
| \Psi _{\rm in } \rangle = | \mathcal{F} \rangle | b \rangle |0 \rangle |0 \rangle |0 \rangle | 0 \rangle | 0 \rangle | 0 \rangle |0 \rangle ,
\end{align}
where we write down the state of each quantum register explicitly.  

\paragraph*{Case 1:} If $| b \rangle = |1 \rangle $, we flip the last qubit register from $|0 \rangle $ to $|1 \rangle $ controlled on the ancilla qubit $| b \rangle = |1 \rangle $. In this way, we prepare the output state as 
\begin{align}
| \Psi _{\rm out } \rangle = | \mathcal{F} \rangle | 1 \rangle |0 \rangle |0 \rangle |0 \rangle | 0 \rangle | 0 \rangle | 0 \rangle |1 \rangle . \label{eq:null_case_1}
\end{align}

\paragraph*{Case 2:} If $| b \rangle = | 0 \rangle $, we start with the input $ | \Psi _{\rm in} \rangle =  | \mathcal{F} \rangle | 0 \rangle |0 \rangle |0 \rangle |0 \rangle | 0 \rangle | 0 \rangle | 0 \rangle |0 \rangle  $ and operate the registers as follows.

\begin{enumerate}
\item Prepare a uniform superposition of the indices $i=0,1,2, \cdots , \mathcal{D} - 1$ on the third register. The overall state of the quantum register $ | \Psi _{\rm in} \rangle $ becomes
\begin{align}
| \Psi _1 \rangle = \frac{1}{\sqrt{ \mathcal{D} }} \sum _{i=0}^{ \mathcal{D} -1} | \mathcal{F} \rangle | 0 \rangle |i \rangle |0 \rangle |0 \rangle | 0 \rangle | 0 \rangle | 0 \rangle |0 \rangle  .
\end{align}

\item For a specific index $i$ and $  | \mathcal{F} \rangle $, we apply the $O_{\rm F}$ oracle [Eq. \eqref{eq:O_F_prime}] on the first (leftmost), third, fourth, fifth, sixth, and seventh registers to obtain
\begin{align}
| \Psi _2 \rangle = \frac{1}{\sqrt{ \mathcal{D} }} \sum _{i=0}^{ \mathcal{D} -1} | \mathcal{F} \rangle | 0 \rangle | { 0} \rangle | P_i \rangle |\mathcal{F}'_i \rangle | Q_i \rangle | y_i \rangle | 0 \rangle |0 \rangle  .
\end{align}

\item Controlled on $ | y _i \rangle = | 0 \rangle $, we apply the $O_{\rm H} $ oracle [Eq. \eqref{eq:O_H_prime}] to the first, fourth, fifth and sixth registers. The kernel $ \langle \mathcal{F}_i', Q_i | H' | \mathcal{F} , P_i \rangle  $ is computed in an ancilla register that is initiated as $|0 \rangle $. Controlled on the value $ \langle \mathcal{F}_i', Q_i | H' | \mathcal{F} , P_i \rangle  $, we rotate the last (rightmost) single qubit (initialized as $|0 \rangle $) in $|\Psi _2 \rangle $ as
\begin{align}
| 0 \rangle \mapsto \sqrt{ \frac{  \langle \mathcal{F}_i', Q_i | \mathcal{H} | \mathcal{F} , P_i \rangle  }{ \Lambda  _m }  }  | 0 \rangle + \sqrt{ 1-  \frac{  | \langle \mathcal{F}_i', Q_i | \mathcal{H} | \mathcal{F} , P_i \rangle | }{\Lambda  _m  }   } | 1 \rangle , \label{eq:single_qubit_rot}
\end{align}
where the operation of the square root is defined below [Eq. \eqref{eq:def_sqrt}]. We choose $\Lambda _m $ to be larger than the largest absolute value of the two-body kernel: $ \Lambda _m \geq | \langle \mathcal{F}_i', Q_i | \mathcal{H} | \mathcal{F} , P_i \rangle | = |\langle p_iq_i | H | r_is_i\rangle | $ for any $i \in [0, \mathcal{D}-1]$, such that $ 0 \leq  \frac{ | \langle \mathcal{F}_i', Q_i | \mathcal{H} | \mathcal{F} , P_i \rangle | }{\Lambda _m }  \leq 1 $ is satisfied. Then, uncompute the ancilla register that computes $ \langle \mathcal{F}_i', Q_i | \mathcal{H} | \mathcal{F} , P_i \rangle  $ by another controlled query to $O_{\rm H}$ after the above single-qubit rotation [Eq. \eqref{eq:single_qubit_rot}] is completed.  

In the case of $|y_i \rangle =| 1 \rangle $, we anticipate that $| \mathcal{F}_i' \rangle $ does not connect to $| \mathcal{F} \rangle $ via the many-nucleon Hamiltonian, and $ \langle \mathcal{F}_i', Q_i | \mathcal{H} | \mathcal{F} , P_i \rangle =0 $ (due to the violation of, e.g., the symmetry, the Pauli principle, etc.). In this case, we flip the last qubit from $|0 \rangle $ to $ |1 \rangle $ controlled on $|y_i \rangle =| 1 \rangle $.

After this step, without loss of generality, the full state of the quantum registers becomes
\begin{multline}
| \Psi  _{\rm out} \rangle = \frac{1}{\sqrt{ \mathcal{D} }} \sum _{i=0}^{ \mathcal{D} -1} \Bigg[ \sqrt{ \frac{  \langle \mathcal{F}_i', Q_i | \mathcal{H} | \mathcal{F} , P_i \rangle  }{ \Lambda  _m }  }   | \mathcal{F} \rangle | 0 \rangle | { 0} \rangle | P_i \rangle |\mathcal{F}'_i \rangle | Q_i \rangle | y_i \rangle | 0 \rangle | 0 \rangle  \Bigg] \\
+ \frac{1}{\sqrt{ \mathcal{D} }} \sum _{i=0}^{ \mathcal{D} -1} \Bigg[ \sqrt{ 1-  \frac{  | \langle \mathcal{F}_i', Q_i | \mathcal{H} | \mathcal{F} , P_i \rangle | }{\Lambda  _m  }   } | \mathcal{F} \rangle | 0 \rangle | { 0} \rangle | P_i \rangle |\mathcal{F}'_i \rangle | Q_i \rangle | y_i \rangle | 0 \rangle  | 1 \rangle \Bigg]. \label{eq:App_Psi_5_prime}  
\end{multline}

It is understood that the necessary condition for the first term of the above equation to exist is $| y_i \rangle = | 0 \rangle $. When $ | y_i \rangle  = | 1 \rangle $, we should take $ \langle \mathcal{F}_i', Q_i | \mathcal{H} | \mathcal{F} , P_i \rangle = 0 $ and flip the last qubit register from $|0\rangle $ to $|1 \rangle $. In our construction, we make use an additional, redundant qubit, the second last qubit which is initialized as $|0 \rangle $, to guarantee the orthogonalization between the terms with $|y_i \rangle =| 0\rangle $ and those with $|y_i \rangle =| 1 \rangle $ [see discussion of the swap operator [Eq. \eqref{eq:swap_operation}] below].

\end{enumerate}

In viewing the fact that only nonvanishing kernels $  \langle \mathcal{F}_i', Q_i | \mathcal{H} | \mathcal{F} , P_i \rangle  $ contribute to the summation in the first term of $ | \Psi _{\rm out} \rangle $. Therefore, we can simplify Eq. \eqref{eq:App_Psi_5_prime} as 
\begin{multline}
| \Psi _{\rm out} \rangle = \frac{1}{\sqrt{ \mathcal{D} }} \sum _{i \in \mathcal{I}(\mathcal{F})}  \Bigg[ \sqrt{ \frac{  \langle \mathcal{F}_i', Q_i | \mathcal{H} | \mathcal{F} , P_i \rangle  }{ \Lambda _m }  } | \mathcal{F} \rangle | 0 \rangle |{ 0} \rangle | P_i \rangle |\mathcal{F}'_i \rangle | Q_i \rangle | y_i \rangle | 0 \rangle |0 \rangle  \Bigg] \\ 
	+ \frac{1}{\sqrt{ \mathcal{D} }} \sum _{i=0}^{\mathcal{D}-1} \Bigg[ \sqrt{ 1 -  \frac{ | \langle \mathcal{F}_i', Q_i | \mathcal{H} | \mathcal{F} , P_i \rangle | }{\Lambda _m }   } | \mathcal{F} \rangle | 0 \rangle | { 0} \rangle | P_i \rangle |\mathcal{F}'_i \rangle | Q_i \rangle | y_i \rangle | 0 \rangle | 1 \rangle \Bigg] \label{eq:App_Psi_5} ,
\end{multline}
where $  \mathcal{I}(\mathcal{F})$ denotes the set of indices $i$ with $ \langle \mathcal{F}_i', Q_i | \mathcal{H} | \mathcal{F} , P_i \rangle \neq 0$. Equation \eqref{eq:App_Psi_5} can also be rewritten as
\begin{align}
| \Psi _{\rm out} \rangle = \sqrt{ \frac{ 1 }{ \mathcal{D} \Lambda _m } } \sum _{i \in \mathcal{I}(\mathcal{F})}  \Bigg[\sqrt{ \langle \mathcal{F}_i', Q_i | \mathcal{H} | \mathcal{F} , P_i \rangle } | \mathcal{F} \rangle | 0 \rangle |{ 0} \rangle | P_i \rangle |\mathcal{F}'_i \rangle | Q_i \rangle | y_i \rangle | 0 \rangle |0 \rangle \Bigg] 
				    + \sqrt{ 1- \frac{ \sigma _{\mathcal{F}}}{ \mathcal{D} \Lambda _m  }  } | \mathcal{F} \rangle |0 \rangle | \zeta _{\mathcal{F}} \rangle |1 \rangle , \label{eq:App_Psi_5_primed}
\end{align}
where we define
\begin{align}
| \zeta _{\mathcal{F}} \rangle =  \sqrt{\frac{1}{ 1- \frac{ \sigma _{\mathcal{F}}}{\mathcal{D} \Lambda _m  } }}  \sum _{i=0}^{\mathcal{D} -1} \Bigg[ \sqrt{ 1 -  \frac{ | \langle \mathcal{F}_i', Q_i | \mathcal{H} | \mathcal{F} , P_i \rangle | }{\Lambda _m } }  | {0} \rangle | P_i \rangle |\mathcal{F}'_i \rangle | Q_i \rangle | y_i \rangle | 0 \rangle  \Bigg] , \label{eq:App_Psi_5_prime_prime}
\end{align}
with $ \sigma _{\mathcal{F}} = \sum _{i=0}^{\mathcal{D} -1} |\langle \mathcal{F}_i', Q_i | \mathcal{H} | \mathcal{F} , P_i \rangle |$. Note that $\sigma _{\mathcal{F}} $ serves as a parameter to normalize $|\Psi _{\rm out} \rangle $ in Eq. \eqref{eq:App_Psi_5_primed}. This can be checked based on the orthogonality relation between any term in the summand of the first term and the second term in Eq. \eqref{eq:App_Psi_5}.

To summarize, we have present the construction of $\mathcal{T} $ that operates as Eq. \eqref{eq:isometry_T_prime} based on one query of $O_{\rm F}$ [Eq. \eqref{eq:O_F_prime}] and two queries of $O_{\rm H}$ [Eq. \eqref{eq:O_H_prime}].

\end{proof}

\subsection{Block-encoding scheme}
\label{sec:relation_to_H_matrix_element}
With the isometry $\mathcal{T}$ [Eq. \eqref{eq:isometry_T_modified}], we can access the Hamiltonian via the block-encoding scheme following the approach in Refs. \cite{DWBerry:2012,AMChilds:2009}.
Indeed, it is straightforward to show that  
\begin{align}
\mathcal{T}^{\dag} S \mathcal{T} = \frac{1 }{ \mathcal{D} \Lambda _m } H \otimes |0 \rangle \langle 0 | + |0 \rangle \langle 0|  \otimes |1 \rangle \langle 1 | , \label{eq:identity_Quantum_walk}
\end{align}
where $ |0 \rangle \langle 0 | $ and $ |1 \rangle \langle 1 | $ (following the symbols ``$\otimes$") project the subspace that corresponds to the state $|b\rangle $ (in either $|0\rangle $ or $|1\rangle $ state) in Eq. \eqref{eq:isometry_T_modified}. $H$ is the many-nucleon Hamiltonian defined in Eq. \eqref{eq:H_total} that acts on the many-nucleon state, as does the operator $|0\rangle \otimes \langle 0| $ in the second term above. $S$ is the swap operator defined in Eq. \eqref{eq:swap_operation} below.

To prove the identity [Eq. \eqref{eq:identity_Quantum_walk}], we compute $ \langle \mathcal{F} | \langle  b | \big( T^{\dag} S T \big) | \mathcal{G} \rangle |b' \rangle  $ with $b,b' =$ 0 or 1. Based on the construction of the isometry $\mathcal{T}$ [Eq. \eqref{eq:App_Psi_5_prime}], we have
\begin{align}
\mathcal{T} | \mathcal{F} \rangle | b= 0 \rangle  =& \frac{1}{\sqrt{\mathcal{D}}} \sum _{i=0}^{ \mathcal{D} -1} \Bigg[ \sqrt{ \frac{ \langle \mathcal{F}_i', Q_i | \mathcal{H} | \mathcal{F} , P_i \rangle  }{ \Lambda _m }  } | \mathcal{F} \rangle | 0 \rangle | { 0} \rangle | P_i \rangle |\mathcal{F}'_i \rangle | Q_i \rangle | y_i \rangle | 0 \rangle |0 \rangle  \Bigg] \nonumber \\
				   & \ \ \ \ \ \ \ \ \ \ \  + \frac{1}{\sqrt{\mathcal{D}}} \sum _{i=0}^{\mathcal{D} -1} \Bigg[ \sqrt{ 1 -  \frac{ | \langle \mathcal{F}_i', Q_i | \mathcal{H} | \mathcal{F} , P_i \rangle | }{ \Lambda _m }   } | \mathcal{F} \rangle | 0 \rangle | { 0} \rangle | P_i \rangle |\mathcal{F}'_i \rangle | Q_i \rangle | y_i \rangle | 0 \rangle | 1 \rangle \Bigg] , \\
\mathcal{T} | \mathcal{F} \rangle | b= 1 \rangle  =& | \mathcal{F} \rangle | 1 \rangle |0 \rangle |0 \rangle |0 \rangle | 0 \rangle | 0 \rangle | 0 \rangle |1 \rangle . 	\label{eq:alternative_b_1}	   
\end{align}

Similarly, we also have
\begin{align}
\mathcal{T} | \mathcal{G} \rangle | b' = 0 \rangle  =& \frac{1}{\sqrt{\mathcal{D}}} \sum _{j=0}^{\mathcal{D} -1} \Bigg[ \sqrt{ \frac{ \langle \mathcal{G}_j', Q_j | \mathcal{H} | \mathcal{G} , P_j \rangle  }{ \Lambda _m }  } | \mathcal{G} \rangle | 0 \rangle | { 0} \rangle | P_j \rangle |\mathcal{G}'_j \rangle | Q_j \rangle | y_j \rangle | 0 \rangle |0 \rangle  \Bigg] \nonumber \\
				   & \ \ \ \ \ \ \ \ \ \ \  + \frac{1}{\sqrt{\mathcal{D}}} \sum _{j=0}^{\mathcal{D}-1} \Bigg[ \sqrt{ 1 -  \frac{ | \langle \mathcal{G}_j', Q_j | \mathcal{H} | \mathcal{G} , P_j \rangle | }{ \Lambda _m }   } | \mathcal{G} \rangle | 0 \rangle | { 0} \rangle | P_j \rangle |\mathcal{G}'_j \rangle | Q_j \rangle | y_j \rangle | 0 \rangle | 1 \rangle \Bigg] , \\
\mathcal{T} | \mathcal{G} \rangle | b' = 1 \rangle  =& | \mathcal{G} \rangle | 1 \rangle |0 \rangle |0 \rangle |0 \rangle | 0 \rangle | 0 \rangle | 0 \rangle |1 \rangle . 
\end{align}

\paragraph*{Case 1.}
We first compute the case with $b=b'=0$. We have
\begin{align}
  \langle \mathcal{F} | \langle 0| ( \mathcal{T} ^{\dag}S \mathcal{T} ) | \mathcal{G} \rangle | 0 \rangle 
=& \Bigg\{ \frac{1}{\sqrt{\mathcal{D}} } \sum _{i=0}^{\mathcal{D}-1} \Bigg[ \sqrt{ \frac{  \langle \mathcal{F}_i', Q_i | \mathcal{H} | \mathcal{F} , P_i \rangle  }{ \Lambda _m }  } | \mathcal{F} \rangle | 0 \rangle | { 0} \rangle | P_i \rangle |\mathcal{F}'_i \rangle | Q_i \rangle | y_i \rangle | 0 \rangle |0 \rangle  \Bigg] \nonumber \\
 & \ \ \ \ \ \ \ \ \ \ \ \ + \frac{1}{\sqrt{\mathcal{D}}} \sum _{i=0}^{\mathcal{D}-1} \Bigg[ \sqrt{ 1 -  \frac{ | \langle \mathcal{F}_i', Q_i | \mathcal{H} | \mathcal{F} , P_i \rangle | }{ \Lambda _m }   } | \mathcal{F} \rangle | 0 \rangle | { 0} \rangle | P_i \rangle |\mathcal{F}'_i \rangle | Q_i \rangle | y_i \rangle | 0 \rangle | 1 \rangle \Bigg]
  \Bigg\}^{\dag} \nonumber \\
& S \Bigg\{ \frac{1}{\sqrt{D}} \sum _{j=0}^{\mathcal{D}-1} \Bigg[ \sqrt{ \frac{   \langle \mathcal{G}_j', Q_j | \mathcal{H} | \mathcal{G} , P_j \rangle  }{\Lambda _m }  } | \mathcal{G} \rangle | 0 \rangle | { 0} \rangle | P_j \rangle |\mathcal{G}'_j \rangle | Q_j \rangle | y_j \rangle | 0 \rangle |0 \rangle  \Bigg] \nonumber \\
 & \ \ \ \ \ \ \ \ \ \ \ \  + \frac{1}{\sqrt{\mathcal{D}}} \sum _{j=0}^{\mathcal{D}-1} \Bigg[ \sqrt{ 1 -  \frac{ | \langle \mathcal{G}_j', Q_j | \mathcal{H} | \mathcal{G} , P_j \rangle | }{\Lambda _m }   } | \mathcal{G} \rangle | 0 \rangle | { 0} \rangle | P_j \rangle |\mathcal{G}'_j \rangle | Q_j \rangle | y_j \rangle | 0 \rangle | 1 \rangle \Bigg]
   \Bigg\} .
\end{align}
We define the swap operator $S$ as
\begin{align}
S | r_0 \rangle |r_1 \rangle |r_2 \rangle |r_3 \rangle |r_4 \rangle |r_5 \rangle |r_6 \rangle |r_7 \rangle |r_8 \rangle   = | r_4 \rangle |r_8 \rangle |r_2 \rangle |r_5 \rangle |r_0 \rangle |r_3 \rangle |r_7 \rangle |r_6 \rangle |r_1 \rangle  , \label{eq:swap_operation}
\end{align}
where we swap states stored in the registers as $| r_0 \rangle \leftrightarrow | r_4 \rangle $, $| r_1 \rangle \leftrightarrow | r_8 \rangle $, $| r_3 \rangle \leftrightarrow | r_5 \rangle $, and $| r_6 \rangle \leftrightarrow | r_7 \rangle $. With the swap operations, we obtain
\begin{align}
\langle \mathcal{F} | \langle 0| ( \mathcal{T} ^{\dag}S \mathcal{T} ) | \mathcal{G} \rangle | 0 \rangle 
=& \Bigg\{ \frac{1}{\sqrt{\mathcal{D}}} \sum _{i=0}^{\mathcal{D}-1} \Bigg[ \sqrt{ \frac{  \langle \mathcal{F}_i', Q_i | \mathcal{H} | \mathcal{F} , P_i \rangle  }{ \Lambda _m }  } | \mathcal{F} \rangle | 0 \rangle | { 0} \rangle | P_i \rangle |\mathcal{F}'_i \rangle | Q_i \rangle | y_i \rangle | 0 \rangle |0 \rangle  \Bigg] \nonumber \\
 & \ \ \ \ \ \ \ \ \ \ \ \ + \frac{1}{\sqrt{\mathcal{D}}} \sum _{i=0}^{\mathcal{D}-1} \Bigg[ \sqrt{ 1 -  \frac{ | \langle \mathcal{F}_i', Q_i | \mathcal{H} | \mathcal{F} , P_i \rangle | }{ \Lambda _m }   } | \mathcal{F} \rangle | 0 \rangle | { 0} \rangle | P_i \rangle |\mathcal{F}'_i \rangle | Q_i \rangle | y_i \rangle | 0 \rangle | 1 \rangle \Bigg]
  \Bigg\}^{\dag} \nonumber \\
&  
   \Bigg\{ \frac{1}{\sqrt{\mathcal{D}}} \sum _{j=0}^{\mathcal{D}-1} \Bigg[ \sqrt{ \frac{   \langle \mathcal{G}_j', Q_j | \mathcal{H} | \mathcal{G} , P_j \rangle  }{ \Lambda _m }  } 
   | \mathcal{G}'_j \rangle | 0 \rangle | { 0} \rangle | Q_j \rangle |\mathcal{G} \rangle | P_j \rangle | 0 \rangle | y_j \rangle |0 \rangle  \Bigg] \nonumber \\
 & \ \ \ \ \ \ \ \ \ \ \ \  + \frac{1}{\sqrt{\mathcal{D}}} \sum _{j=0}^{\mathcal{D}-1} \Bigg[ \sqrt{ 1 -  \frac{ | \langle \mathcal{G}_j', Q_j | \mathcal{H} | \mathcal{G} , P_j \rangle | }{ \Lambda _m }   } 
   | \mathcal{G}'_j \rangle | 1 \rangle | { 0} \rangle | Q_j \rangle |\mathcal{G} \rangle | P_j \rangle | 0 \rangle | y_j \rangle | 0 \rangle \Bigg]
   \Bigg\} .
\end{align}

Implementing the orthonormality condition, the above equation can be simplified as
\begin{align}
\langle \mathcal{F} | \langle 0| ( \mathcal{T} ^{\dag}S \mathcal{T} ) | \mathcal{G} \rangle | 0 \rangle = & \frac{ 1 }{\mathcal{D} \Lambda _m } \sum _{i=0}^{\mathcal{D}-1} \sum _{j=0}^{\mathcal{D}-1} \Big( \sqrt{  \langle \mathcal{F}_i', Q_i | \mathcal{H} | \mathcal{F} , P_i \rangle  }\Big)^{\ast} \sqrt{ \langle \mathcal{G}_j', Q_j | \mathcal{H} | \mathcal{G} , P_j \rangle } \delta _{\mathcal{F}, \mathcal{G}'_j} \delta _{P_i,Q_j} \delta _{\mathcal{F}_i', \mathcal{G}} \delta _{Q_i,P_j} \delta _{y_i,0} \delta _{y_j ,0} .
 \label{eq:matrix_element_QuaWalk}
\end{align}
We note that the swap $| r_6 \rangle \leftrightarrow | r_7 \rangle $ guarantees that only those terms with $|y_i \rangle = |y_j \rangle = |0 \rangle $ contribute; this swap operation eliminates the contributions from terms with other possible combinations of $|y_i \rangle $ and $|y_j \rangle $. The summation indices $i$ and $j$ enumerates the monomials in the second-quantized Hamiltonian [Eq. \eqref{eq:H_modified_H_prime}]. By observation, the above equation enumerates the contributions from two types of monomials (see notations in Eq. \eqref{eq:H_modified_XXXXL}):
\begin{enumerate}
\item $ b_{Q_i}^{\dag} b_{P_i} $, which corresponds to the kernel $  \langle  \mathcal{F}'_i | b_{Q_i}^{\dag} b_{P_i} | \mathcal{F} \rangle $ with $P_i \mapsto \{r_i, s_i \}$ ($r_i< s_i$) and $Q_i \mapsto \{p_i, q_i \}$ ($p_i< q_i$);

\item $ b_{Q_j}^{\dag} b_{P_j} $, which corresponds to the kernel $  \langle  \mathcal{G}'_j | b_{Q_j}^{\dag} b_{P_j} | \mathcal{G} \rangle $ with  $ P_j \mapsto \{r_j, s_j \} $ ($r_j< s_j$) and $Q_j \mapsto \{p_j, q_j \} $ ($p_j < q_j$).
\end{enumerate}
The delta functions in Eq. \eqref{eq:matrix_element_QuaWalk} (resulted from the orthonormality relations of the register states) enforce the conditions that $| \mathcal{F} \rangle = | \mathcal{G}'_j \rangle $, $|\mathcal{F}'_i \rangle = |\mathcal{G} \rangle $, and retain monomials that are conjugate transpose to each other, i.e., $ (b_{Q_i}^{\dag} b_{P_i})^{\dag} = b_{Q_j}^{\dag} b_{P_j} $ with $ P_i =Q_j $ and $Q_i = P_j$. Note that we define the mapping $P_i \mapsto \{r_i, s_i \}$ ($r_i< s_i$) and $Q_j \mapsto \{p_j, q_j \}$ ($p_j< q_j$) such that $P_i =Q_j$ if and only if $r_i=p_j$ and $s_i=q_j$.

For the trivial case with $\langle \mathcal{F} | H | \mathcal{G} \rangle = 0 $, then Eq. \eqref{eq:matrix_element_QuaWalk} produces $0$ in the right-hand side, which is as expected. As for the non-trivial case with $\langle \mathcal{F} | H | \mathcal{G} \rangle \neq 0 $, there exists at least one choice of $i \in \mathcal{I}_{\mathcal{F}} $ such that the right-hand side of Eq. \eqref{eq:matrix_element_QuaWalk} is non-vanishing. In this case, we can rewrite Eq. \eqref{eq:matrix_element_QuaWalk} as
\begin{align}
\langle \mathcal{F} | \langle 0| ( \mathcal{T} ^{\dag}S \mathcal{T} ) | \mathcal{G} \rangle | 0 \rangle = \frac{ 1 }{ \mathcal{D} \Lambda _m } \sum _{i=0}^{\mathcal{D}-1} \Big( \sqrt{  \langle \mathcal{G}, Q_i | \mathcal{H} | \mathcal{F} , P_i \rangle  }\Big)^{\ast} \sqrt{ \langle \mathcal{F}, P_i | \mathcal{H} | \mathcal{G} , Q_i \rangle } .  \label{eq:case_00}
\end{align}

It is also noteworthy that we need to choose an appropriate convention for the square-root operation, especially for the cases where $ { \langle \mathcal{F}, P_i | \mathcal{H} | \mathcal{G} , Q_i \rangle } $ has negative or complex values. In this case, we define 
\begin{align}
{ \langle \mathcal{F}, P_i | \mathcal{H} | \mathcal{G} , Q_i \rangle } = |{ \langle \mathcal{F}, P_i | \mathcal{H} | \mathcal{G} , Q_i \rangle } | \exp [ i \theta _{ \mathcal{F},P_i ; \mathcal{G},Q_i} ],
\end{align}
with $ \theta _{ \mathcal{F},P_i ; \mathcal{G},Q_i} = { \arg}[\langle \mathcal{F}, P_i | \mathcal{H} | \mathcal{G} , Q_i \rangle] \in (-\pi, \pi ] $. Then the quantity $\sqrt{ { \langle \mathcal{F}, P_i | \mathcal{H} | \mathcal{G} , Q_i \rangle } } $ can be uniquely defined as:
\begin{align}
\sqrt{ { \langle \mathcal{F}, P_i | \mathcal{H} | \mathcal{G} , Q_i \rangle } } =& \sqrt{| { \langle \mathcal{F}, P_i | \mathcal{H} | \mathcal{G} , Q_i \rangle } |} \exp [ i  \theta _{ \mathcal{F},P_i ; \mathcal{G},Q_i} /2 ] , \label{eq:def_sqrt}
\end{align} 
With the standard operation of Hermitian matrix elements
\begin{align}
\Big( \sqrt{ \langle \mathcal{G}, Q_i | \mathcal{H} | \mathcal{F} , P_i \rangle } \Big)^{\ast} = \sqrt{ \langle \mathcal{F} , P_i | \mathcal{H} | \mathcal{G}, Q_i \rangle  } ,
\end{align}
we have 
\begin{align}
\Big( \sqrt{ \langle \mathcal{G}, Q_i | \mathcal{H} | \mathcal{F} , P_i \rangle } \Big)^{\ast} \sqrt{ \langle \mathcal{F}, P_i | \mathcal{H} | \mathcal{G} , Q_i \rangle } = | { \langle \mathcal{F}, P_i | \mathcal{H} | \mathcal{G} , Q_i \rangle } |  \exp [ i  \theta _{ \mathcal{F},P_i ; \mathcal{G},Q_i} ] .
\end{align}
With the definition of the square-root operation [Eq. \eqref{eq:def_sqrt}], Eq. \eqref{eq:case_00} becomes
\begin{align}
\langle \mathcal{F} | \langle 0| ( \mathcal{T}^{\dag}S \mathcal{T} ) | \mathcal{G} \rangle | 0 \rangle = \frac{ 1 }{\mathcal{D} \Lambda _m } \sum _{i=0}^{\mathcal{D}-1} { \langle \mathcal{F}, P_i | \mathcal{H} | \mathcal{G} , Q_i \rangle }  = \frac{ 1 }{\mathcal{D} \Lambda _m } \langle   \mathcal{F} | H | \mathcal{G} \rangle , \label{eq:walk_identity}
\end{align}
where we have used the relation in Eq. \eqref{eq:equivalence_restore}. 
The summation in the above equation enumerates all the nonvanishing matrix elements $ { \langle \mathcal{F}, P_i | H' | \mathcal{G} , Q_i \rangle } $ which contribute to the Hamiltonian matrix element $ \langle \mathcal{F} | H | \mathcal{G} \rangle $.

\paragraph*{Case 2.}
For the cases with $(b,b')=(0,1)$ and $(1,0)$, analogous calculation yields
\begin{align}
\langle \mathcal{F} | \langle  b | \big( \mathcal{T}^{\dag} S \mathcal{T} \big) | \mathcal{G} \rangle |b' \rangle =0 \label{eq:walk_identity_prime} .
\end{align}

\paragraph*{Case 3.}
For the case with $(b,b')=(1,1)$, we can also calculate 
\begin{align}
\langle \mathcal{F} | \langle 1| ( \mathcal{T}^{\dag}S \mathcal{T} ) | \mathcal{G} \rangle | 1 \rangle = \delta _{\mathcal{F} ,0 } \delta _{\mathcal{G} ,0 } ,
\end{align}
with the application of Eq. \eqref{eq:alternative_b_1} and Eq. \eqref{eq:swap_operation}.

Combining {\it Case 1}, {\it Case 2}, and {\it Case 3}, we verify Eq. \eqref{eq:identity_Quantum_walk}.


\newpage










\begin{thebibliography}{100}

\bibitem{Carlson:1997qn}
J.~Carlson and R.~Schiavilla,
Rev. Mod. Phys. \textbf{70}, 743-842 (1998)
doi:10.1103/RevModPhys.70.743

\bibitem{Carlson:2014vla}
J.~Carlson, S.~Gandolfi, F.~Pederiva, S.~C.~Pieper, R.~Schiavilla, K.~E.~Schmidt and R.~B.~Wiringa,
Rev. Mod. Phys. \textbf{87}, 1067 (2015)
doi:10.1103/RevModPhys.87.1067
[arXiv:1412.3081 [nucl-th]].

\bibitem{Carlson:2017ebk}
J.~Carlson, M.~P.~Carpenter, R.~Casten, C.~Elster, P.~Fallon, A.~Gade, C.~Gross, G.~Hagen, A.~C.~Hayes and D.~W.~Higinbotham, \textit{et al.}
Prog. Part. Nucl. Phys. \textbf{94}, 68-124 (2017)
doi:10.1016/j.ppnp.2016.11.002

\bibitem{Feynman:1982fey}
R. P. Feynman, Int. J. Theor. Phys. {\bf 21}, 467 (1982). doi:10.1007/BF02650179

\bibitem{NielsenANDChuang:2001}
Michael A. Nielsen and Isaac L. Chuang, {\it Quantum Computation and Quantum Information} (Cambridge University Press, Cambridge, UK, 2001).

\bibitem{Dumitrescu:2018njn}
E.~F.~Dumitrescu, A.~J.~McCaskey, G.~Hagen, G.~R.~Jansen, T.~D.~Morris, T.~Papenbrock, R.~C.~Pooser, D.~J.~Dean and P.~Lougovski,
Phys. Rev. Lett. \textbf{120}, no.21, 210501 (2018)
doi:10.1103/PhysRevLett.120.210501
[arXiv:1801.03897 [quant-ph]].

\bibitem{Roggero:2020sgd}
A.~Roggero, C.~Gu, A.~Baroni and T.~Papenbrock,
Phys. Rev. C \textbf{102}, no.6, 064624 (2020)
doi:10.1103/PhysRevC.102.064624
[arXiv:2009.13485 [quant-ph]].

\bibitem{Kiss:2022kkz}
O.~Kiss, M.~Grossi, P.~Lougovski, F.~Sanchez, S.~Vallecorsa and T.~Papenbrock,
[arXiv:2205.00864 [nucl-th]].

\bibitem{Klco:2021lap}
N.~Klco, A.~Roggero and M.~J.~Savage,
Rept. Prog. Phys. \textbf{85}, no.6, 064301 (2022)
doi:10.1088/1361-6633/ac58a4
[arXiv:2107.04769 [quant-ph]].

\bibitem{Du:2020glq}
W.~Du, J.~P.~Vary, X.~Zhao and W.~Zuo,
Phys. Rev. A \textbf{104}, no.1, 012611 (2021)
doi:10.1103/PhysRevA.104.012611
[arXiv:2006.01369 [nucl-th]].

\bibitem{Baroni:2021xtl}
A.~Baroni, J.~Carlson, R.~Gupta, A.~C.~Y.~Li, G.~N.~Perdue and A.~Roggero,
Phys. Rev. D \textbf{105}, no.7, 074503 (2022)
doi:10.1103/PhysRevD.105.074503
[arXiv:2111.02982 [quant-ph]].

\bibitem{Stetcu:2021cbj}
I.~Stetcu, A.~Baroni and J.~Carlson,
Phys. Rev. C \textbf{105}, 064308 (2022)
doi:10.1103/PhysRevC.105.064308
[arXiv:2110.06098 [nucl-th]].

\bibitem{Romero:2022blx}
A.~M.~Romero, J.~Engel, H.~L.~Tang and S.~E.~Economou,
Phys. Rev. C \textbf{105}, 064317 (2022)
doi:10.1103/PhysRevC.105.064317
[arXiv:2203.01619 [nucl-th]].

\bibitem{Preskill:2018preskill}
John Preskill, 
Quantum {\bf 2}, 79 (2018).

\bibitem{Barrett:2013nh}
B.~R.~Barrett, P.~Navratil and J.~P.~Vary,
Prog. Part. Nucl. Phys. \textbf{69}, 131-181 (2013)
doi:10.1016/j.ppnp.2012.10.003

\bibitem{Navratil:2000ww}
P.~Navratil, J.~P.~Vary and B.~R.~Barrett,
Phys. Rev. Lett. \textbf{84}, 5728-5731 (2000)
doi:10.1103/PhysRevLett.84.5728
[arXiv:nucl-th/0004058 [nucl-th]].

\bibitem{Navratil:2000gs}
P.~Navratil, J.~P.~Vary and B.~R.~Barrett,
Phys. Rev. C \textbf{62}, 054311 (2000)
doi:10.1103/PhysRevC.62.054311

\bibitem{Aharonov:2003aha}
D. Aharonov and A. Ta-Shma, in {\it Proceedings of the 35th Annual ACM Symposium on Theory of Computing, STOC
`03} (Association for Computing Machinery, New York, 2003), pp. 20–29.

\bibitem{Childs:2003am}
A. M. Childs, R. Cleve, E. Deotto, E. Farhi, S. Gutmann, and D. A. Spielman, in {\it Proceedings of the Thirty-Fifth Annual ACM Symposium on Theory of Computing, STOC `03} (Association for Computing Machinery, New York, 2003), pp. 59–68.

\bibitem{Berry:2007dwb}
Dominic W. Berry, Graeme Ahokas, Richard Cleve, and Barry C. Sanders,
7 Commun. Math. Phys. {\bf 270} 359 (2007).

\bibitem{AMChilds:2009}
Andrew M. Childs,
Commun. Math. Phys. {\bf 294}, 581-603 (2010)
[arXiv:0810.0312 [quant-ph]].

\bibitem{DWBerry:2012}
Dominic W. Berry, Andrew M. Childs,
Quantum Inf. Comput. {\bf 12}, 29 (2012)
[arXiv:0910.4157 [quant-ph]].

\bibitem{AMChilds:2013}
D. W. Berry, A. M. Childs, R. Cleve, R. Kothari, and R. D. Somma,
in {\it Proceedings of the 46th Annual ACM Symposium on Theory of Computing} (Association for Computing Machinery, New York, 2014), p. 283. 
[arXiv:1312.1414].

\bibitem{Berry:2015prlDWB}
Dominic W. Berry, Andrew M. Childs, Richard Cleve, Robin Kothari, and Rolando D. Somma,
Phys. Rev. Lett. {\bf 114}, 090502 (2015).

\bibitem{Berry:2015IEEE}
D. W. Berry, A. M. Childs, and R. Kothari, in {\it 2015 IEEE 56th Annual Symposium on Foundations of Computer Science, Berkeley, CA} (IEEE, Piscataway, NJ, 2015), pp. 792–809.

\bibitem{Low:2017}
Guang Hao Low, and Isaac L. Chuang,
Phys. Rev. Lett. {\bf 118}, 010501
(2017).

\bibitem{Low:2019}
G. H. Low and I. L. Chuang,
Quantum {\bf 3}, 163 (2019).

\bibitem{Low:2018IntPic}
Guang Hao Low, and Nathan Wiebe,
arXiv:1805.00675 [quant-ph] (2018).

\bibitem{Berry:2020}
D. W. Berry, A. M. Childs, Y. Su, X. Wang, and N. Wiebe,
Quantum {\bf 4}, 254 (2020).

\bibitem{Kirby:2021ajp}
W.~M.~Kirby, S.~Hadi, M.~Kreshchuk and P.~J.~Love,
Phys. Rev. A \textbf{104}, no.4, 042607 (2021)
doi:10.1103/PhysRevA.104.042607
[arXiv:2105.10941 [quant-ph]].

\bibitem{Chakraborty:2018}
Shantanav Chakraborty, Andr\'as Gily\'en, and Stacey Jeffery,
In {\it Proceedings of the 46th International Colloquium on Automata, Languages, and Programming (ICALP 2019)}, pp. 33:1-33:14. [arXiv:1804.01973 [quant-ph]]

\bibitem{Lin:2022lectureNote}
Lin Lin,
{\it Lecture Notes on Quantum Algorithms for Scientific Computation} (2022), arXiv:2201.08309 [quant-ph].

\bibitem{Lipkin:1958zza}
H.~J.~Lipkin,
Phys. Rev. \textbf{109}, 2071-2072 (1958)
doi:10.1103/PhysRev.109.2071

\bibitem{Gloeckner:1974sst}
D.~H.~Gloeckner and R.~D.~Lawson,
Phys. Lett. B \textbf{53}, 313-318 (1974)
doi:10.1016/0370-2693(74)90390-6

\bibitem{Childs:LCU2012}
Andrew M. Childs, Nathan Wiebe,
Quantum Information and Computation {\bf 12} 901 (2012).

\bibitem{Maris:2012du}
P.~Maris, H.~M.~Aktulga, M.~A.~Caprio, U.~Catalyurek, E.~G.~Ng, D.~Oryspayev, H.~Potter, E.~Saule, M.~Sosonkina and J.~P.~Vary, \textit{et al.}
J. Phys. Conf. Ser. \textbf{403}, 012019 (2012)
doi:10.1088/1742-6596/403/1/012019

\bibitem{Abe:2021sky}
T.~Abe, P.~Maris, T.~Otsuka, N.~Shimizu, Y.~Utsuno and J.~P.~Vary,
Phys. Rev. C \textbf{104}, no.5, 054315 (2021)
doi:10.1103/PhysRevC.104.054315
[arXiv:2106.15114 [nucl-th]].

\bibitem{JW:1928}
P. Jordan and E. Wigner, Z. Phys. {\bf 47}, 631 (1928).

\bibitem{DAbrams:1997}
Daniel S. Abrams, Seth Lloyd,
Phys. Rev. Lett. {\bf 79} 2586 (1997).

\bibitem{Somma:2002}
R. Somma, G. Ortiz, J. E. Gubernatis, E. Knill, and R. Laflamme, 
Phys. Rev. A {\bf 65}, 042323 (2002).

\bibitem{Jensen:2017}
Morten Hjorth-Jensen, Maria Paola Lombardo, Ubirajara van Kolck,
{\it An Advanced Course in Computational Nuclear Physics: Bridging the Scales from Quarks to Neutron Stars (Lecture Notes in Physics, 936)},  1st edition, (Springer, 2017).

\bibitem{Kreshchuk:2020dla}
M.~Kreshchuk, W.~M.~Kirby, G.~Goldstein, H.~Beauchemin and P.~J.~Love,
Phys. Rev. A \textbf{105}, no.3, 032418 (2022)
doi:10.1103/PhysRevA.105.032418
[arXiv:2002.04016 [quant-ph]].

\bibitem{Childs:2010Lim}
Andrew M. Childs, Robin Kothari,
Quantum Information and Computation, {\bf 10}, 669 (2010).
[arXiv:0908.4398 [quant-ph]].

\bibitem{Babbush:2018bubbush}
Ryan Babbush, Dominic W. Berry, Yuval R. Sanders, Ian D. Kivlichan, Artur Scherer, Annie Y. Wei, Peter J. Love, and Al\'an Aspuru-Guzik,
Quantum Sci. Technol. {\bf 3} 015006 (2018).

\bibitem{Farhi:2001Science}
E. Farhi, J. Goldstone, S. Gutmann, J. Lapan, A. Lundgren and D. Preda, Science {\bf 292}, 472 (2001),
[arXiv:quant-ph/0104129].

\bibitem{Albash:2018RMP}
Tameem Albash and Daniel A. Lidar, 
Rev. Mod. Phys. {\bf 90} 015002 (2018).

\bibitem{Messiah:1962QM}
Albert Messiah, {\it Quantum mechanics: Volume II} (North-Holland Publishing Company Amsterdam, 1962).


\bibitem{Kitaev:1995}
A. Yu. Kitaev, 
arXiv:quant-ph/9511026 (1995).


\bibitem{Abrams:1998pd}
D.~S.~Abrams and S.~Lloyd,
Phys. Rev. Lett. \textbf{83}, 5162-5165 (1999)
doi:10.1103/PhysRevLett.83.5162
[arXiv:quant-ph/9807070 [quant-ph]].

\bibitem{Farhi:2011QiC}
E. Farhi, J. Goldstone, D. Gosset, S. Gutmann, H. B. Meyer, and P. Shor, 
Quantum Info. Comput. {\bf 11} 3 (2011). 
[arXiv: 0909.4766 [quant-ph]].

\bibitem{Wecker:2015pra}
D. Wecker, M. B. Hastings, N. Wiebe, B. K. Clark, C. Nayak, and M. Troyer, Phys. Rev. A {\bf 92}, 062318 (2015).

\bibitem{Du:2021ctr}
W.~Du, J.~P.~Vary, X.~Zhao and W.~Zuo,
[arXiv:2105.08910 [nucl-th]].

\bibitem{Guzik:2005Sci}
A. Aspuru-Guzik, A. Dutoi, P. Love and M. Head-Gordon, Science {\bf 309}, 1704 (2005).

\bibitem{GGC:2014}
Jin-Shi Xu, Man-Hong Yung, Xiao-Ye Xu, Sergio Boixo, Zheng-Wei Zhou, Chuan-Feng Li, Alán Aspuru-Guzik, and Guang-Can Guo,
Nature Photonics {\bf 8}, 113 (2014).

\bibitem{Choi:2020pdg}
K.~Choi, D.~Lee, J.~Bonitati, Z.~Qian and J.~Watkins,
Phys. Rev. Lett. \textbf{127}, no.4, 040505 (2021)
doi:10.1103/PhysRevLett.127.040505
[arXiv:2009.04092 [quant-ph]].

\bibitem{Qian:2021wya}
Z.~Qian, J.~Watkins, G.~Given, J.~Bonitati, K.~Choi and D.~Lee,
[arXiv:2110.07747 [quant-ph]].

\bibitem{Bee-Lindgren:2022nqb}
M.~Bee-Lindgren, Z.~Qian, M.~DeCross, N.~C.~Brown, C.~N.~Gilbreth, J.~Watkins, X.~Zhang and D.~Lee,
[arXiv:2208.13557 [quant-ph]].

\bibitem{DLeePC}
D. Lee, private communication. 

\bibitem{Mitarai:2019}
Kosuke Mitarai and Keisuke Fujii,
Phys. Rev. Research {\bf 1}, 013006 (2019).

\bibitem{Siwach:2022ugy}
P.~Siwach and P.~Arumugam,
Phys. Rev. C {\bf 105}, 064318
[arXiv:2206.08510 [quant-ph]].

\bibitem{Lin:2020Linlin}
Lin Lin and Yu Tong,
Quantum {\bf 4}, 372 (2020).

\bibitem{Chuang:2021chuangIssac}
John M. Martyn, Zane M. Rossi, Andrew K. Tan, and Isaac L. Chuang,
PRX Quantum {\bf 2}, 040203 (2021).

\bibitem{Dong:2022Linlin}
Yulong Dong, Lin Lin, and Yu Tong,
PRX Quantum {\bf 3}, 040305 (2022).

\bibitem{RBabbush:2016}
Ryan Babbush, Dominic W. Berry, Ian D. Kivlichan, Annie Y. Wei, Peter J. Love, and Al\'an Aspuru-Guzik,
New J. Phys. {\bf 18} 033032 (2016).

\bibitem{Vedral:1996}
V. Vedral, A. Barenco, and A. Ekert, Phys. Rev. A 54, 147 (1996).

\bibitem{Abhari:2014}
A. Javadi Abhari, S. Patil, D. Kudrow, J. Heckey, A. Lvov, F. T. Chong, and M. Martonosi, in {\it Proceedings of the 11th ACM Conference on Computing Frontiers, CF ’14} (Association for Computing Machinery, New York, 2014), pp. 1–10.

\bibitem{Wiebe:2011wbe}
Nathan Wiebe, Dominic W. Berry, Peter H{\o}yer, and Barry C Sanders,
J. Phys. A: Math. Theor. {\bf 44} 445308 (2011).

\bibitem{Hastings:2015hast}
M. B. Hastings, D. Wecker, B. Bauer, M. Troyer,
Quantum Inf. Comput. {\bf 15} 1 (2015).

\bibitem{Slater:1929}
J. C. Slater,
Phys. Rev. {\bf 34} 1293 (1929).

\bibitem{Condon:1930}
E. U. Condon,
Phys. Rev. {\bf 36} 1121 (1930).

\bibitem{Babbush:2018BasisChoice}
Ryan Babbush, Nathan Wiebe, Jarrod McClean, James McClain, Hartmut Neven, and Garnet Kin-Lic Chan
Phys. Rev. X {\bf 8}, 011044 (2018).

\end{thebibliography}
\end{document}